\documentclass[traditabstract]{aa}

\usepackage{graphicx}
\usepackage{txfonts}
\usepackage{longtable}
\begin{document}
   \title{The CN isotopic ratios in comets
   \thanks{Figures \ref{plotdeV}-\ref{plottuttle}
 and Table~\ref{kstars} are only available in electronic form via
             http://www.edpsciences.org}%
\fnmsep\thanks{%
Based on observations made with ESO Telescopes 
at the La Silla Paranal Observatory under programmes ID
268.C-5570,
270.C-5043, 073.C-0525, 274.C-5015
and 075.C-0355(A)}}

\author{J.~Manfroid
          \inst{1}\fnmsep\thanks{JM is Research Director FNRS, 
EJ is Research Associate FNRS and DH is Senior Research Associate FNRS}
          \and
E.~Jehin
\inst{1}\fnmsep{$^{\star\star}$}\and
D.~Hutsem\'ekers 
\inst{1}\fnmsep{$^{\star\star}$}\and
A.~Cochran 
\inst{2}\and
J.-M.~Zucconi
\inst{3}\fnmsep%
\thanks{Deceased on 14 May 2009.}
\and
C.~Arpigny     
 \inst{1}\and
R.~Schulz  
 \inst{4}\and
J.A.~St\"uwe  
\inst{5}\and
I.~Ilyin
 \inst{6}
}

   \offprints{J. Manfroid}

\institute{Institut d'Astrophysique et de G\'eophysique,
Universit\'e de Li\`ege, All\'ee du 6 ao\^ut 17, B-4000 Li\`ege
\and
Department of Astronomy and McDonald Observatory, University 
of Texas at Austin, C-1400, Austin, USA
\and
Observatoire de Besan\c{c}on, F25010 Besan\c{c}on Cedex, France
\and
ESA/RSSD, ESTEC, P.O. Box 299, NL-2200 AG Noordwijk, 
The Netherlands
\and
Leiden Observatory, NL-2300 RA Leiden, The Netherlands
\and
AIP, An der Sternwarte 16, D-14482 Potsdam, Germany
}

   \date{}

 
  \abstract{%
Our aim is to determine the isotopic ratios {$^{12}$C/$^{13}$C}{}\ and {$^{14}$N/$^{15}$N}{}\ in a variety of
comets and link   these measurements to the formation and evolution of the solar
system. 
The {$^{12}$C/$^{13}$C}{}\ and {$^{14}$N/$^{15}$N}{}\ isotopic ratios are measured { for} the CN radical
by means of high-resolution optical spectra 
of the R branch of the  B-X (0,0) violet band. { 23 comets from different dynamical classes}
have been observed, sometimes at various heliocentric
{ and nucleocentric}
distances, in
order to estimate possible variations of the isotopic ratios in parent molecules.
The {$^{12}$C/$^{13}$C}{}\ and {$^{14}$N/$^{15}$N}{}\ isotopic ratios in CN are remarkably  constant
(average values of, respectively, $91.0\pm3.6$ and $147.8\pm5.7$) 
within our measurement errors, for all comets whatever their
origin or heliocentric distance. 
While the carbon isotopic ratio does agree with the terrestrial value (89),
the nitrogen ratio is a factor of two lower than the terrestrial
value (272), indicating a fractionation in the early solar system, 
or in the protosolar nebula, common
to all the comets of our sample.  This points towards a common
origin of the comets 
independently of their birthplaces,
and   a relationship between HCN and CN.
} 

   \maketitle
%

\section{Introduction}
\label{sec:intro}

Comets are believed to have condensed in the outer solar nebula
and to contain relatively unaltered
material from that period. The analysis of the composition
of such material and particularly the determination of isotopic 
ratios should answer questions regarding the origin and 
the nature of the solar nebula.

This task is  challenging.
In-situ measurements in the coma with mass spectrometers are difficult, 
because  
heavy isotopes can be masked by hydrides,
e.g., $^{13}$C and  $^{12}$CH, $^{15}$N and $^{14}$NH. 
Values exist from in-situ measurements for
the dust coma of comet 1P/Halley, for which the carbon isotopic ratio, {$^{12}$C/$^{13}$C}{},     
varied by orders of magnitude from one grain to another (Solc et al.\ \cite{solc}; 
Jessberger and Kissel \cite{1991ASSL..167.1075J}) -- but no data 
were available for nitrogen.

Isotopic ratios can be determined remotely by high-resolution
spectroscopy of molecular bands.
The emission
lines of rare isotopes are weak
and they  have to be
distinguished from emission lines of other isotopes, other species (e.g.,
NH$_2$ in the case of the C$_2$ Swan bands)
and from the background.
Likewise, blends are possible at
sub-millimetre range: e.g., SO$_2$ blends with the H$^{13}$CN (4-3)
line (Lis et al.\ \cite{1997Icar..130..355L}).
Hence, high
spectral resolution and high signal-to-noise ratios are needed.
The accuracy of these determinations also depends
on the models
representing the emission by the various
isotopologues, as well as other species appearing
in the same domain.

Up to now, isotopic ratios have been published
for H, C, N, O and S (Jehin et al.\ \cite{jehinemp}).
The Rosetta mission to 67P/Churyumov-Gerasimenko will
provide a detailed appraisal of the nature and isotopic composition
of { the material} present at the surface of one comet. 
The present work addresses the C and N ground-based determinations
that can now be routinely obtained { for}  moderately bright comets.

\section{Historical background}
\label{sec:historical}

For carbon and nitrogen isotopes, the first observed molecules were
CN and C$_2$ in the optical and HCN in the sub-millimetre range. 
The first determinations of the {$^{12}$C/$^{13}$C}{}\ ratio were obtained for bright comets
from 
the {$^{12}$C}{}{$^{13}$C}{}\ (1,0) Swan bandhead at 4145~\AA\ (Stawikowski \& Greenstein \cite{stawi};
Owen \cite{owen}). 
This feature was chosen because
it is clearly separated from the corresponding {$^{12}$C}{}{$^{12}$C}{}\ 
bandhead. However, these
measurements
were affected by blending with NH$_2$ emission.
The cleaner CN B-X (0-0) band was used for 
the first time for comet 1P/Halley (Wyckoff \& Wehinger  \cite{wyckoffa}, 
Wyckoff et al.\ \cite{wyckoffc},
Kleine et al.\ \cite{kleine}), resulting in positive determinations
of {$^{12}$C/$^{13}$C}{}. 
{ By 2000, estimates of the  {$^{12}$C/$^{13}$C}{}\  ratio were   available for ten        
comets.} 
Although the same CN band lends itself to the determination of the nitrogen ratio,
only lower limits were obtained.

The nitrogen isotopic ratio {$^{14}$N/$^{15}$N}{}\  was measured for the first 
time in comet C/1995 O1 (Hale-Bopp) in 1997 in the HCN 
as well as the CN band. 
{ The {$^{12}$C/$^{13}$C}{}\ ratio was determined simultaneously. 
The {$^{12}$C/$^{13}$C}{}\ and {$^{14}$N/$^{15}$N}{}\ values derived from 
the HCN band at sub-millimetre range 
(respectively $111\pm12$ and $323\pm46$, Jewitt et al. \cite{jewitt};
$109\pm22$ and $330\pm98$,
 Ziurys et al. \cite{ziurys}),
were consistent with the telluric values (89, 272, Anders and Grevesse \cite{anders}), although with large 
uncertainties.
The {$^{12}$C/$^{13}$C}{}\  values derived from CN ({$^{12}$C/$^{13}$C}{}$=100\pm35$) were terrestrial, but
the nitrogen isotopic ratio  
({$^{14}$N/$^{15}$N}{}$=140\pm35$)
was widely discordant with the terrestrial and HCN values (Arpigny et al.\ \cite{arpi}, \cite{arpib}).
}

Examination of published spectra of several other bright comets
showed that {$^{12}$C$^{15}$N}{}\ was probably 
{ also} overabundant. The unexpected strength of these lines
was a reason for discarding  them as {$^{12}$C$^{15}$N}{}\
(see, e.g., features labelled k,m and o in the Halley spectrum
displayed in Fig.~5 of Kleine et al.\ \cite{kleine},
or lines c and d in the C/1990 K1 spectrum in Fig.~8 of Wyckoff et al.\ \cite{wyckoff}).

This discrepancy between CN and HCN, a presumed parent,
was eventually solved  with
the observation of CN and HCN in comet 17P/Holmes 
and the reanalysis of the  Hale-Bopp sub-millimeter data
(Bockel\'ee-Morvan et al.\ \cite{bockelee})
establishing that HCN 
has the same non-terrestrial nitrogen isotopic composition as CN.

The seemingly conflicting results and
the evidence for an anomalous value of {$^{14}$N/$^{15}$N}{}\  had led us to  initiate
an observing campaign 
with the UVES high-resolution
spectrograph (Dekker et al.\ \cite{2000SPIE.4008..534D}) 
mounted on the 8m Kueyen telescope of the ESO VLT
(Chile),
in order to gather data on different comets
presenting a variety of origins and
physical conditions. 
The CN  $B^2 \Sigma
^+-X^2 \Sigma ^+$ (0,0) violet band (near 3880 \AA ) was our main target.

Independently, in the Northern hemisphere, 
observations of many comets  have been made since 1995 
with the { 2DCoud{\'e}} high-resolution spectrograph of the 2.7 m
telescope of the McDonald Observatory (Tull et al.\ \cite{1995PASP..107..251T}) 
and contributed a large proportion of our
CN data. 

\section{Our sample of comets and previous results}
\label{sec:sample}
So far, { 23 comets} have been observed
(counting the fragments of 73P/Schwassmann-Wachmann 3 as one comet).
They are listed in Table~\ref{obscom}, with basic orbital characteristics
including   
the Tisserand parameter $T_{\rm J}$ relative to Jupiter :
\begin{equation}
T_{\rm J}=a_{\rm J}/a+2 \sqrt{(1-e^2)\,a/a_{\rm J}}\, \cos(i).
\end{equation}
$a$ and $a_{\rm J}$ are the semi-major axes of the orbits of the 
comet and Jupiter,
$e$ the eccentricity and $i$ the inclination of the comet's orbit. 
The Tisserand parameter is useful in comet taxonomy. 
Several classifications have been elaborated. 
We give in Table~\ref{obscom} the classifications according to Levison (\cite{levison}) 
and Horner et al. (\cite{horner}).  
The former proposes a major division at $T_{\rm J}=2$ between isotropic and ecliptic orbits.   
Further subdivision in our sample leads to Halley-type { ($T_{\rm J}<2$ and $a<40$~AU),
 external ($T_{\rm J}<2$ and  $40<a<10000$), new ($T_{\rm J}<2$ and $a>10000$)} 
and Jupiter-family ($2<T_{\rm J}<3$) comets. 
{ The comets of the first three groups probably come from the Oort cloud.
The origin of the JF comets is less certain.}
In the Horner at al.\ scheme, for comets with perihelion { closer  than} 4 AU, 
the  classification 
reduces to four major types, Encke-type (E), short-period (SP), intermediate-period (I)
and long-period (L), corresponding to aphelion divisions at 4, 35 and 1000 AU.
The objects are further differentiated according to the Tisserand parameter,
with class I to IV defined by the boundaries 2.0, 2.5 and 2.8.

Our sample does not contain Levison's Encke-type 
objects (defined as $T_{\rm J}>3$, $a<a_{\rm J}$).
Considering Horner's classification, SP$_{\rm II}$ is missing 
and SP$_{\rm III}$ is barely realized by the fragments of 73P/SW-3.
The sample shows a majority of the 
long period objects (according to Horner's scheme)  
divided equally in external and new comets (Levison's scheme). 
There are two comets 
from the Halley family (SP$_{\rm I}$) and 
{ four from the Jupiter family (one SP$_{\rm III}$ and three SP$_{\rm IV}$),} 
considering the two fragments of  73P as a single member.

We could not derive isotopic abundances for all 23 observed comets
mainly because { of the insufficient} signal-to-noise ratio of the CN band,
but 18 comets yielded positive measurements of both C and N ratios. 
Results have already been published partially for 10 of them.
Our first target for UVES was comet C/2000 WM1 (LINEAR)
which showed isotopic ratios { compatible with} those we had
found in Hale-Bopp (Arpigny et al. \cite{arpi}).
This isotopic anomaly 
and the discrepancy with the N ratio determined from HCN 
in comet Hale-Bopp (Jewitt et al. \cite{jewitt}, Ziurys et al. \cite{ziurys})
{ led} us to suggest the existence of parent(s) of
CN other than HCN, with an even lower N isotopic ratio. Organic compounds like those 
found in interplanetary dust particles were proposed as possible candidates
(Arpigny et al. \cite{arpib}). 

Our analysis of data gathered at McDonald observatory on 
122P/de Vico
and 153P/Ikeya-Zhang 
yielded isotopic ratios similar to
Hale-Bopp and C/2000 WM1 (LINEAR) (Jehin et al.\ \cite{jehin}). 

Most observations were done at small heliocentric distances since this 
yields the { brightest 
$m_r=m-5 \log (\Delta)$, i.e., the magnitude that would be seen               
from a distance of 1 AU} -- 
a good indication of
the flux entering the slit of the spectrometer. 
However, we gathered data on comets over a wider range of heliocentric distance
in the hope of detecting variations of the isotopic ratios linked to 
different abundances of the parent species.
It is likely that at large distances, the CN radical originates mainly 
from the photodissociation of HCN (see, e.g., Rauer et al.\
\cite{2003A&A...397.1109R}).
The  isotopic ratios were     determined
in comets C/1995 O1 (Hale-Bopp), C/2001 Q4 (NEAT) and
C/2003 K4 (LINEAR) at  heliocentric distances of, respectively,
2.7, 3.7 and 2.6 AU (Manfroid et al.\ \cite{manfroid}) as well as
close to  1 AU.
No significant differences were found { 
as a function of heliocentric distance, and the values were consistent with those}
obtained for other comets.               
The apparent discrepancy between the nitrogen isotopic ratio 
measurements { in} CN and HCN over such a  range of { irradiation} conditions seemed to
rule out HCN as a major parent
of the cometary CN radicals.

The isotopic 
ratios were then determined for the first time in a Jupiter-family (SP$_{\rm IV}$) comet,
88P/Howell, 
and in the chemically peculiar Oort Cloud comet
C/1999 S4 (LINEAR) (Hutsem\'ekers et al.\ \cite{hutsemekers}). 
The carbon and nitrogen isotopic ratios agreed       
within the uncertainties with those already obtained.

The Deep Impact mission to comet 9P/Tempel~1 provided 
an  extraordinary opportunity to observe from the 
ground the cometary material coming from relatively deep (a few meters) layers 
of the nucleus. The impact 
resulted in the release of sub-surface material 
from the comet nucleus (A'Hearn et al. \cite{2005Sci...310..258A}), 
which formed a jet structure that 
expanded within the coma and was observable for several days 
after the impact. Observations of the activity and 
composition of the comet after the impact showed that the 
new material was compositionally different 
from that seen before impact and that the mass ratio of dust 
to gas  in the ejecta was much larger than before
(Meech et al.\ \cite{2005Sci...310..265M}).
This suggested that the isotopic abundances could be different too.

Observations of comet 9P/Tempel 1 were carried out before,
during, and after the collision with the
optical spectrometers UVES and HIRES mounted on the Kueyen telescope of
the ESO VLT (Chile) and {the Keck I telescope} 
on Mauna Kea (Hawaii), respectively
(Jehin et al.\ \cite{jehina}). 
They show that the material
released by the impact probably has the same carbon and nitrogen isotopic
composition as the surface material, once again in line with the values
derived so far.           

Whether the material came from layers deep 
enough to be unaffected by space weathering remains unknown. 
If indeed the Deep Impact event led to the release of pristine 
material representative of the unaltered matter preserved in the interior 
of a comet nucleus since its formation, the measurement of 
the same isotopic composition in CN before and after impact 
would provide evidence for the {$^{12}$C/$^{13}$C}{}\    
and {$^{14}$N/$^{15}$N}{}\  isotopic ratios 
being primordial
in the parent molecule that produces CN. 
However, infrared measurements obtained before and 
after impact showed an enhanced C$_2$H$_6$/H$_2$O ratio, but an 
unchanged HCN/H$_2$O ratio relative to the quiescent (pre-impact) 
source (DiSanti \& Mumma \cite{2008SSRv..tmp...70D}). 
It may therefore be that non-altered CN-bearing material had not been reached by the impact.
{ However, the processes that  could affect the isotopic ratios
are not the same as those affecting the relative amounts of volatiles
near the surface. It is also possible that the nucleus is not
differentiated in HCN.}

The 2007 outburst of the Jupiter Family comet 17P/Holmes allowed us 
to reconcile the HCN and CN data 
(Bockel\'ee-Morvan et al.\ \cite{bockelee}).
Observation of $J=3-2$ rotational lines
of H{$^{12}$C$^{14}$N}{}, H{$^{12}$C$^{15}$N}{}\ and H{$^{13}$C$^{14}$N}{}\ { yielded} {$^{14}$N/$^{15}$N}{}$=139\pm 26$ while the measurements
of the CN violet band { gave}  {$^{14}$N/$^{15}$N}{}$=165\pm 40$.  
Moreover it appeared that the HCN
ratios that had been published for Hale-Bopp were seriously overestimated. 
A reanalysis of the data { gave}  $138<${$^{14}$N/$^{15}$N}{}$<239$, and {$^{12}$C/$^{13}$C}{}$=94\pm8$.
The isotopic ratios derived from CN and HCN are thus compatible within 
the error bars, and 
in line with those we measured in other comets.

In the following sections, we  
present a homogeneous compilation of the
whole data set and revisit the previous results. 
Isotopic ratios of 8 comets are given for the first time.
This completely
new analysis uses, in particular, the latest
fluorescence model including a more realistic handling of the collisional effects.

\section{Observations}
\label{sec:observations}

\begin{figure}
   \resizebox{\hsize}{!}{\includegraphics[angle=0]{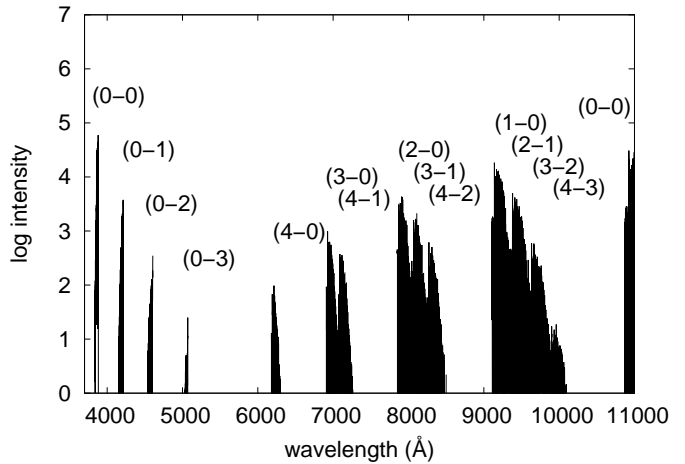}}
      \caption{Synthetic spectrum of {$^{12}$C$^{14}$N}{}{} over the whole optical domain ($r=1$~AU, $\dot{r}=0$). 
    The $y$-axis
     shows the decimal logarithm of the relative intensities { (arbitrary units)}.
   The blue bands below 5000 \AA\ belong to the  B-X system, 
   the red bands to the A-X system.
 The model is described in  Section~\ref{appModel}.}
         \label{plotentire}
  \end{figure}

{ In order to study the carbon and nitrogen isotopes in cometary spectra, we chose to observe the (0,0) violet band (B-X) of CN.} 
Figure \ref{plotentire} shows a typical (synthetic) 
{$^{12}$C$^{14}$N}{}\ spectrum over the whole optical domain. 
The spectra of the heavier isotopologues have very similar distributions.
The red bands are sparser than the blue ones, so the 
actual flux emitted in each band cannot be easily inferred from
the figure. They are given in Table~\ref{tableflux}.
Most of the optical emission occurs in the violet
(0-0) band at 3880~\AA\ and in the red 
($v^{\prime}$-$v^{\prime\prime}=1$) 
bands
between 9000 and 10000~\AA. 
However, the red spectrum is more dispersed over
many more lines.

\begin{table}
\caption{\label{tableflux} Relative flux emitted in the various
 CN bands shown in Fig.~\ref{plotentire}}
\vspace{.5cm}
\begin{tabular}{cr} 
\hline\hline\rule{0mm}{3ex}
$v^{\prime}-v^{\prime\prime}$&Flux\\
\hline\rule{0mm}{3ex}
0 &   0.41717    \\
--1 & 0.02996    \\
--2 & 0.00192    \\
--3 & 0.00007    \\
4 &   0.00113    \\
3 &   0.01697    \\
2 &   0.11758    \\
1 &   0.41520   \\
\hline
\end{tabular}
\end{table}

To better appreciate which lines are the most useful, the brightest domains of 
the spectra of the three species { ({$^{12}$C$^{14}$N}{},{$^{13}$C$^{14}$N}{}\ and {$^{12}$C$^{15}$N}{})} are shown in 
Figs.~\ref{plotCN3863_3871}--\ref{plotCN9221_9229}.

Comparison with the atlas of comet de Vico (Cochran \& Cochran \cite{2002Icar..157..297C})
shows several of these isotopic lines among the unidentified features.

The B-X (0,0) band appears to be the most interesting
for our purposes.
The lines are strong, and at that wavelength the 
solar spectrum is relatively weak. 
The red A-X bands 
are affected by emission from other molecules (mainly NH$_2$ and C$_2$ -- see, e.g., the de Vico atlas),
a stronger solar background  and
complex absorption by atmospheric features (H$_2$O, O$_2$).
The B-X (0,1) band at 420 nm 
is also fainter and, hence, less useful for our purposes.
Ideally, all these bands could be included in a global analysis
to provide the best estimate of the isotopic ratios as well as the
physical parameters in the coma but, in the present work, 
we limit ourselves to the bright (0-0) violet band.

High-resolution spectra 
are needed because the faint emission of the rare species 
are close to the strong lines { coming} from the abundant ones and because the best
contrast to   the underlying dust-scattered solar continuum must be
obtained.
Only the R branch is of interest because the wavelength shifts in the P branch are 
small compared to the typical spectral resolution and 
Doppler broadening of the lines.

The echelle spectrographs used in the present study were 
\begin{itemize}
\item The UVES spectrograph at the Nasmyth focus of the 8m Kueyen telescope (ESO/VLT, Paranal);
\item SOFIN, the Cassegrain spectrograph at the 2.56m NOT telescope (La Palma, Spain);
\item The 2dCoud\'e echelle spectrograph at the 2.7 m Harlan F. Smith telescope of the 
McDonald Observatory (Texas);
\item {The} HIRES spectrograph of the 10m Keck I telescope (Hawaii).      
\end{itemize}

The circumstances of the various observing runs are summarized in Table~\ref{observcirc}, while the
data relative to individual spectra are given in Table~\ref{kstars}. 

\section{Data reduction}
\label{sec:reduction}

We used the echelle package of the IRAF software (NOAO) to calibrate
the spectra and to extract the relevant order(s) of the CN violet band. 
In the case of the VLT spectra, the UVES configuration is such 
that the CN band appears in 
two orders. 
{ The SOFIN and 2DCoud{\'e} spectra do not suffer from the
same problem : all of the band is in a single order. 
The UVES }
wavelength calibration was done on the CN lines themselves
in order to avoid possible shifts between
the science spectra and calibration lamp spectra. Indeed, the UVES
calibration errors measured between the CN and lamp solutions { show} 
an rms scatter of 0.66~km~s$^{-1}$.

The echelle orders were combined using an optimal scheme 
taking into account the 
low levels expected for the isotopic lines -- i.e., a regime where the
signal-to-noise ratio is directly proportional to the signal
-- and simultaneously correcting the echelle ripple (blaze function).
The wavelength domain is very small, $\sim10$~\AA, so that a normalization is
done in such a way that flat-field spectra extracted in the same
conditions are constant. 
This procedure yields  the ``flat-field normalized" 
spectrum $o_{\rm ff}$ and 
the S/N ratio expected for a low-level constant signal, $f(\lambda)$. 
The S/N ratios of faint lines of intensity $I$ can be directly compared using 
\begin{equation}
\frac{{\rm SN}_i}{{\rm SN}_j}=\frac{I_i f(\lambda_i)}{I_j f(\lambda_j)}\,.
\label{eqSN1}
\end{equation}
Alternatively, a ``S/N-normalized'' spectrum $o_{\rm sn}(\lambda)$ 
can be produced, such that the 
intensity of faint lines is directly proportional to their S/N ratio
\begin{equation}
o_{\rm sn}(\lambda)=o_{\rm ff}(\lambda) f(\lambda)\,.
\label{eqSN}
\end{equation}

  \begin{figure}
   \resizebox{\hsize}{!}{\includegraphics[angle=0]{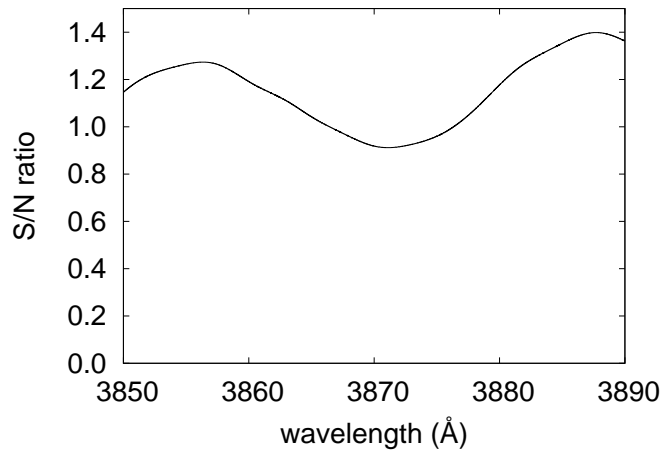}}
      \caption{Variations of of the signal-to-noise ratio $f(\lambda)$ across the UVES
      CN spectra  for low-level signals.}
         \label{plotSNnorm}
  \end{figure}

A typical $f(\lambda)$ is presented { in} Fig.~\ref{plotSNnorm}. The
minimum falls on the CN band, in the overlapping region between two echelle orders, 
and corresponds approximately to a $2^{-1/2}$ loss.
Unfortunately, changing the relevant UVES settings so as to bring
the CN R band in the middle of one order
was not possible.

The dust-reflected sunlight and, in some  cases, the
twilight contamination and the lunar sky background 
underlying the cometary emissions, were removed from the $o_{\rm ff}$
by fitting a solar reference spectrum 
to each of these components. Moon or twilight sky spectra were often used
for this purpose as they directly provided correct line profiles, 
but other good high-resolution solar spectra can be used, 
such as the UVES reference solar spectrum produced by ESO
or the atlas by Delbouille et al.~(\cite{delbouille}), available online
at the BASS2000 site (http://bass2000.obspm.fr/solar\_spect.php).
 
Appropriate Doppler shifts had to be applied
since, at the resolution of the echelle spectra, the observer, the Moon, the Sun and the comet
have non-negligible relative velocities. 

\begin{itemize}
\item The Moon spectrum  and the lunar sky background 
show a radial velocity equal to the sum of the heliocentric
and topocentric velocities of the Moon;

\item The dust-reflected sunlight
shows a radial velocity equal to the sum of the heliocentric
and topocentric velocities of the comet;  

\item The twilight component is shifted by the topocentric
velocity of the Sun;   

\item The cometary emission is   
shifted by the topocentric velocity of the comet.
   
\end{itemize}

The relevant kinematic data were retrieved from the JPL Horizon ephemeris generator. 
Because of the Earth rotation, the topocentric velocities show 
the most rapid variations. Generally, the use of average 
values is sufficient even when correcting the longest exposures. 
However, the
variations of the twilight are so fast that 
the Doppler shift corresponds to a slightly different time 
than the other components (earlier in the evening, later in the morning).
The spectra of the Moon rising or setting during long exposures
sometimes needs similar adjustments. 

The transfer functions (line profiles) of those components are not identical. 
The profiles of the various solar components were fitted by convolution
with a Gaussian kernel. This transformation was generally negligible, 
except for widely discordant resolution (e.g., some SOFIN spectra of Hale-Bopp
with too long exposures at too low elevations in the Cassegrain focus).
The intensity fit was done by
comparing the intensity of the solar photospheric lines in each component.

After removal of the dust, twilight and moonlight spectra, no solar
feature subsists. The remaining contamination consists
of an extremely weak quasi-featureless continuum due to 
a residual of improperly subtracted  scattered light in the instrument, and to the
underlying sky background.
This continuum is fitted  by  a spline  through the 
emission-free intervals of the spectrum and is subtracted from the spectrum. 

The signal levels on the CCD amount at best to only a few electrons
per pixel above the bias level, except in the major $R$ lines of the
abundant species where it is typically  $10^2$ to $10^3$ electrons.

The linearity of the spectrographs is very good,
and well maintained. For instance
at ESO, the linearity of the UVES detector between 0 and $10^5$~ADU is guaranteed
to within about 1\% by the frequent verifications performed
by the staff. 
Whenever possible, however,
we checked the linearity directly
on our data by comparing cometary lines and stellar spectra 
appearing in overlapping 
orders with vastly different sensitivities.
The comparison of specific line ratios to their theoretical values 
yields the same conclusion.

\section{Fluorescence model of CN}
\label{appModel}

The structure of the CN radical and the details of the fluorescence
calculations are presented elsewhere 
(Zucconi \& Festou, \cite{zucconi})
and will
not be repeated here. Instead, we will only focus on some features and
on their improvements.

The CN fluorescence is very sensitive to the Swings
 effect (Swings \cite{swings}) and the
computation requires accurate line wavelengths and solar fluxes.
Wavelengths of the $^{12}$C$^{14}$N B-X system (the violet system)
have been computed using the spectroscopic constants of 
Ram et al.\ (\cite{ram}), leading to an accuracy
better than 1~m\AA. On the other hand we have used the Kurucz atlas
(Kurucz \cite{kurucz})  to evaluate the solar intensities in the violet
bands. 

{ The influence of the A-X system on the shape of the violet
bands cannot be neglected because the pumping via this system also
affects the population of the ground state levels.}

Unfortunately, we do
not have high-resolution absolute spectra of the Sun in the red and near
infrared.
Ground-based atlases contain many strong water absorption
lines and are { unsuitable}. Hence           we decided to use the solar
fluxes of 
Labs \&  Neckel (\cite{labs}).
Although the Swings effect is thus
ignored in the red bands, we obtain a better fit of the spectra than
when using the Kurucz atlas in the red and near infrared. This could
be due to the fact that there is a large number of CN lines in those
bands, minimizing the Swings effect.
There is no extensive laboratory study of the violet bands of
$^{13}$C$^{14}$N and $^{12}$C$^{15}$N and only the most intense bands
have been observed. We derived the isotope line
wavelengths from the $^{12}$C$^{14}$N lines using the isotopic shift
formula. The predicted wavelengths agree with the faint isotope lines
observed in our spectra. We also checked that the computed
wavelengths agree with the available data for the ground vibrational
state (H\"ubner et al.\ \cite{hubner}).

Dipole moment, radiative lifetimes and Franck-Condon factors for the
violet bands are those listed in Zucconi \&  Festou (\cite{zucconi}) and we
adjusted the vibrational transition rate and the 
lifetime $\tau_{\rm A}$ of the A
state to get the best fit of our observations.
The value adopted for $\tau_{\rm A}$ $(v^{\prime}=0)$, $10^{-5}$~s, 
is very close to
the average between various experimental measurements and theoretical
estimates (e.g., Lu et al.\ \cite{1992ApJ...395..710L}, 
Bauschlicher et al.\ \cite{Bauschlicher}, and references
therein). It is also in excellent agreement with the value based on a
recent analysis of the solar spectrum (Sauval et al.\ 2009).

Because we probe the central part of the coma, collisions 
{ may play a
non negligible role} and we had to include this
effect in our model. Ideally this would lead to 
{ consideration of} the density
variation in the field of view and        the dependence of
collisional cross sections as a function of the relative velocity of
the colliders and of the rotational quantum number. 

The general formula expressing the transition probability between two
levels is given by
\begin{equation}
g_l C_{l\rightarrow u} = g_u C_{u\rightarrow l} \exp(-(E_u-E_l)/kT)
\label{eqtransprob}
\end{equation}
where $u$ and $l$ denote the upper and lower levels of the transition
respectively. $g_u$ and $g_l$ are the statistical weights of the levels
and $E_u$ and $E_l$ are their corresponding { energies}.

In order to reduce the number of free parameters in the model, we have
chosen to use a constant transition probability $Q\equiv C_{u\rightarrow l}$ between two
rotational states. We have also assumed that the collisional 
interactions are mainly dipolar, so
that $Q$ is non zero only for $\Delta J=0,\pm1$. 
We have found that the
best fit is  obtained with $Q$ independent of $\Delta J$.
It should be noted that
the temperature $T$ appearing here is not the thermodynamic
temperature of the medium because the CN radical is not in 
{ 
thermal equilibrium but rather in a
statistical equilibrium involving both radiative and collisional
processes. $T$ should be taken as a temperature characterizing the
velocity distribution of the colliding particles responsible for
the rotational excitation of the radicals. It is similar to the
kinetic temperatures defined in studies of the interstellar medium
or of gaseous nebulae.}

The ``average" temperatures required to achieve the best agreement
between observed and synthetic spectra are appreciably higher than those
indicated in the interpretation of radioastronomical spectra (20--100~K
near 1~AU from the Sun) of the mother molecules H$_2$O, CH$_3$OH, CO in several
comets (Bockel\'ee-Morvan \& Crovisier \cite{1987A&A...187..425B}, 
Bockel\'ee-Morvan et al.\ \cite{1994A&A...287..647B},
Biver et al.\ \cite{2002EM&P...90....5B}). The difference may be related to the fact that we
are dealing here with daughter products, which are formed at greater
nucleocentric distances in the coma. We note further that collisions
with electrons may play a role in the excitation of the lower rotational
levels of CN, as suggested in the case of water by Xie \& Mumma (\cite{1992ApJ...386..720X})
and of methanol by Bockel\'ee-Morvan et al.\ (\cite{1994A&A...287..647B}) . The cross-sections
for rotational excitation of CN by electron impact are indeed quite
appreciable at low energies (Allison \& Dalgarno \cite{1971A&A....13..331A}).
{ We note that ions (essentially H$_3$O$^+$ and H$_2$O$^+$, the most abundant ones)
might in principle also contribute to excite CN rotationally. However,
their relevant cross-sections, although expected to be rather high
(Xie \& Mumma (\cite{1992ApJ...386..720X}), are unknown and the corresponding excitation rates
 would anyhow be appreciably smaller than those of electrons because of
 the much larger mass of the ions and their consequently $\sim200$ times lower
 velocities.}

A study under way will take into consideration the
structure of the coma, in particular the distributions with 
nucleocentric distance of the
densities, temperatures, and velocities, solving the statistical
equilibrium equations point by point, with both radiative and collisional
terms. The hope is that some information may then be derived on the nature
and physical properties of the different colliders.

\begin{figure*}
\includegraphics[width=17cm]{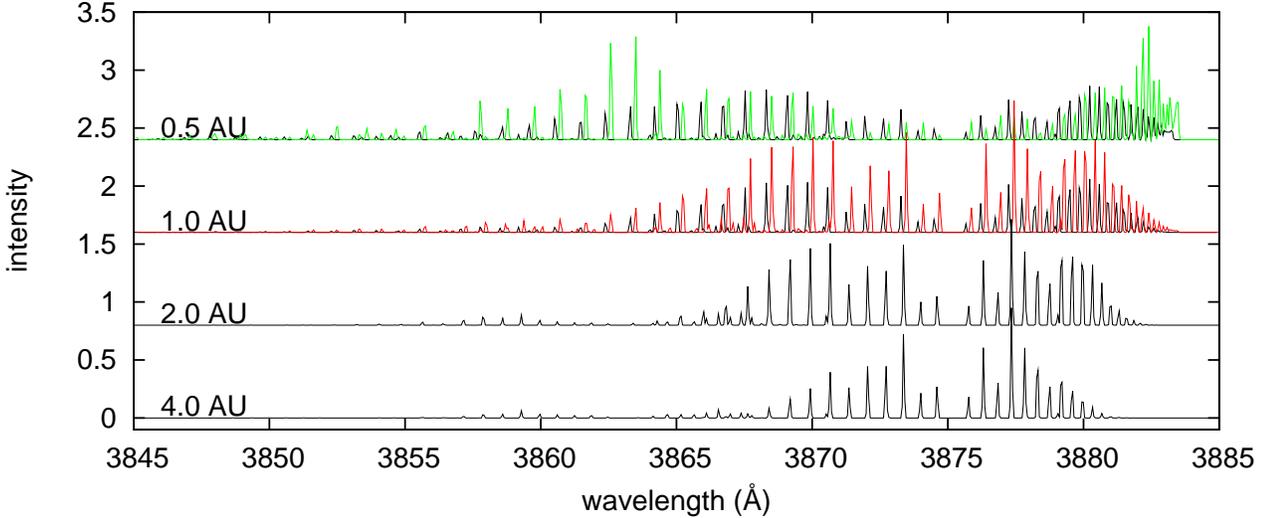}
\caption{Synthetic spectra   of {$^{12}$C$^{14}$N}{}\ at various heliocentric distances. 
{ The intensity scale is arbitrary. The $R$-branch of the (0--0) band extends shortward of 3875~\AA\ 
and the $P$-branch
longward of 3875~\AA. Weak lines of 
the $P$-branch of the (1--1) band are visible
at wavelengths shorter than 3872.5~\AA.}
The black spectra correspond to a collisionless coma with $\dot{r}=0$.
The green spectrum shows the Swings effect for $\dot{r}=20$\,km\,s$^{-1}$.
The red spectrum corresponds to strong collisional effects ($Q=0.01$~s$^{-1}$,$T=300$ and $\dot{r}=0$).
The red and green spectra have been slightly shifted along the wavelength axis
for clarity.
}
  \label{plotCNred}
\end{figure*}
\begin{figure*}
\includegraphics[width=17cm]{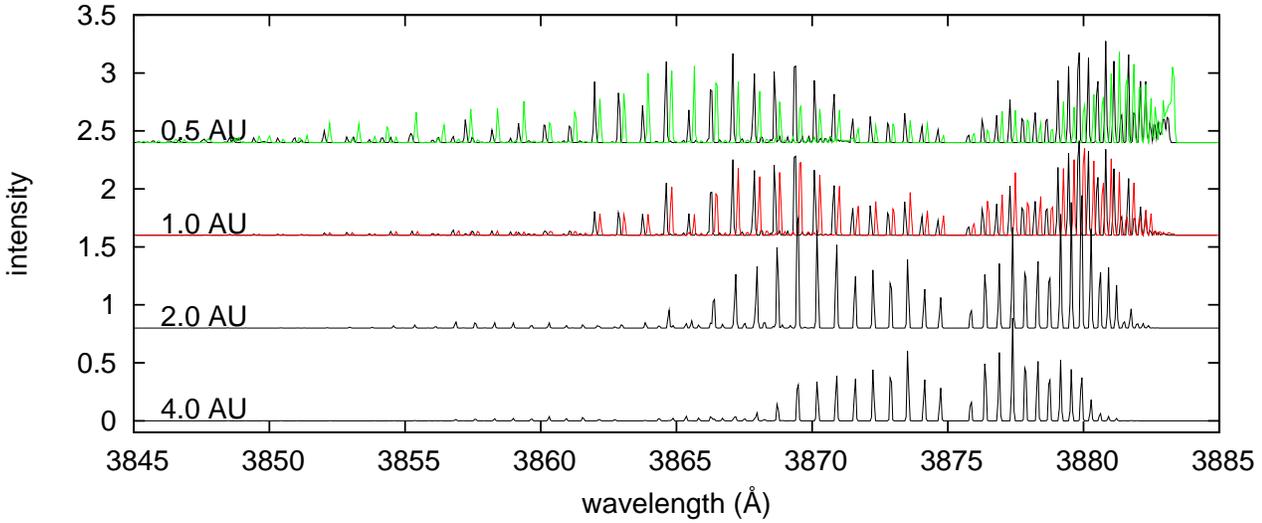}
\caption{Same as Fig.~\ref{plotCNred} for {$^{12}$C$^{15}$N}{}. 
}
  \label{plotCN15red}
\end{figure*}

\begin{figure}
\resizebox{2\hsize}{!}{\includegraphics[angle=-90]{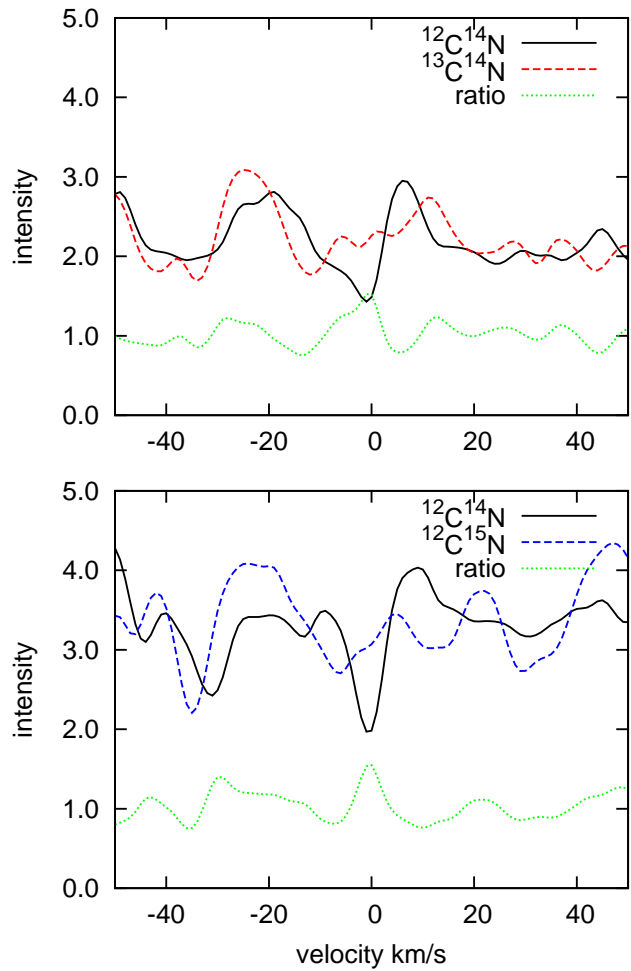}}
\vspace{5mm}
\caption{Swings effect in the R lines used to estimate the {$^{12}$C/$^{13}$C}{}\ (top)
and {$^{14}$N/$^{15}$N}{}\ ratios. 
The solid line corresponds to the emission of {$^{12}$C$^{14}$N}{}; the red line to the emission
of an equal amount of the rare isotopologue;
the green line is the ratio {$^{12}$C/$^{13}$C}{}\ or {$^{14}$N/$^{15}$N}{}. { Intensity is in arbitrary units.}
Observing comets close to the perihelion
is thus particularly interesting because of the weaker contamination by the       
lines of {$^{12}$C$^{14}$N}{}.  }
  \label{plotswings}
\end{figure}

The effects of the heliocentric distance $r$ and velocity $\dot{r}$
and of the collisions on the fluorescence spectra 
of {$^{12}$C$^{14}$N}{}\ and {$^{12}$C$^{15}$N}{}\ are shown in Figs.~\ref{plotCNred},
\ref{plotCN15red}
and \ref{plotswings}.
Synthetic spectra of the
various isotopologues {$^{12}$C$^{14}$N}{}{}, {$^{13}$C$^{14}$N}{}{} and {$^{12}$C$^{15}$N}{}{} 
had to be derived  for each observing circumstance $(\dot{r}\,,\, r)$.
{The Swings effect shown in Fig.~\ref{plotCNred} 
is especially important since the 
solar photospheric spectrum shows strong CN absorption lines which reduce considerably
the excitation at $\dot{r}=0$ (see also Fig.~\ref{plotswings}).
Because of the wavelength separation
between {$^{12}$C$^{14}$N}{}\ and {$^{12}$C$^{15}$N}{}\ 
the $\dot{r}=0$ spectrum in Fig.~\ref{plotCN15red} is not so severely 
depressed. Observing comets close to the perihelion
is thus particularly interesting because of the weaker contamination by the 
lines of {$^{12}$C$^{14}$N}{}. 
}

Ideally a complete 3-D model should have been used, with a range of 
values for $T$ and $Q$ inside the coma.
As a first approximation, we used a single cell model, with
only one value for each parameter; this appears to adequately reproduce the spectra.
The parameters  were  derived iteratively, { through a minimization
procedure over the $T$ and $Q$ parameters},
so as to best reproduce the intensity of all observed {$^{12}$C$^{14}$N}{}\ lines. 
{ The median rms deviation of the ratio between the individual observed and 
modeled {$^{12}$C$^{14}$N}{}\  lines is 0.05, computed over $\sim 30$ lines. 
For the best spectra the median rms is 0.02. 
The same values of $T$ and $Q$ were  used to compute the {$^{13}$C$^{14}$N}{}\ and {$^{12}$C$^{15}$N}{}\
spectra. 
We expect the models of the isotopologues to be equally good,
assuming that the collisional 
cross sections of all isotopologues are of the same order.}

\begin{figure}
   \resizebox{\hsize}{!}{\includegraphics{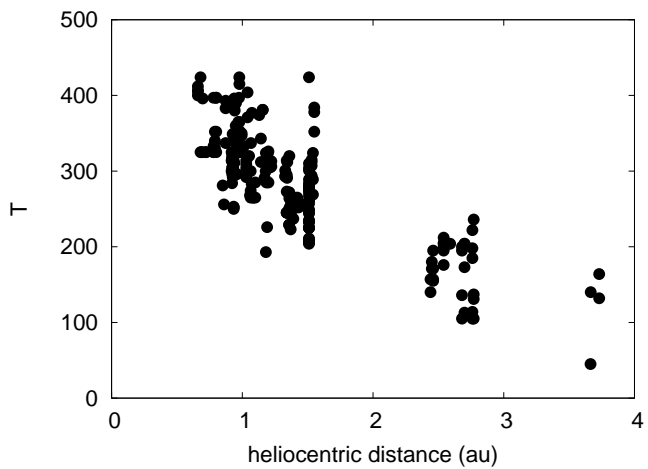}}
      \caption{Variations of the collisional temperature $T$ with the heliocentric distance.}
         \label{plotfigTr}
  \end{figure}

The adopted values for  $T$ and $Q$ are given in Table~\ref{kstars}.
Figure~\ref{plotfigTr}  shows the trend of $T$ 
with the heliocentric distance. The large scatter around $r=1.5$~AU 
is due to the Tempel~1 data. The Deep Impact experiment produced a noticeable 
but short-lived increase in the collisional excitation temperature.

\section{Analysis}
\label{sec:analysis}
Very high-resolution synthetic spectra are  computed with the IRAF
{\it mk1dspec} task, using the fluorescence model described in  Section~\ref{appModel}.
{ The observed line profile (psf, point spread function) is wider and smoother 
than the synthetic one and it is possible to find a convolution transforming the latter
into the former. 
The {\it psfmatch} task does it, and it} 
is  then applied to each spectrum 
in order to match
the synthetic psf of {$^{12}$C$^{14}$N}{}{} to the observed 
instrumental profile for a series
of clean lines. 
The intensities are  normalized simultaneously through the same procedure. 
{$^{13}$C$^{14}$N}{}{} and {$^{12}$C$^{15}$N}{}{} synthetic spectra are produced by applying the
same {\it psfmatch} kernel. 

The final individual spectra (observed and synthetic) 
were combined (by comets and observing runs) in small groups (designated by subscript $k$) 
using an optimal weighting scheme in order to
maximize the overall signal-to-noise ratio of the observed spectra.
This yields a library of observed and synthetic 
spectra $O_k(\lambda)$  and
$S_{k,1}(\lambda)$ (for {$^{12}$C$^{14}$N}{}{}), $S_{k,2}(\lambda)$ ({$^{13}$C$^{14}$N}{}{}) 
and $S_{k,3}(\lambda)$ ({$^{12}$C$^{15}$N}{}{}) -- the last two
corresponding to isotopic ratios of 1.

The 
observed spectrum $O_k$ and the synthetic ones $S_{k,i}$
must then be compared in order to obtain the isotopic
ratio of the comet corresponding to group $k$. 
We  have to estimate 
$\alpha_k=${$^{13}$C$^{14}$N}{}/{$^{12}$C$^{14}$N}{}{}
and 
$\beta_k=${$^{12}$C$^{15}$N}{}/{$^{12}$C$^{14}$N}{}{}  such that

\begin{equation}
O_k(\lambda)=S_{k,1}(\lambda) +  \alpha_k S_{k,2}(\lambda) +  \beta_k S_{k,3}(\lambda)\,.
\label{eq1}
\end{equation}

This is done in an interactive procedure (see Appendix~\ref{appEstimating}).

The { values of the isotopic ratios (Table~\ref{table4c}) are} obtained  over
the full set of isotopic lines, unblended with {$^{12}$C$^{14}$N}{}. Calculations made with
independent subsets helped us to estimate the errors. We { also took} 
into account the inaccuracies of the model, which we estimated from
the residuals obtained when fitting the {$^{12}$C$^{14}$N}{}\ lines. 
Since the uncertainties are not dominated by random errors, the quoted
values are conservative estimates which represent the outer limits
beyond which reasonable fits of the rare species cannot be accepted.

Weighted mean values and estimated errors are given in Table~\ref{table6},
using the procedure explained in Appendix~\ref{appAverages}.
They are computed for a variety
of groups:  all data, all data excluding 73P/Schwassmann-Wachmann 3 B and C, 
and each of the
comet families as defined by  Levison (JF, HF, ext, new) or  Horner et al. (SP, I, L).
There are no significant differences between the groups except perhaps for the JF
or SP$_{\rm III}$ families which contain -- exclusively for SP$_{\rm III}$ -- 
the 73P/Schwassmann-Wachmann 3 fragments.
The constancy of the isotopic ratios is interesting in view of the significant
differences in molecular abundances exhibited by some of the comets under study
(see, e.g., Biver et al.\ \cite{biver}).

\begin{figure*}
   \includegraphics[width=17.0cm]{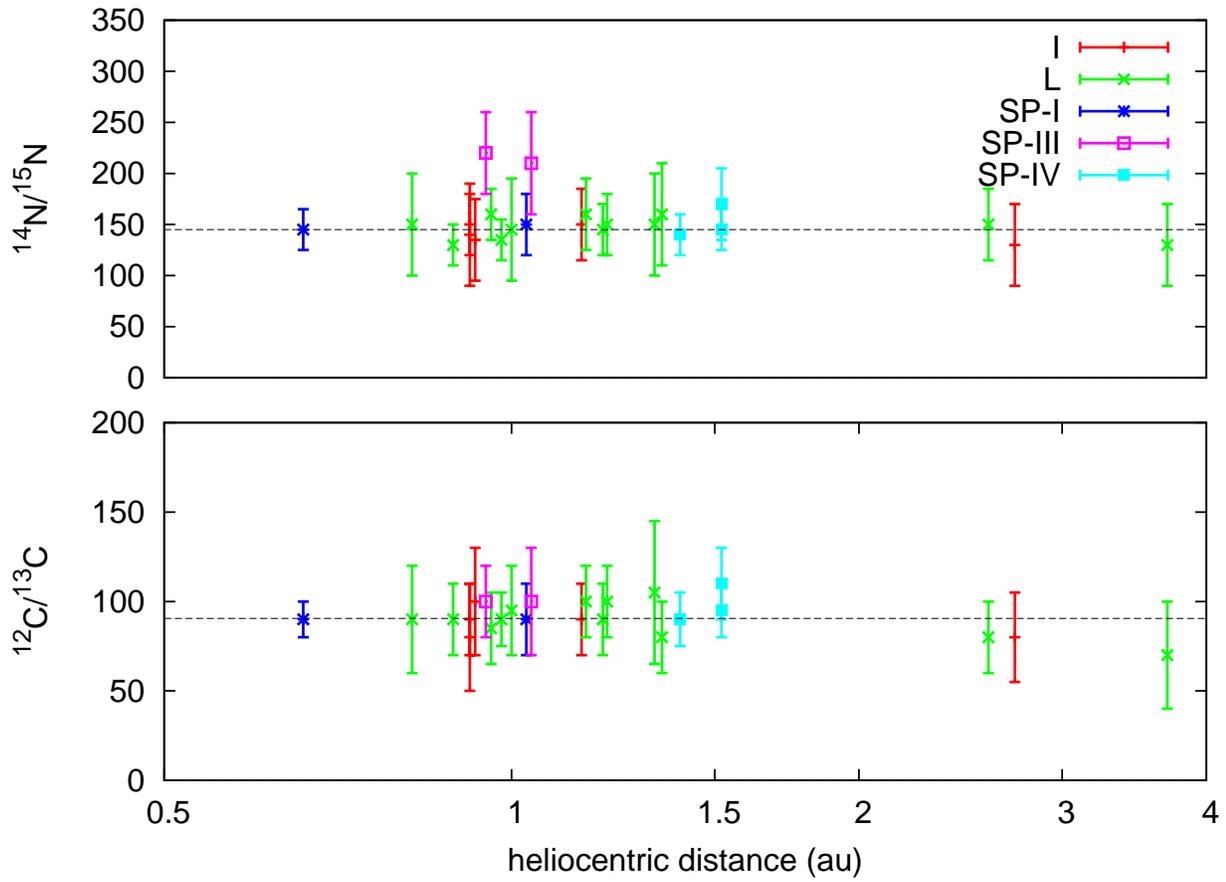}
\vspace{10mm}
      \caption{{$^{12}$C/$^{13}$C}{}\ and {$^{14}$N/$^{15}$N}{}\ versus the heliocentric distance.
      The horizontal lines show the mean values (91.0 and 147.8, respectively). Different symbols refer 
      to different comet families.}
         \label{plotr}
  \end{figure*}

{ Figure~\ref{plotr} shows no correlation of the ratios with 
the heliocentric distance, confirming our previous results
(Manfroid et al.\ \cite{manfroid}).
Now that the CN and HCN isotopic ratios are known to be similar 
(Bockel\'ee-Morvan et al.\ \cite{bockelee}),
this absence of correlation is no longer a strong argument
against HCN being a major parent of CN.
Other possible major parents likely have similar isotopic ratios.} 

\begin{figure}
\vspace{15mm}
   \includegraphics[width=8.0cm]{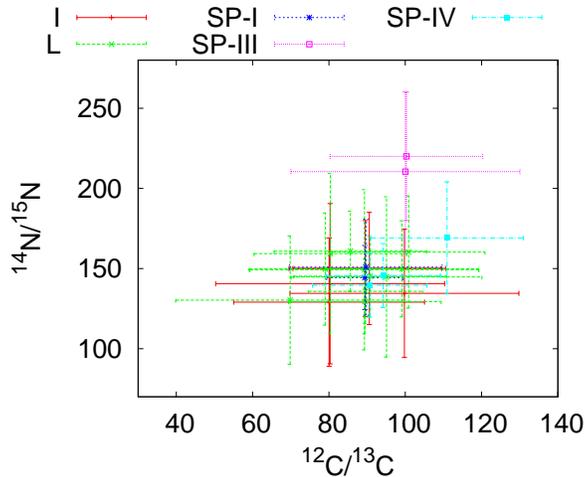}
\vspace{10mm}
      \caption{{$^{14}$N/$^{15}$N}{}\ versus {$^{12}$C/$^{13}$C}{}.
      Different symbols refer 
      to different comet families. { Small shifts have been introduced
      to reduce overlapping.}}
\vspace{10mm}
         \label{plotr2}
  \end{figure}

There is also no significant correlation between {$^{12}$C/$^{13}$C}{}\ and {$^{14}$N/$^{15}$N}{}\ (Fig.~\ref{plotr2}).

Finally, we looked for possible variations of the isotopic ratios with distance
to the nucleus. Such a change would reveal a mixture of CN parents with different
isotopic ratios and different lifetimes (see Table~\ref{table4d}). 
The best  spectra for this study are those 
of C/2001 Q4 (NEAT) (run no. 24, 5-7 May 2004). Spectra of sufficient S/N ratio
were taken at a projected distance of roughly 0, 25\,000 and 50\,000 km. 
We found the same isotopic ratios for the three groups. 
The spatial coverage is relatively small compared 
to the scale lengths involved, but
we could not afford the very long exposure times needed to reach 
larger distances.
Spectra of C/2002 T7 (LINEAR) taken at large nucleocentric distance, but with 
poorer S/N ratios and over a longer time interval,
show a similar behaviour. { Again, the lack  of correlation
is compatible with HCN being a major parent of CN, and excludes
parents with different isotopic ratios.} 

\begin{figure*}
   \includegraphics[width=15.5cm]{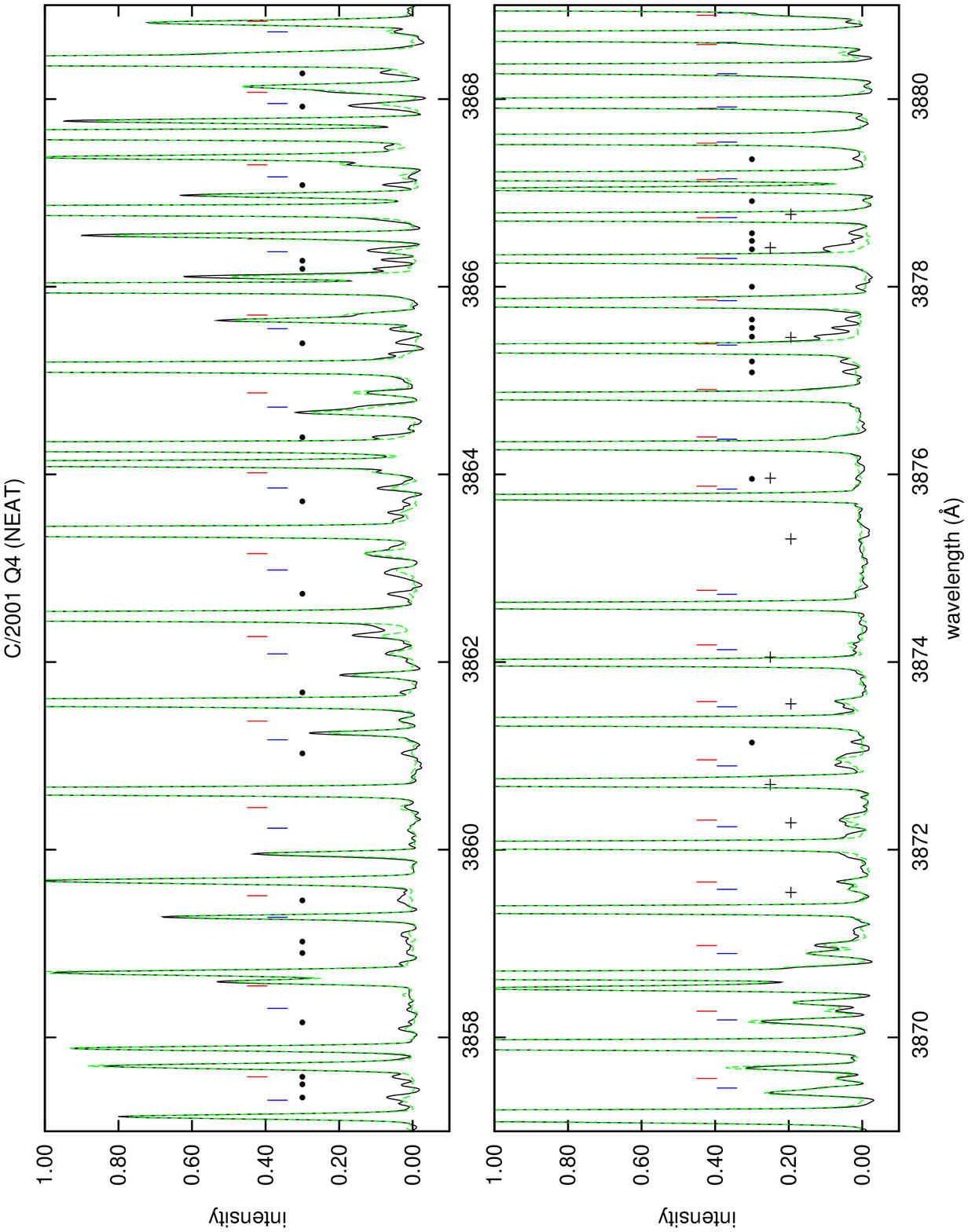}
\vspace{5mm}
      \caption{Spectra of comet C/2001 Q4 (NEAT) obtained close to the nucleus
(solid line) and at about 50000 km (green dashes). The positions of spatially peaked
 features appearing only in the nucleus spectrum are shown by dots. CH lines
 are indicated by $+$ symbols.
 The upper (red) vertical ticks indicate the position
        of the main lines of {$^{13}$C$^{14}$N}{}\, the lower (blue) ticks indicate the position
        of the main lines of {$^{12}$C$^{15}$N}{}. { The intensity scale is in relative units.} } 
         \label{plotoff}
  \end{figure*}

The comparison of spectra obtained close to  
the nucleus and far from it reveals the presence of 
contaminating emission lines
with short scale length
in the domain of the B-X (0,0) band. The main ones are listed 
in Table~\ref{peaked_lines} (see also Fig.~\ref{plotoff}).
Some of them can be attributed to CH, and are within the 
P branch of CN, i.e., in
a region not used for the isotope study, but most are still unidentified.
Though an obvious candidate in this
domain (Gausset et al. \cite{1965ApJ...142...45G}),
C$_3$ does not fit the observed spectrum.

Interestingly, one of the brightest {$^{12}$C$^{15}$N}{}\ lines (R10), which is blended
by an   unidentified feature at $\lambda=3867.92$~\AA\ close to 
the nucleus, appears to be
useable at larger { nucleocentric} distances.
\section{Conclusions}
\label{sec:conclusions}
We have derived homogeneously the isotopic ratios {$^{12}$C/$^{13}$C}{}\ and {$^{14}$N/$^{15}$N}{}\ from
high-quality CN spectra of 18 comets. An improved
fluorescence model taking into account the collisional effects has been used. 
This explains small differences with respect to our previously
published values for some  of these objects { (122P/de Vico, Hale-Bopp,
C/1999 S4, C/2000 WM1, C/2003 K4)}.
The new measurements usually agree with our previous estimates.
With the possible exception of 73P/Schwassmann-Wachmann~3, 
most comets observed so far show the 
same {$^{12}$C/$^{13}$C}{}\  and {$^{14}$N/$^{15}$N}{}\  
isotopic ratios of the CN molecule, irrespective 
of the comet type and the heliocentric distance at which the observations 
were obtained.

The carbon isotopic ratio appears to be well established. 
It agrees with the few cometary values determined from HCN and C$_2$.
It is also consistent with the solar system value of 89 (Anders \& Grevesse
\cite{anders}). 
It is somewhat higher than the local ISM value  ($68\pm15$ at galactic radius $R=7.9$~kpc, 
Milam et al.\ \cite{milam})
and  higher still
than the value of about 63 at the birth place distance of the Sun at $R=6.6$~kpc 
(Wielen \& Wilson \cite{wielen}). This  can be explained by the galactic {$^{13}$C}{}\ enrichment
by low mass stars over the last 4.6 Gyr.
The dispersion of cometary {$^{12}$C/$^{13}$C}{}\ is small when compared to the
large variations observed in CHON grains of 1P/Halley (Jessberger \& Kissel 
\cite{1991ASSL..167.1075J}).
On the other hand, dust samples of the Jupiter family comet 
81P/Wild~2 collected by Stardust have 
a solar composition (McKeegan et al.\ \cite{mckeegan}). 
There must have been little or no fractionation of carbon
in the protosolar cloud and in the solar nebula.

The nitrogen ratio in comets is much lower than the Earth atmosphere value (147 vs 272,
Anders \& Grevesse \cite{anders}), and the bulk of meteorites.
It is even much lower than 
the primordial value in the solar system (presumed to be that of Jupiter,
$435\pm58$, Owen et al.\ \cite{owen2001}), or the 
local ISM value ($450\pm22$ at $R=8$~kpc, 
Dahmen et al.\ 
\cite{dahmen})
or  
the value of about 415 at the birth place distance of the Sun at $R=6.6$~kpc 
(Wielen \& Wilson \cite{wielen}).

Primitive meteorites and IDPs are characterized by large {$^{15}$N}{}\ 
excesses (Floss et al.\ \cite{floss}), 
as are hotspots in 81P/Wild~2 dust samples (McKeegan et al.\ \cite{mckeegan}).
A special mechanism of fractionation (superfractionation) involving
interstellar gas-phase chemistry in very cold clouds has been proposed
as the source of cometary and meteoritic 
{$^{15}$N}{}\ anomalies (Rodgers \& Charnley  \cite{rodgers}) as it
can lead to enhancement by a factor of up to 10.


There is no evidence that the sub-surface material, 
from which CN was released in comet 9P/Tempel~1 { as a result of the Deep Impact event,}
was 
different from the surface material. 
On the other hand, the 73P/Schwassmann-Wachmann 3
data seem to yield a marginally higher {$^{14}$N/$^{15}$N}{}\ ratio (Jehin et al.\ \cite{jehinb}).
Because of the history of successive fragmentations of 73P/Schwassmann-Wachmann 3,
{ we might have measured} recently exposed material. The lower abundance
of {$^{12}$C$^{15}$N}{}\ may indicate pristine material. However, other peculiarities
of  73P/Schwassmann-Wachmann 3 may point to a unique chemical composition 
{( Dello Russo et al.\ \cite{2007Natur.448..172R},
Kobayashi et al.\ \cite{kobayashi}). 
A nuclear spin temperature of about 40~K for NH$_3$
(Jehin et al.\ \cite{jehinb})
and $>37$~K for H$_2$O (Dello Russo et al.\ \cite{2007Natur.448..172R})
instead of the more usual 30~K found in other comets for several molecules
(see, e.g., Kawakita \cite{kawa2005})
may also indicate a peculiar comet.}

Additional measurements of the nitrogen isotopic ratio in CN and
other nitrogen-bearing species in additional comets,
especially chemically peculiar ones, 
are needed to shed more light into these issues. 


\begin{table*}
\caption{\label{obscom} {\bf Observed comets.}}
\vspace{.5cm}
{\footnotesize
\begin{tabular}{lllllllllll} 
\hline \hline 
Comet & $T_{\rm P}$  & $e$ &$a$           & $q$&Aph&$i$ & $P$ & $T_{\rm J}$ & \multicolumn{2}{c}{Type}\\
 &(yyyy-mm-dd)& &(AU) &(AU) &(AU) &(\degr) &(yr) & &L&H\\
\hline 
122P/de Vico$^a$ & 1995-10-06 &  0.96271 & 17.7 & 0.66 & 34.7 & 85 & 74.3 & 0.37 & HF & SP$_{\rm I}$ \\
C/1996 B2 (Hyakutake)& 1996-05-01 & 0.99990 & 2296 & 0.23 & 4591 & 125 & 110000   & -0.34 & EXT & L \\
C/1995 O1 (Hale-Bopp)$^{b,c}$ & 1997-04-01 & 0.99509 & 186.0 & 0.91 & 371.1 & 89 & 2537 & 0.04 & EXT & I \\
55P/Tempel-Tuttle & 1998-02-28 &   0.90555 & 10.3 & 0.98 & 19.7 & 162 & 33.2 & -0.64 & HF & SP$_{\rm I}$ \\
C/1999 H1 (Lee) & 1999-07-11 & 0.99974 & 2773   & 0.71 & 5545 & 149 & 146000   & -0.90 & EXT & L \\
C/1999 S4 (LINEAR) $^d$  & 2000-07-26 & 1.00010 & --  & 0.77 & --     & 149 & --    & --    & NEW & L \\
C/1999 T1 (McNaught-Hartley) & 2000-12-13 & 0.99986 & 8502   & 1.17 & 17000   & 80 & 780000   & 0.23 & EXT & L \\
C/2001 A2 (LINEAR) & 2001-05-25 & 0.99969 & 2530   & 0.78 & 5060   & 36 & 127000   & 0.88 & EXT & L \\
C/2000 WM1 (LINEAR)$^b$ & 2002-01-23 & 1.00025 & --       & 0.56 & --       & 73 & --      & --   & NEW & L \\
153P/Ikeya-Zhang $^a$ & 2002-03-19 &   0.99010 & 51.2 & 0.51 & 101.9 & 28 & 366.5 & 0.88 & EXT & IP \\
C/2002 X5 (Kudo-Fujikawa) & 2003-01-29 & 0.99984 & 1175   & 0.19 & 2350   & 94 & 40300   & -0.03 & EXT & L \\
C/2002 V1 (NEAT) & 2003-02-18 & 0.99990 & 1010   & 0.10 & 2021   & 82 & 32100   & 0.06 & EXT & L \\
C/2002 Y1 (Juels-Holvorcem) & 2003-04-13 & 0.99715 & 250.6 & 0.71 & 500.5 & 104 & 3967   & -0.23 & EXT & I \\
88P/Howell  $^d$ & 2004-04-13 &   0.56124 & 3.1 & 1.37 & 4.9 & 4 & 5.5 & 2.95 & JF & SP$_{\rm IV}$ \\
C/2002 T7 (LINEAR) & 2004-04-23 & 1.00048 & --    & 0.61 & --       & 161 & --      & --    & NEW & L \\
C/2001 Q4 (NEAT) $^c$& 2004-05-16 & 1.00069 & --  & 0.96 & --       & 100 & --      & --    & NEW & L \\
C/2003 K4 (LINEAR) $^c$ & 2004-10-14 & 1.00030 & --  & 1.02 & --   & 134 & --      & --    & NEW & L \\
9P/Tempel 1 $^e$ & 2005-07-05 &  0.51749 & 3.1 & 1.51 & 4.7 & 11 & 5.5 & 2.97 & JF & SP$_{\rm IV}$ \\
73P-B/Schwassmann-Wachmann 3  $^f$ & 2006-06-08 &  0.69350 & 3.1 & 0.94 & 5.2 & 11 & 5.4 & 2.78 & JF & SP$_{\rm III}$ \\
73P-C/Schwassmann-Wachmann 3  $^f$& 2006-06-07 &  0.69338 & 3.1 & 0.94 & 5.2 & 11 & 5.4 & 2.78 & JF & SP$_{\rm III}$ \\
C/2006 M4 (SWAN) & 2006-09-29 & 1.00018 & --       & 0.78 & --       & 112 & --      & --    & NEW & L \\
17P/Holmes  $^g$& 2007-05-04& 0.43242 & 3.6 & 2.05 & 5.18 & 19 & 6.9 &  2.86   & JF  & SP$_{\rm IV}$ \\
8P/Tuttle  $^h$& 2008-01-27 & 0.8198 & 5.7 & 1.03 & 10.4 & 55 & 13.6    &  1.60   & HF  &SP$_{\rm I}$ \\
C/2007 N3 (Lulin)&2009-01-10 & 1.00004& -- & 1.21&--&178&--&--& NEW&L\\
\hline \hline\\
\end{tabular}

\hspace{6mm}\begin{minipage}{0.85\textwidth} 
a Jehin et al.\ \cite{jehin} --
b Arpigny et al.\ \cite{arpi}, \cite{arpib} --
c Manfroid et al.\  \cite{manfroid} --
d Hutsem\'ekers et al.\ \cite{hutsemekers} -- 
e Jehin et al.\ \cite{jehina} --

f Jehin et al.\  \cite{jehinb} --
g Bockel\'ee-Morvan et al.\  \cite{bockelee} --
h Bockel\'ee-Morvan et al.\  \cite{bockeleeb}
\vspace{.2cm}

The elements are obtained from the JPL 
Solar System Dynamics website. 
$T_{\rm P}$ is the epoch of the perihelion,
$e$ the eccentricity, $q$ the perihelion distance in AU, 
Aph the aphelion distance, $i$ the inclination on the ecliptic,
$P$ the period, $T_{\rm J}$ the Tisserand parameter 
relative to Jupiter, the last columns give 
the classification according to Levison
(\cite{levison}) and Horner et al. (\cite{horner}).
The subdivision I-IV based on $T_{\rm J}$ is given for the SP comets only,
all others being of the sub-type I. 
\end{minipage}
}
\end{table*}

\begin{table*}
\caption{\label{observcirc} {\bf Observational circumstances.}}
\vspace{.5cm}
  \includegraphics[angle=0,width=17.5cm]{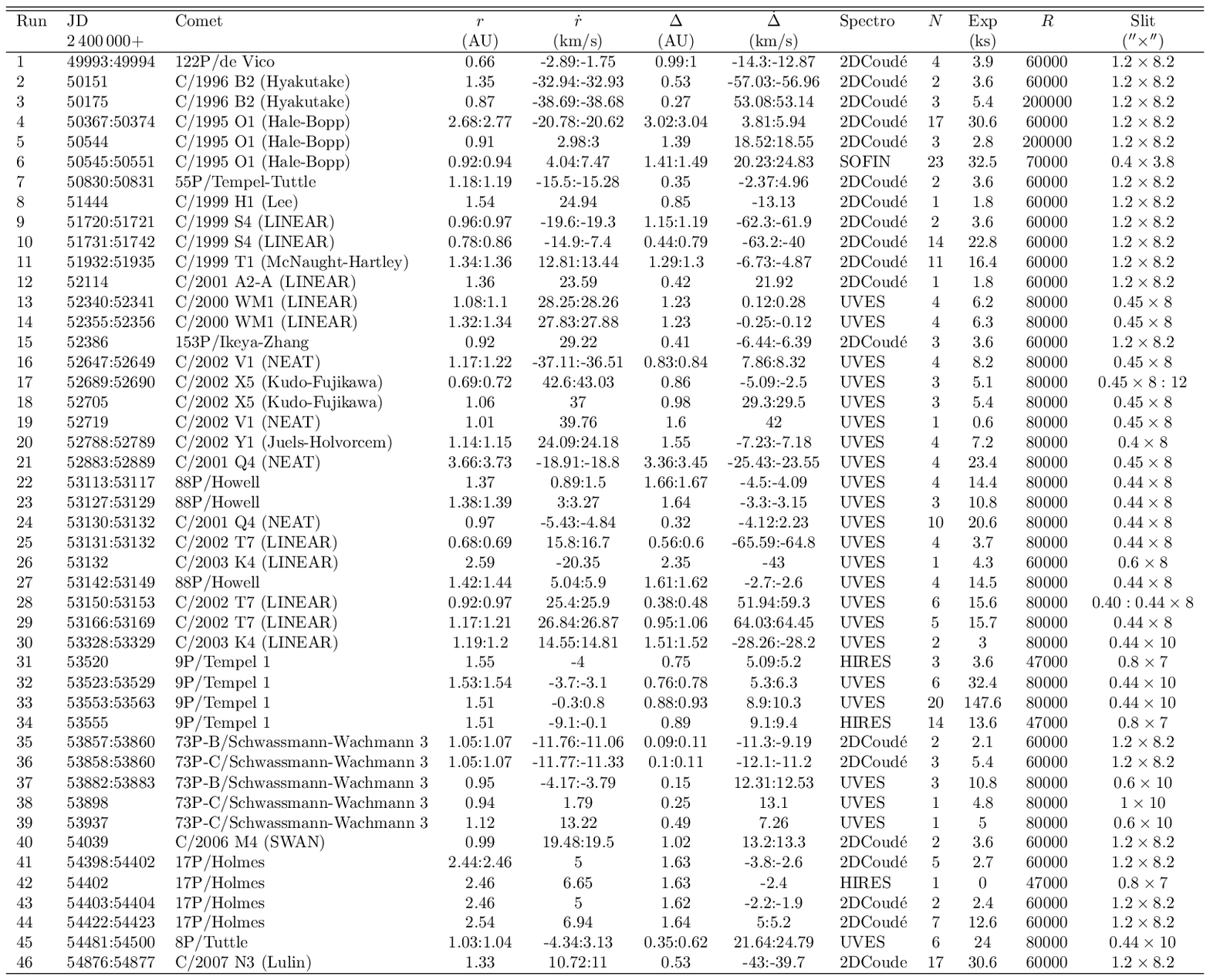}
\vspace{3mm}

\hspace{6mm}\begin{minipage}[t]{0.85\textwidth} 
For convenience the data are grouped in observing runs. 
$r$ is the heliocentric distance in astronomical units (AU), 
$\dot{r}$ the heliocentric radial velocity, 
$\Delta$ is the geocentric distance, 
$\dot{\Delta}$ the topocentric radial velocity,
'Spectro' the spectrograph used,
$N$ is the number of individual spectra, 
`Exp' the total exposure time in thousands of seconds, 
$R=\Delta\lambda/\lambda$ the resolving power.
The nominal size of
the entrance slit of the spectrograph is
given in arcsec. Specific values relative to individual spectra 
can be found in Table~\ref{kstars}.
\end{minipage}
\end{table*}
\begin{table*}
\caption{\label{table4c} {\bf Isotopic ratios in CN.}}
\begin{tabular}{lllll} \hline \hline \rule[-1.5mm]{0mm}{6mm}
Runs & Comet & {$^{12}$C/$^{13}$C}{} &  {$^{14}$N/$^{15}$N}{} &Reference\\
\hline \rule{0mm}{5mm}
1  &   122P/de Vico &$         90\pm10 $&$ 145 \pm 20 $&{\it Jehin et al.\ \cite{jehin}}\\
2 & C/1996 B2 (Hyakutake) &$          \ge60 $&$ \ge60  $\\
3 & C/1996 B2 (Hyakutake) &$          \ge50 $&$ \ge50 $\\
4 &  C/1995 O1 (Hale-Bopp) &$          80\pm 25 $&$ 130  \pm 40 $&{\it Manfroid et al.\  \cite{manfroid}}\\
5 &  C/1995 O1 (Hale-Bopp) &$          90\pm 20 $&$ 150 \pm 30$&{\it Manfroid et al.\  \cite{manfroid}}\\
6 &  C/1995 O1 (Hale-Bopp) &$         100\pm 30 $&$ 135 \pm 40 $&{\it Manfroid et al.\  \cite{manfroid}}\\
7 & 55P/Tempel-Tuttle &$              - $&$  - $\\
8 & C/1999 H1 (Lee) &$               \ge60 $&$ \ge60 $\\
9,10 & C/1999 S4 (LINEAR) &$          90\pm 30 $&$ 150 \pm 50 $&{\it Hutsem\'ekers et al.\ \cite{hutsemekers}} \\
11 & C/1999 T1 (McNaught-Hartley) &$  80\pm 20 $&$ 160 \pm 50$\\
12 & C/2001 A2-A (LINEAR) &$         \ge60 $&$ \ge60 $\\
13,14 & C/2000 WM1 (LINEAR) &$       100\pm 20 $&$ 150 \pm 30 $&{\it Arpigny et al.\ \cite{arpi}, \cite{arpib}}\\
15 & 153P/Ikeya-Zhang &$              80\pm 30 $&$ 140 \pm 50 $&{\it Jehin et al.\ \cite{jehin}}\\
17,18 & C/2002 X5 (Kudo-Fujikawa) &$  90\pm 20 $&$ 130 \pm 20$\\
16,19 & C/2002 V1 (NEAT) &$             100\pm 20 $&$ 160 \pm 35$\\
20 & C/2002 Y1 (Juels-Holvorcem) &$   90\pm 20 $&$ 150 \pm 35$\\
21 & C/2001 Q4 (NEAT) &$              70\pm 30 $&$ 130 \pm 40 $& Manfroid et al.\  \cite{manfroid}\\
22,23 &  88P/Howell  &$               90\pm 15 $&$ 140  \pm 20 $&Hutsem\'ekers et al.\ \cite{hutsemekers}\\
24 & C/2001 Q4 (NEAT) &$              90\pm 15 $&$ 135 \pm 20 $& Manfroid et al.\  \cite{manfroid}\\
25,28,29 & C/2002 T7 (LINEAR) &$      85\pm 20 $&$ 160 \pm 25$\\
26 & C/2003 K4 (LINEAR)  &$           80\pm 20 $&$ 150 \pm 35 $&{\it Manfroid et al.\  \cite{manfroid}}\\
30 & C/2003 K4 (LINEAR)  &$           90\pm 20 $&$ 145 \pm 25 $&{\it Manfroid et al.\  \cite{manfroid}}\\
31,34 & 9P/Tempel 1  &$              110\pm 20 $&$ 170 \pm 35 $&Jehin et al.\ \cite{jehina}\\
32,33 & 9P/Tempel 1  &$               95\pm 15 $&$ 145 \pm 20 $&Jehin et al.\ \cite{jehina}\\
35 & 73P-B/Schwassmann-Wachmann 3 &$-$&$-$\\
36 & 73P-C/Schwassmann-Wachmann 3 &$-$&$-$\\
37 & 73P-B/Schwassmann-Wachmann 3 &$ 100\pm 30 $&$ 210 \pm 50$&Jehin et al.\  \cite{jehinb}\\
38,39 & 73P-C/Schwassmann-Wachmann 3 &$    100\pm 20 $&$ 220 \pm 40$&Jehin et al.\  \cite{jehinb}\\
40 & C/2006 M4 (SWAN) &$              95\pm 25 $&$ 145 \pm 50$\\
41-44 & 17P/Holmes &$                 90\pm 20 $&$ 165 \pm 35$\\
41,43,44 & 17P/Holmes &$             \ge10 $&$ \ge10 $\\
42 & 17P/Holmes &$                    95\pm 20 $&$ 165 \pm 40$&Bockel\'ee-Morvan et al.\  \cite{bockelee}\\
45 & 8P/Tuttle &$   90\pm 20  $&$ 150\pm 30 $&Bockel\'ee-Morvan et al.\  \cite{bockeleeb}\\
46 & C/2007 N3 (Lulin) &$105\pm 40 $&$ 150 \pm 50 $\\
\hline
\end{tabular}

\vspace{3mm}
\hspace{6mm}
\begin{minipage}[t]{0.65\textwidth} 
For convenience the data are grouped in observing runs as defined in Table~\ref{observcirc}.
References are given when these are not first determinations, in italics if the values
have been revised.
\end{minipage}
\end{table*}

\begin{table}
\caption{\label{peaked_lines} Lines not identified as CN
within the 3857--3880~\AA\  range. Most are 
not yet identified.}
\vspace{.5cm}
\begin{tabular}{ll} 
\hline\hline\rule{0mm}{3ex}
$\lambda$ (\AA)&Identification\\
\hline\rule{0mm}{3ex}
3857.36\\
3857.50\\
3857.58\\
3858.16\\
3858.90\\
3859.02\\
3859.46\\
3861.03\\
3861.68\\
3862.73\\
3863.71\\
3864.41\\
3865.40 \\
3866.19\\
3866.28\\
3867.08\\
3867.92\\
3868.27\\
3873.14\\
3875.95&  CH R$_1$ $J^{\prime\prime}=3.5$\\
3877.09 \\
3877.20 \\
3877.47&  CH R$_2$ $J^{\prime\prime}=1.5$\\
3877.56 \\
3877.65 \\
3878.00\\
3878.40&  CH R$_1$ $J^{\prime\prime}=2.5$\\
3878.49 \\
3878.57 \\
3878.91 \\
3879.36 \\
\hline
\end{tabular}
\end{table}

\begin{table}
\caption{\label{table6} {\bf Average isotopic ratios.}
The weighted mean is computed for a variety 
of groups :  all data, all data excluding 73P/Schwassmann-Wachmann 3 B and C,
and each of the { categories} of Levison (JF, HF, EXT, NEW) or  
Horner et al.\ (SP, I, L).}

\begin{tabular}{llll} \hline \hline \rule[-1.5mm]{0mm}{6mm}
Group& {$^{12}$C/$^{13}$C}{} &  {$^{14}$N/$^{15}$N}{}  \\
\hline
All comets&$91.1\pm3.7$&$147.8\pm5.6$\\
All without S-W 3&$90.6\pm3.8$&$145.2\pm5.6$\\
JF &$97.2\pm7.6$&$156.8\pm12.2$\\
JF without S-W&$96.3\pm8.6$&$146.4\pm12.4$\\
Non JF (Oort)&$89.1\pm4.2$&$144.0\pm6.5$\\
HF&$90.0\pm8.4$&$146.5\pm15.1$\\
EXT &$89.0\pm6.9$&$141.3\pm10.7$\\
NEW &$89.1\pm6.8$&$145.3\pm9.5$\\
SP$_{\rm I}$&$90.0\pm8.4$&$146.5\pm15.1$\\
SP$_{\rm III}$&$100.0\pm15.1$&$216.1\pm27.3$\\
SP$_{\rm IV}$&$96.3\pm8.6$&$146.4\pm12.4$\\
I&$88.2\pm9.3$&$142.8\pm14.8$\\
L&$89.3\pm5.8$&$143.8\pm8.1$\\
\hline
\end{tabular}
\end{table}

\begin{table*}
\caption{\label{table4d} 
{\bf 
Isotopic ratios in CN at various 
nucleocentric distances.}
The C/2002 T7 (LINEAR) data { include} spectra taken
at up to 180\,000 km but are strongly weighted 
for the range $24\,000-45\,000$ km.}
\begin{tabular}{llll} \hline \hline \rule[-1.5mm]{0mm}{6mm}
Comet &Distance& {$^{12}$C/$^{13}$C}{} &  {$^{14}$N/$^{15}$N}{}  \\
&(km)\\
\hline \rule{0mm}{5mm}
C/2001 Q4 (NEAT) &$0-3\,000$&$             90\pm 20 $&$ 135 \pm 25 $\\
C/2001 Q4 (NEAT) &$25\,000$&$              95\pm 20 $&$ 135 \pm 25 $\\
C/2001 Q4 (NEAT) &$50\,000$&$              90\pm 20 $&$ 145 \pm 25 $\\
C/2002 T7 (LINEAR) &$0-2\,000$&$      85\pm 20 $&$ 155 \pm 25$\\
C/2002 T7 (LINEAR) &$24\,000-180\,000$&$      80\pm 25 $&$ 165 \pm 30$\\
\hline
\end{tabular}

\end{table*}

\onllongtab{8}{
\begin{longtable}{llllrlrrrrrcc}
\caption{\label{kstars} Individual spectra. 
$r$ is the heliocentric distance in astronomical units (AU),
$\Delta$  the geocentric distance,
Spectro the spectrograph used,
$MJD=JD-2400000.5$ the Modified Julian Day,
Run the run number,
Exp the exposure time in  seconds,
$R$ the spectral resolution.
Slit and Offset give the size of
the entrance slit of the spectrograph and the offset from the nucleus.
$T$ and $Q$ are the parameters 
used for the collisional effects in the synthetic spectra.
 }\\
\hline\hline
Comet& $r$ &$\Delta$&$m_r$&Spectro & MJD&Run&Exp& $R$&Slit&Offset& $T$&$5+\log Q$\\
     &(AU) & (AU)     &   &   & &&(s)&    &($''\times''$)&($''$) &(K) &(s$^{-1}$) \\
\hline
\endfirsthead
\caption{continued.}\\
\hline\hline
Comet& $r$ &$\Delta$&$m_r$&Spectro & MJD&Run&Exp& $R$&Slit&Offset& $T$&$5+\log Q$\\
     &(AU) & (AU)     &   &   & &&(s)&    &($''\times''$)&($''$) &(K) &(s$^{-1}$) \\
\hline
\endhead
\hline
\endfoot
deVico&0.66&1.00&5.60&2DCoud{\'e}&49993.488&1&600&60000&$1.20\times8.20$&0&410&1.9\\
deVico&0.66&1.00&5.60&2DCoud{\'e}&49993.502&1&600&60000&$1.20\times8.20$&0&410&1.8\\
deVico&0.66&0.99&5.62&2DCoud{\'e}&49994.475&1&1500&60000&$1.20\times8.20$&0&410&1.8\\
deVico&0.66&0.99&5.62&2DCoud{\'e}&49994.495&1&1200&60000&$1.20\times8.20$&100&400&1.7\\
Hyakutake&1.36&0.54&6.55&2DCoud{\'e}&50151.467&2&1800&60000&$1.20\times8.20$&0&230&2.8\\
Hyakutake&1.36&0.54&6.55&2DCoud{\'e}&50151.490&2&1800&60000&$1.20\times8.20$&0&230&2.7\\
Hyakutake&0.87&0.28&4.79&2DCoud{\'e}&50175.090&3&1800&200000&$1.20\times8.20$&0&390&3.1\\
Hyakutake&0.87&0.28&4.78&2DCoud{\'e}&50175.113&3&1800&200000&$1.20\times8.20$&0&340&1.1\\
Hyakutake&0.87&0.28&4.78&2DCoud{\'e}&50175.134&3&1800&200000&$1.20\times8.20$&0&380&3.1\\
HB&2.77&3.02&3.19&2DCoud{\'e}&50367.086&4&1800&60000&$1.20\times8.20$&0&140&1.0\\
HB&2.77&3.02&3.19&2DCoud{\'e}&50367.111&4&1800&60000&$1.20\times8.20$&0&130&0.7\\
HB&2.77&3.02&3.19&2DCoud{\'e}&50367.141&4&1800&60000&$1.20\times8.20$&0&--&--\\
HB&2.77&3.02&3.19&2DCoud{\'e}&50367.165&4&1800&60000&$1.20\times8.20$&0&240&0.8\\
HB&2.76&3.03&3.19&2DCoud{\'e}&50368.066&4&1800&60000&$1.20\times8.20$&0&--&--\\
HB&2.76&3.03&3.19&2DCoud{\'e}&50368.091&4&1800&60000&$1.20\times8.20$&0&--&--\\
HB&2.76&3.03&3.19&2DCoud{\'e}&50368.115&4&1800&60000&$1.20\times8.20$&0&190&0.7\\
HB&2.76&3.03&3.19&2DCoud{\'e}&50368.138&4&1800&60000&$1.20\times8.20$&0&110&0.1\\
HB&2.76&3.03&3.19&2DCoud{\'e}&50368.162&4&1800&60000&$1.20\times8.20$&0&220&0.9\\
HB&2.70&3.04&3.08&2DCoud{\'e}&50373.062&4&1800&60000&$1.20\times8.20$&0&--&--\\
HB&2.70&3.04&3.08&2DCoud{\'e}&50373.086&4&1800&60000&$1.20\times8.20$&0&--&--\\
HB&2.70&3.04&3.08&2DCoud{\'e}&50373.110&4&1800&60000&$1.20\times8.20$&0&170&1.0\\
HB&2.70&3.04&3.08&2DCoud{\'e}&50373.134&4&1800&60000&$1.20\times8.20$&0&200&0.1\\
HB&2.68&3.04&3.08&2DCoud{\'e}&50374.057&4&1800&60000&$1.20\times8.20$&0&200&0.8\\
HB&2.68&3.04&3.08&2DCoud{\'e}&50374.080&4&1800&60000&$1.20\times8.20$&0&--&--\\
HB&2.68&3.04&3.08&2DCoud{\'e}&50374.103&4&1800&60000&$1.20\times8.20$&0&--&--\\
HB&2.68&3.04&3.08&2DCoud{\'e}&50374.126&4&1800&60000&$1.20\times8.20$&0&140&0.9\\
HB&0.92&1.40&-1.12&2DCoud{\'e}&50544.068&5&120&200000&$1.20\times8.20$&0&300&3.2\\
HB&0.92&1.40&-1.12&2DCoud{\'e}&50544.079&5&900&200000&$1.20\times8.20$&0&310&1.9\\
HB&0.92&1.40&-1.12&2DCoud{\'e}&50544.112&5&1800&200000&$1.20\times8.20$&0&--&--\\
HB&0.92&1.42&-1.15&SOFIN&50545.863&6&1814&70000&$0.45\times3.80$&0&280&3.0\\
HB&0.92&1.42&-1.15&SOFIN&50545.886&6&2106&70000&$0.45\times3.80$&0&320&3.2\\
HB&0.92&1.43&-1.17&SOFIN&50546.832&6&600&70000&$0.45\times3.80$&0&300&3.0\\
HB&0.92&1.43&-1.17&SOFIN&50546.839&6&460&70000&$0.45\times3.80$&0&300&3.2\\
HB&0.92&1.43&-1.17&SOFIN&50546.865&6&3838&70000&$0.45\times3.80$&0&310&2.9\\
HB&0.93&1.44&-1.19&SOFIN&50547.836&6&1800&70000&$0.45\times3.80$&0&300&2.9\\
HB&0.93&1.44&-1.19&SOFIN&50547.862&6&1800&70000&$0.45\times3.80$&0&300&3.2\\
HB&0.93&1.44&-1.19&SOFIN&50547.887&6&1800&70000&$0.45\times3.80$&0&290&3.0\\
HB&0.93&1.45&-1.20&SOFIN&50548.837&6&1200&70000&$0.45\times3.80$&0&320&3.1\\
HB&0.93&1.45&-1.20&SOFIN&50548.856&6&1200&70000&$0.45\times3.80$&0&320&3.2\\
HB&0.93&1.45&-1.20&SOFIN&50548.874&6&1200&70000&$0.45\times3.80$&0&300&3.1\\
HB&0.93&1.45&-1.10&SOFIN&50548.893&6&1200&70000&$0.45\times3.80$&0&320&3.0\\
HB&0.93&1.47&-1.12&SOFIN&50549.837&6&1200&70000&$0.45\times3.80$&0&250&2.7\\
HB&0.93&1.47&-1.12&SOFIN&50549.855&6&1200&70000&$0.45\times3.80$&0&250&2.8\\
HB&0.93&1.47&-1.12&SOFIN&50549.875&6&1200&70000&$0.45\times3.80$&0&250&3.0\\
HB&0.94&1.48&-1.14&SOFIN&50550.836&6&1200&70000&$0.45\times3.80$&0&300&3.0\\
HB&0.94&1.48&-1.14&SOFIN&50550.857&6&1516&70000&$0.45\times3.80$&0&300&2.6\\
HB&0.94&1.48&-1.14&SOFIN&50550.877&6&1200&70000&$0.45\times3.80$&0&400&3.2\\
HB&0.94&1.48&-1.14&SOFIN&50550.896&6&1199&70000&$0.45\times3.80$&0&380&3.0\\
HB&0.94&1.49&-1.16&SOFIN&50551.834&6&1200&70000&$0.45\times3.80$&0&390&2.2\\
HB&0.94&1.49&-1.16&SOFIN&50551.856&6&1200&70000&$0.45\times3.80$&0&380&3.2\\
HB&0.94&1.49&-1.16&SOFIN&50551.874&6&1199&70000&$0.45\times3.80$&0&350&2.7\\
HB&0.94&1.49&-1.16&SOFIN&50551.892&6&1200&70000&$0.45\times3.80$&0&340&3.0\\
TT&1.19&0.36&10.24&2DCoud{\'e}&50830.104&7&1800&60000&$1.20\times8.20$&0&230&1.7\\
TT&1.18&0.36&10.23&2DCoud{\'e}&50831.158&7&1800&60000&$1.20\times8.20$&0&190&1.9\\
Lee&1.54&0.85&8.74&2DCoud{\'e}&51444.219&8&1800&60000&$1.20\times8.20$&0&270&1.2\\
1999S4&0.97&1.19&7.81&2DCoud{\'e}&51720.377&9&1800&60000&$1.20\times8.20$&0&340&1.9\\
1999S4&0.96&1.16&7.87&2DCoud{\'e}&51721.362&9&1800&60000&$1.20\times8.20$&0&300&1.9\\
1999S4&0.86&0.80&8.39&2DCoud{\'e}&51731.387&10&1800&60000&$1.20\times8.20$&0&260&2.0\\
1999S4&0.85&0.76&8.50&2DCoud{\'e}&51732.390&10&1800&60000&$1.20\times8.20$&0&280&1.9\\
1999S4&0.80&0.52&8.91&2DCoud{\'e}&51739.388&10&1800&60000&$1.20\times8.20$&0&400&1.7\\
1999S4&0.80&0.52&8.91&2DCoud{\'e}&51739.414&10&1800&60000&$1.20\times8.20$&0&--&--\\
1999S4&0.80&0.52&8.91&2DCoud{\'e}&51739.439&10&1800&60000&$1.20\times8.20$&0&350&1.1\\
1999S4&0.80&0.52&8.91&2DCoud{\'e}&51739.459&10&1200&60000&$1.20\times8.20$&0&400&1.7\\
1999S4&0.79&0.49&8.84&2DCoud{\'e}&51740.423&10&1200&60000&$1.20\times8.20$&0&330&0.1\\
1999S4&0.79&0.46&8.86&2DCoud{\'e}&51741.443&10&1800&60000&$1.20\times8.20$&0&340&1.2\\
1999S4&0.79&0.46&8.86&2DCoud{\'e}&51741.463&10&1200&60000&$1.20\times8.20$&0&350&1.0\\
1999S4&0.78&0.44&8.87&2DCoud{\'e}&51742.359&10&1800&60000&$1.20\times8.20$&0&330&1.1\\
1999S4&0.78&0.44&8.87&2DCoud{\'e}&51742.384&10&1800&60000&$1.20\times8.20$&0&330&0.9\\
1999S4&0.78&0.44&8.87&2DCoud{\'e}&51742.419&10&1800&60000&$1.20\times8.20$&0&400&0.8\\
1999S4&0.78&0.44&8.87&2DCoud{\'e}&51742.443&10&1800&60000&$1.20\times8.20$&0&330&0.1\\
1999S4&0.78&0.44&8.88&2DCoud{\'e}&51742.463&10&1200&60000&$1.20\times8.20$&0&330&1.2\\
1999T1&1.34&1.31&7.11&2DCoud{\'e}&51932.520&11&1200&60000&$1.20\times8.20$&0&--&--\\
1999T1&1.34&1.31&7.11&2DCoud{\'e}&51932.537&11&1200&60000&$1.20\times8.20$&0&--&--\\
1999T1&1.34&1.30&7.12&2DCoud{\'e}&51933.449&11&1800&60000&$1.20\times8.20$&0&300&2.0\\
1999T1&1.34&1.30&7.12&2DCoud{\'e}&51933.476&11&1800&60000&$1.20\times8.20$&0&310&1.1\\
1999T1&1.34&1.30&7.12&2DCoud{\'e}&51933.505&11&1800&60000&$1.20\times8.20$&0&--&--\\
1999T1&1.34&1.30&7.12&2DCoud{\'e}&51933.549&11&700&60000&$1.20\times8.20$&0&270&1.9\\
1999T1&1.36&1.30&7.13&2DCoud{\'e}&51935.439&11&1800&60000&$1.20\times8.20$&0&--&--\\
1999T1&1.36&1.30&7.13&2DCoud{\'e}&51935.468&11&1800&60000&$1.20\times8.20$&0&320&0.9\\
1999T1&1.36&1.30&7.13&2DCoud{\'e}&51935.494&11&1800&60000&$1.20\times8.20$&0&270&1.9\\
1999T1&1.36&1.30&7.13&2DCoud{\'e}&51935.522&11&1800&60000&$1.20\times8.20$&0&250&-0.4\\
1999T1&1.36&1.30&7.13&2DCoud{\'e}&51935.545&11&700&60000&$1.20\times8.20$&0&260&1.9\\
2001A2-A&1.36&0.43&8.45&2DCoud{\'e}&52114.341&12&1800&60000&$1.20\times8.20$&0&270&1.8\\
2000WM1&1.08&1.24&6.83&UVES&52340.362&13&1550&80000&$0.45\times8.00$&10&270&0.1\\
2000WM1&1.08&1.24&6.83&UVES&52340.381&13&1550&80000&$0.45\times8.00$&10&270&0.1\\
2000WM1&1.10&1.24&6.83&UVES&52341.368&13&1550&80000&$0.45\times8.00$&10&270&0.1\\
2000WM1&1.10&1.24&6.83&UVES&52341.387&13&1550&80000&$0.45\times8.00$&6&290&0.1\\
2000WM1&1.33&1.24&8.43&UVES&52355.362&14&1550&80000&$0.45\times8.00$&2&290&1.0\\
2000WM1&1.33&1.24&8.43&UVES&52355.380&14&1550&80000&$0.45\times8.00$&2&300&1.0\\
2000WM1&1.34&1.24&8.43&UVES&52356.363&14&1610&80000&$0.45\times8.00$&1&290&1.1\\
2000WM1&1.34&1.24&8.43&UVES&52356.382&14&1610&80000&$0.45\times8.00$&1&310&1.0\\
IZ&0.92&0.42&6.39&2DCoud{\'e}&52386.431&15&1200&60000&$1.20\times8.20$&0&310&2.1\\
IZ&0.92&0.42&6.39&2DCoud{\'e}&52386.447&15&1200&60000&$1.20\times8.20$&0&310&1.2\\
IZ&0.92&0.42&6.39&2DCoud{\'e}&52386.464&15&1200&60000&$1.20\times8.20$&52&320&1.9\\
2002X5&0.70&0.87&4.81&UVES&52689.013&17&2000&80000&$0.45\times8.00$&0&400&1.9\\
2002X5&0.72&0.86&4.81&UVES&52690.007&17&1100&80000&$0.45\times12.00$&20&330&1.2\\
2002X5&0.72&0.86&4.81&UVES&52690.020&17&2000&80000&$0.45\times12.00$&20&330&1.2\\
2002X5&1.06&0.99&9.32&UVES&52705.017&18&1800&80000&$0.45\times8.00$&0&300&1.5\\
2002X5&1.06&0.99&9.32&UVES&52705.039&18&1800&80000&$0.45\times8.00$&3&270&1.2\\
2002X5&1.06&0.99&9.32&UVES&52705.060&18&1800&80000&$0.45\times8.00$&3&270&1.2\\
2002V1&1.22&0.83&7.89&UVES&52647.037&16&2100&80000&$0.45\times8.00$&0&310&1.2\\
2002V1&1.22&0.83&7.89&UVES&52647.062&16&2100&80000&$0.45\times8.00$&0&310&1.2\\
2002V1&1.18&0.84&7.87&UVES&52649.031&16&2100&80000&$0.45\times8.00$&0&320&1.2\\
2002V1&1.18&0.84&7.87&UVES&52649.056&16&1983&80000&$0.45\times8.00$&0&300&1.2\\
2002V1&1.01&1.60&5.47&UVES&52719.985&19&600&80000&$0.45\times8.00$&0&330&1.9\\
2002Y1&1.14&1.56&7.13&UVES&52788.394&20&1800&80000&$0.40\times8.00$&0&310&1.2\\
2002Y1&1.14&1.56&7.13&UVES&52788.416&20&1800&80000&$0.40\times8.00$&0&340&1.6\\
2002Y1&1.16&1.55&7.14&UVES&52789.393&20&1800&80000&$0.40\times8.00$&3&380&1.2\\
2002Y1&1.16&1.55&7.14&UVES&52789.415&20&1800&80000&$0.40\times8.00$&3&380&1.1\\
Howell&1.37&1.68&11.07&UVES&53113.372&22&3600&80000&$0.44\times8.00$&0&260&1.0\\
Howell&1.37&1.67&11.08&UVES&53114.364&22&3600&80000&$0.44\times8.00$&0&250&1.2\\
Howell&1.37&1.67&8.78&UVES&53115.368&22&3600&80000&$0.44\times8.00$&0&220&1.0\\
Howell&1.37&1.67&8.79&UVES&53117.371&22&3600&80000&$0.44\times8.00$&0&260&1.1\\
Howell&1.38&1.65&8.81&UVES&53127.372&23&3600&80000&$0.44\times8.00$&0&260&1.2\\
Howell&1.39&1.65&8.81&UVES&53128.363&23&3600&80000&$0.44\times8.00$&0&250&1.2\\
Howell&1.39&1.64&8.82&UVES&53129.372&23&3600&80000&$0.44\times8.00$&0&240&1.2\\
Howell&1.42&1.63&8.84&UVES&53142.358&29&3600&80000&$0.44\times8.00$&0&270&1.2\\
Howell&1.43&1.62&8.95&UVES&53146.371&29&3600&80000&$0.44\times8.00$&0&250&1.2\\
Howell&1.44&1.62&8.95&UVES&53147.391&29&2499&80000&$0.44\times8.00$&0&260&1.2\\
Howell&1.44&1.62&8.95&UVES&53149.361&29&4900&80000&$0.44\times8.00$&0&260&1.2\\
2002T7&0.68&0.61&4.87&UVES&53131.406&25&1080&80000&$0.44\times8.00$&5&330&1.2\\
2002T7&0.68&0.61&4.87&UVES&53131.421&25&1080&80000&$0.44\times8.00$&5&420&1.8\\
2002T7&0.69&0.57&5.02&UVES&53132.412&28&800&80000&$0.44\times8.00$&110&--&--\\
2002T7&0.69&0.57&5.02&UVES&53132.424&28&800&80000&$0.44\times8.00$&110&--&--\\
2002T7&0.93&0.38&5.18&UVES&53150.983&30&3208&80000&$0.44\times8.00$&0&350&2.1\\
2002T7&0.94&0.41&5.01&UVES&53151.976&30&2677&80000&$0.40\times8.00$&0&310&2.6\\
2002T7&0.94&0.42&5.00&UVES&53152.036&30&1800&80000&$0.40\times8.00$&0&340&2.1\\
2002T7&0.96&0.45&4.84&UVES&53152.970&30&3900&80000&$0.44\times8.00$&70&330&1.1\\
2002T7&0.97&0.48&4.79&UVES&53153.973&30&487&80000&$0.40\times8.00$&70&370&0.6\\
2002T7&0.97&0.48&4.78&UVES&53153.986&30&3600&80000&$0.44\times8.00$&70&340&0.9\\
2002T7&1.17&0.95&6.10&UVES&53166.967&31&3600&80000&$0.44\times8.00$&70&290&1.1\\
2002T7&1.18&0.99&6.12&UVES&53167.967&31&3000&80000&$0.44\times8.00$&245&310&0.9\\
2002T7&1.18&0.99&6.11&UVES&53168.012&31&3000&80000&$0.44\times8.00$&245&290&0.1\\
2002T7&1.20&1.03&6.24&UVES&53168.983&31&3120&80000&$0.44\times8.00$&245&290&0.1\\
2002T7&1.22&1.07&6.36&UVES&53169.983&31&3000&80000&$0.44\times8.00$&245&310&-0.3\\
2001Q4&3.73&3.45&9.51&UVES&52883.293&21&4500&80000&$0.45\times8.00$&0&160&0.1\\
2001Q4&3.73&3.45&9.51&UVES&52883.349&21&4500&80000&$0.45\times8.00$&0&130&0.7\\
2001Q4&3.67&3.37&9.56&UVES&52889.236&21&7200&80000&$0.45\times8.00$&0&140&0.8\\
2001Q4&3.66&3.37&9.56&UVES&52889.320&21&7200&80000&$0.45\times8.00$&0&--&--\\
2001Q4&0.98&0.32&6.06&UVES&53130.958&24&120&80000&$0.44\times8.00$&2&350&2.0\\
2001Q4&0.98&0.32&6.06&UVES&53130.960&24&120&80000&$0.44\times8.00$&3&330&1.2\\
2001Q4&0.98&0.32&6.06&UVES&53130.962&24&120&80000&$0.44\times8.00$&13&420&0.9\\
2001Q4&0.98&0.32&6.06&UVES&53130.965&24&120&80000&$0.44\times8.00$&108&--&--\\
2001Q4&0.98&0.32&6.06&UVES&53130.967&24&600&80000&$0.44\times8.00$&108&330&-0.0\\
2001Q4&0.98&0.32&6.06&UVES&53130.975&24&7200&80000&$0.44\times8.00$&217&330&0.1\\
2001Q4&0.98&0.32&6.06&UVES&53131.066&24&2185&80000&$0.44\times8.00$&13&420&1.0\\
2001Q4&0.98&0.32&6.06&UVES&53131.952&26&1782&80000&$0.44\times8.00$&110&330&0.1\\
2001Q4&0.98&0.32&6.06&UVES&53131.991&26&6300&80000&$0.44\times8.00$&216&400&0.8\\
2001Q4&0.97&0.32&6.06&UVES&53132.065&26&2144&80000&$0.44\times8.00$&13&330&1.2\\
2003K4&2.59&2.35&9.14&UVES&53132.343&27&4380&60000&$0.60\times8.00$&0&200&0.1\\
2003K4&1.19&1.53&5.78&UVES&53328.347&32&1500&80000&$0.44\times10.00$&0&320&1.7\\
2003K4&1.20&1.51&5.80&UVES&53329.344&32&1500&80000&$0.44\times10.00$&0&330&1.7\\
Tempel1&1.55&0.75&10.71&HIRES&53520.363&33&1200&47000&$0.86\times7.00$&0&350&1.2\\
Tempel1&1.55&0.75&10.71&HIRES&53520.378&33&1200&47000&$0.86\times7.00$&0&380&1.0\\
Tempel1&1.55&0.75&10.71&HIRES&53520.392&33&1200&47000&$0.86\times7.00$&0&380&1.1\\
Tempel1&1.54&0.76&10.69&UVES&53523.016&34&5400&80000&$0.44\times10.00$&0&320&1.2\\
Tempel1&1.54&0.76&10.69&UVES&53523.083&34&5400&80000&$0.44\times10.00$&0&290&1.2\\
Tempel1&1.53&0.78&10.65&UVES&53528.025&34&5400&80000&$0.44\times10.00$&0&310&1.2\\
Tempel1&1.53&0.78&10.65&UVES&53528.091&34&5400&80000&$0.44\times10.00$&0&290&1.2\\
Tempel1&1.53&0.78&10.63&UVES&53529.033&34&5400&80000&$0.44\times10.00$&0&320&1.2\\
Tempel1&1.53&0.78&10.63&UVES&53529.102&34&5400&80000&$0.44\times10.00$&0&310&1.2\\
Tempel1&1.51&0.89&10.36&UVES&53553.955&35&7200&80000&$0.44\times10.00$&1&300&0.8\\
Tempel1&1.51&0.89&10.36&UVES&53554.041&35&7200&80000&$0.44\times10.00$&1&280&0.1\\
Tempel1&1.51&0.89&10.34&UVES&53554.954&35&7200&80000&$0.44\times10.00$&1&--&--\\
Tempel1&1.51&0.89&10.34&UVES&53555.041&35&7200&80000&$0.44\times10.00$&1&210&-0.4\\
Tempel1&1.51&0.89&10.34&HIRES&53555.238&36&720&47000&$0.86\times7.00$&0&420&3.2\\
Tempel1&1.51&0.89&10.34&HIRES&53555.250&36&600&47000&$0.86\times7.00$&0&310&3.2\\
Tempel1&1.51&0.89&10.34&HIRES&53555.258&36&600&47000&$0.86\times7.00$&0&210&1.8\\
Tempel1&1.51&0.89&10.34&HIRES&53555.267&36&900&47000&$0.86\times7.00$&0&200&1.7\\
Tempel1&1.51&0.89&10.34&HIRES&53555.278&36&900&47000&$0.86\times7.00$&0&260&2.2\\
Tempel1&1.51&0.89&10.34&HIRES&53555.289&36&900&47000&$0.86\times7.00$&0&240&1.8\\
Tempel1&1.51&0.89&10.34&HIRES&53555.300&36&900&47000&$0.86\times7.00$&0&210&1.2\\
Tempel1&1.51&0.89&10.34&HIRES&53555.311&36&900&47000&$0.86\times7.00$&0&210&1.2\\
Tempel1&1.51&0.89&10.34&HIRES&53555.322&36&900&47000&$0.86\times7.00$&0&230&1.2\\
Tempel1&1.51&0.89&10.34&HIRES&53555.333&36&900&47000&$0.86\times7.00$&0&230&1.2\\
Tempel1&1.51&0.89&10.34&HIRES&53555.344&36&900&47000&$0.86\times7.00$&0&--&--\\
Tempel1&1.51&0.89&10.34&HIRES&53555.355&36&900&47000&$0.86\times7.00$&0&230&1.2\\
Tempel1&1.51&0.89&10.34&HIRES&53555.372&36&1800&47000&$0.86\times7.00$&0&250&1.0\\
Tempel1&1.51&0.89&10.34&HIRES&53555.393&36&1800&47000&$0.86\times7.00$&0&250&1.2\\
Tempel1&1.51&0.90&10.33&UVES&53555.955&37&7200&80000&$0.44\times10.00$&1&260&-0.3\\
Tempel1&1.51&0.90&10.33&UVES&53556.043&37&7800&80000&$0.44\times10.00$&2&260&0.9\\
Tempel1&1.51&0.90&10.41&UVES&53557.007&37&9600&80000&$0.44\times10.00$&1&270&1.1\\
Tempel1&1.51&0.90&10.41&UVES&53557.121&37&4800&80000&$0.44\times10.00$&0&280&-0.0\\
Tempel1&1.51&0.91&10.40&UVES&53557.955&37&7500&80000&$0.44\times10.00$&0&240&0.1\\
Tempel1&1.51&0.91&10.40&UVES&53558.044&37&7500&80000&$0.44\times10.00$&0&--&--\\
Tempel1&1.51&0.91&10.39&UVES&53558.952&37&7500&80000&$0.44\times10.00$&0&280&0.9\\
Tempel1&1.51&0.92&10.39&UVES&53559.041&37&7500&80000&$0.44\times10.00$&0&--&--\\
Tempel1&1.51&0.92&10.38&UVES&53559.954&37&7500&80000&$0.44\times10.00$&0&280&1.2\\
Tempel1&1.51&0.92&10.48&UVES&53560.044&37&7500&80000&$0.44\times10.00$&0&300&0.8\\
Tempel1&1.51&0.93&10.46&UVES&53560.952&37&7800&80000&$0.44\times10.00$&0&290&1.2\\
Tempel1&1.51&0.93&10.46&UVES&53561.045&37&7800&80000&$0.44\times10.00$&0&--&--\\
Tempel1&1.51&0.93&10.45&UVES&53561.953&37&7200&80000&$0.44\times10.00$&0&280&1.2\\
Tempel1&1.51&0.93&10.55&UVES&53562.041&37&7200&80000&$0.44\times10.00$&0&270&0.9\\
Tempel1&1.51&0.94&10.54&UVES&53562.955&37&7200&80000&$0.44\times10.00$&0&280&1.0\\
Tempel1&1.51&0.94&10.54&UVES&53563.041&37&7200&80000&$0.44\times10.00$&0&230&0.1\\
SW3-B&1.05&0.10&11.02&2DCoud{\'e}&53857.209&38&300&60000&$1.20\times8.20$&0&320&2.5\\
SW3-B&1.07&0.12&10.65&2DCoud{\'e}&53860.322&40&1800&60000&$1.20\times8.20$&0&280&2.5\\
SW3-B&0.95&0.15&11.60&UVES&53882.367&41&4800&80000&$0.60\times10.00$&0&390&1.9\\
SW3-B&0.95&0.16&11.50&UVES&53883.353&41&3609&80000&$0.60\times10.00$&2&350&1.2\\
SW3-B&0.95&0.16&11.50&UVES&53883.398&41&2400&80000&$0.60\times10.00$&2&360&1.2\\
SW3-C&1.06&0.10&11.07&2DCoud{\'e}&53858.190&39&1800&60000&$1.20\times8.20$&0&270&2.5\\
SW3-C&1.07&0.11&10.95&2DCoud{\'e}&53859.212&39&1800&60000&$1.20\times8.20$&0&340&1.9\\
SW3-C&1.07&0.12&10.79&2DCoud{\'e}&53860.183&39&1800&60000&$1.20\times8.20$&0&380&2.5\\
SW3-C&0.94&0.25&11.39&UVES&53898.369&42&4800&80000&$1.00\times10.00$&0&340&1.2\\
SW3-C&1.13&0.50&14.52&UVES&53937.345&43&5000&80000&$0.60\times10.00$&0&370&1.7\\
SWAN&1.00&1.02&5.25&2DCoud{\'e}&54039.052&44&1800&60000&$1.20\times8.20$&0&350&1.9\\
SWAN&1.00&1.02&5.25&2DCoud{\'e}&54039.074&44&1800&60000&$1.20\times8.20$&0&350&1.8\\
Holmes&2.44&1.63&1.43&2DCoud{\'e}&54398.400&45&60&60000&$1.20\times8.20$&20&160&3.2\\
Holmes&2.44&1.63&1.43&2DCoud{\'e}&54398.406&45&600&60000&$1.20\times8.20$&20&140&2.5\\
Holmes&2.45&1.63&1.43&2DCoud{\'e}&54401.290&45&300&60000&$1.20\times8.20$&20&180&0.9\\
Holmes&2.45&1.63&1.43&2DCoud{\'e}&54401.300&45&600&60000&$1.20\times8.20$&20&170&1.0\\
Holmes&2.46&1.63&1.43&2DCoud{\'e}&54402.362&45&1200&60000&$1.20\times8.20$&20&200&0.9\\
Holmes&2.46&1.63&1.43&HIRES&54402.573&46&730&47000&$0.86\times7.00$&0&160&2.1\\
Holmes&2.46&1.62&1.45&2DCoud{\'e}&54403.362&47&1200&60000&$1.20\times8.20$&20&160&1.1\\
Holmes&2.46&1.62&1.45&2DCoud{\'e}&54404.383&47&1200&60000&$1.20\times8.20$&20&170&1.0\\
Holmes&2.54&1.64&1.92&2DCoud{\'e}&54422.366&48&1800&60000&$1.20\times8.20$&20&200&1.2\\
Holmes&2.54&1.64&1.92&2DCoud{\'e}&54422.388&48&1800&60000&$1.20\times8.20$&20&180&1.0\\
Holmes&2.54&1.64&1.92&2DCoud{\'e}&54422.410&48&1800&60000&$1.20\times8.20$&20&200&1.9\\
Holmes&2.54&1.64&1.92&2DCoud{\'e}&54422.432&48&1800&60000&$1.20\times8.20$&20&210&0.9\\
Holmes&2.54&1.64&1.92&2DCoud{\'e}&54423.253&48&1800&60000&$1.20\times8.20$&20&210&0.8\\
Holmes&2.54&1.64&1.92&2DCoud{\'e}&54423.275&48&1800&60000&$1.20\times8.20$&20&200&2.0\\
Holmes&2.54&1.64&1.92&2DCoud{\'e}&54423.297&48&1800&60000&$1.20\times8.20$&20&200&0.9\\
Tuttle&1.04&0.36&8.43&UVES&54481.021&49&3600&80000&$0.44\times10.00$&0&400&1.7\\
Tuttle&1.04&0.36&8.42&UVES&54481.071&49&4800&80000&$0.44\times10.00$&0&370&1.7\\
Tuttle&1.03&0.52&8.01&UVES&54493.018&49&3900&80000&$0.44\times10.00$&0&310&1.8\\
Tuttle&1.03&0.52&8.00&UVES&54493.071&49&3900&80000&$0.44\times10.00$&0&290&1.9\\
Tuttle&1.03&0.62&7.73&UVES&54500.017&49&3900&80000&$0.44\times10.00$&0&310&1.9\\
Tuttle&1.03&0.62&7.73&UVES&54500.070&49&3900&80000&$0.44\times10.00$&0&300&1.8\\
Lulin&1.33&0.53&6.97&2DCoud{\'e}&54876.328&50&1800&60000&$1.20\times8.20$&0&280&2.5\\
Lulin&1.33&0.53&6.97&2DCoud{\'e}&54876.351&50&1800&60000&$1.20\times8.20$&0&390&3.1\\
Lulin&1.33&0.53&6.97&2DCoud{\'e}&54876.373&50&1800&60000&$1.20\times8.20$&0&270&2.6\\
Lulin&1.33&0.53&6.97&2DCoud{\'e}&54876.395&50&1800&60000&$1.20\times8.20$&0&300&2.8\\
Lulin&1.33&0.53&6.97&2DCoud{\'e}&54876.418&50&1800&60000&$1.20\times8.20$&0&260&2.5\\
Lulin&1.33&0.53&6.97&2DCoud{\'e}&54876.440&50&1800&60000&$1.20\times8.20$&0&280&2.5\\
Lulin&1.33&0.53&6.97&2DCoud{\'e}&54876.463&50&1800&60000&$1.20\times8.20$&0&290&2.7\\
Lulin&1.33&0.53&6.97&2DCoud{\'e}&54876.486&50&1800&60000&$1.20\times8.20$&0&260&2.5\\
Lulin&1.33&0.53&6.97&2DCoud{\'e}&54876.508&50&1800&60000&$1.20\times8.20$&0&350&2.5\\
Lulin&1.33&0.53&6.87&2DCoud{\'e}&54877.351&50&1800&60000&$1.20\times8.20$&4&270&2.5\\
Lulin&1.33&0.53&6.87&2DCoud{\'e}&54877.374&50&1800&60000&$1.20\times8.20$&4&280&2.6\\
Lulin&1.33&0.53&6.87&2DCoud{\'e}&54877.397&50&1800&60000&$1.20\times8.20$&4&270&2.5\\
Lulin&1.33&0.53&6.87&2DCoud{\'e}&54877.419&50&1800&60000&$1.20\times8.20$&4&270&2.5\\
Lulin&1.33&0.53&6.87&2DCoud{\'e}&54877.441&50&1800&60000&$1.20\times8.20$&4&280&2.5\\
Lulin&1.33&0.53&6.87&2DCoud{\'e}&54877.463&50&1800&60000&$1.20\times8.20$&4&260&2.6\\
Lulin&1.33&0.53&6.87&2DCoud{\'e}&54877.485&50&1800&60000&$1.20\times8.20$&4&270&2.7\\
Lulin&1.33&0.53&6.87&2DCoud{\'e}&54877.507&50&1800&60000&$1.20\times8.20$&4&260&2.5\\
\end{longtable}
}

\begin{acknowledgements}
{ This paper includes} data taken at the McDonald 
Observatory of the University of Texas at Austin,
at the 
W. M. Keck Observatory, which is operated as a scientific 
partnership by  the California Institute of Technology, 
the University of California, and the National Aeronautics and Space Administration,
and at the Nordic Optical
Telescope (NOT), European Northern Observatory, La Palma,
Spain.
IRAF is distributed by the National Optical Astronomy Observatory, 
which is operated by the Association of Universities for Research in Astronomy (AURA) 
under cooperative agreement with the National Science Foundation.
\end{acknowledgements}

\clearpage
\Online 
\begin{appendix}
\section{Estimating the isotopic ratios}
\label{appEstimating}
 
Because of 
uncertainties in the models and systematic errors in the
observations, the parameters $\alpha$ and $\beta$ (Eq.~\ref{eq1}) cannot be 
estimated directly. Additional parameters are required to deal with the
exact central wavelength of the lines, and the exact level
of a possible residual background $C_k(\lambda)$.

The spectra are divided into 
small domains    
surrounding the central 
wavelength $\lambda_i$ of the most intense, unblended 
$R$ lines of {$^{13}$C$^{14}$N}{}{} and {$^{12}$C$^{15}$N}{}{}, i.e., regions where $\alpha S_{k,2}(\lambda)$
or $\beta S_{k,3}(\lambda)\gg S_{k,1}(\lambda)$. 
This is necessary because the accuracy of the {$^{12}$C$^{14}$N}{}{} model 
is not perfect, especially in the wings of intense lines. 
Lines of other molecules must also be avoided, e.g., an
unidentified feature at $\lambda\sim3867.92$~\AA\ 
precludes the use of the R10 line of {$^{12}$C$^{15}$N}{}\ 
close to the nucleus (see Section \ref{sec:analysis}).

The line profile of the strongest lines can be
fitted by a Gaussian, or their
intensity can be estimated by direct integration, providing 
sets of 
$\alpha$ and $\beta$ which can then be averaged.

\begin{figure}
   \resizebox{\hsize}{!}{\includegraphics[angle=-90]{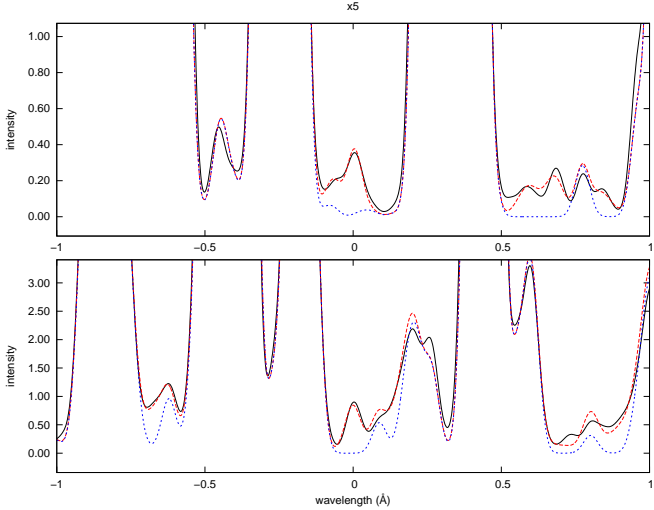}}
\vspace{5mm}
      \caption{Coadded observed (solid line) and synthetic (with and without the
      rare isotopologues) 
     spectra of comet C/2002 X5 (Kudo-Fujikawa).
The upper panel is centered
 on {$^{13}$C$^{14}$N}{}\ lines (in this instance, R1--R6, R14 and R16), 
 the lower one on {$^{12}$C$^{15}$N}{}\ lines (R1--R5, R11--R13, 
 R15--R17).
The large number of useable isotopic lines is allowed by the good quality of
 the spectra used in the combination. { Intensity is in arbitrary units.} 
             }
         \label{plotaddexample}
\end{figure}

However, in order to reduce the number of 
free parameters, we used a different procedure. Instead of fitting separately  
$O(\lambda) - C_i$ over the intervals 
$[\lambda_i-\Delta\lambda/2,\lambda_i+\Delta\lambda/2]$,
we superimpose the profiles by shifting the line centers 
to $\lambda=0$, optimally coadd them, 
and fit the resulting  
profile over the resulting domain 
$[-\Delta\lambda/2,\Delta\lambda/2]$ (see Fig.~\ref{plotaddexample}). Hence, 
dropping the subscript $k$, we write, 
for the {$^{13}$C$^{14}$N}{}{} lines,
\begin{equation}
\sum_i w_i [O(\lambda+\lambda_i) - C_i]
\equiv\sum_i w_i O(\lambda+\lambda_i) - C
=\alpha \sum_i w_i S_2(\lambda+\lambda_i)
\label{eq2}
\end{equation}
and the corresponding formula with $S_3$  and $\beta$ for the
{$^{12}$C$^{15}$N}{}{} lines.    
The $C_i$s merge into a single free parameter $\sum_i w_i C_i\equiv C$.
The weight factor $w_i$ is estimated from the expected intensity
$f(\lambda_i) I_i$
of line $i$ in the SN-normalized spectrum (Eqs.~\ref{eqSN1},\ref{eqSN}).

The width of the observed ``coadded" 
profile  
$\sum_i w_i O(\lambda+\lambda_i)$
is found to be equal to that of the synthetic profile  
$\sum_i w_i S_2(\lambda+\lambda_i)$ (or $\sum_i w_i S_3(\lambda+\lambda_i)$)
and it is symmetric about zero.  This confirms that the
identification of the {$^{13}$C$^{14}$N}{}{} and {$^{12}$C$^{15}$N}{}{} lines
is correct, as well as the
theoretical wavelengths adopted for them.
 
The analysis  is done either by profile fitting around $\lambda=0$ 
or by direct
integration. In the latter case we write  
\begin{equation}
\int_{-\Delta\lambda/2}^{\Delta\lambda/2}  
 [\sum_i w_i O(\lambda+\lambda_i)- C]\delta\lambda=
 \alpha \int_{-\Delta\lambda/2}^{\Delta\lambda/2} 
  \sum_i w_i S_2(\lambda+\lambda_i)\delta\lambda
\label{eq3}
\end{equation}
for the {$^{13}$C$^{14}$N}{}{} lines, and an equivalent formula for the
{$^{12}$C$^{15}$N}{}{} lines. 

The choice of the lines $i$ depends on the quality of the
spectra and on particular circumstances, especially the 
heliocentric distance. Figures~\ref{plotCNred} and \ref{plotCN15red} 
shows that the width of the envelope of the CN band decreases 
at large $r$. The relative intensity of the lines of high 
quantum number drops rapidly. Many lines could be used efficiently 
for the coadded profiles of comets de Vico or  X5 (Fig.~\ref{plotaddexample}),
up to 11 for {$^{12}$C$^{15}$N}{}\ and 8 for {$^{13}$C$^{14}$N}{}. On the contrary, 
at large $r$, a few lines dominate overwhelmingly.
As shown in Section~\ref{sec:analysis}, some blends may 
become less troublesome far from the nucleus.

$R$ lines of {$^{13}$C$^{14}$N}{}{} and {$^{12}$C$^{15}$N}{}{} with low quantum numbers are slightly blended
and also -- depending on the spectral resolution -- 
with the corresponding R line of {$^{12}$C$^{14}$N}{}. An additional difficulty is
the presence of faint lines of the B-X (0-0) band of CH.
This region of the spectrum needs special care (e.g., some iterative procedure)
and may have to be ignored for the lowest quality spectra.
\end{appendix}
\begin{appendix}
\section{Averages}
\label{appAverages}
While weighted { averages} of measurements $x_i$ with errors $e_i$ ($i=1,\ldots,n$)
are easily defined as $m={\sum_i x_i e^{-2}_i}/{\sum_i e^{-2}_i}$, estimating the resulting error
on this value is less obvious. The global data set is far from homogeneous.  Systematic errors
affect the various data sets in different ways. The instrumentation
and the circumstances are never identical.
Two estimates of the standard error $S$ are sometimes used    :

\begin{equation}
S^2_1=\frac{\sum_i (x_i-m)^2  e^{-2}_i}{(n-1)\sum_i e^{-2}_i}
\label{eq5}
\end{equation}
and
\begin{equation}
S^2_2=1/\sum_i e^{-2}_i .
\label{eq6}
\end{equation}

They are not satisfying, particularly for { a small} data set. Equation~\ref{eq5}
does not take properly
into account the individual errors, except for the weighting factors, so that
a few  $x_i$ with large $e_i$ but grouped by chance around $m$ would
give an unrealistically small $S_1$.
On the other hand
Eq.~\ref{eq6} does not take into account the inter-group variations which can
be large in the case of systematic effects. The larger of $S_1$ and $S_2$
may be taken, but we choose a different approach by simulating the  $x_i$
as the average of $n_i$ individual observations $y_{i,j}$ ($j=1,\ldots,n_i$)
with a standard deviation
$e_i$. The weighting is obtained by taking $n_i$ proportional to $e^{-2}_i$
i.e., the standard deviation is the same for each data set, $\sigma_i=\sigma$
and $n_i=\sigma^2 e_i^{-2}$.
It is then possible to combine several $x_i$ by merging their
respective data sets. The mean value is the same as above and the standard deviation
of the mean is given by
\begin{equation}
e=\left(\frac{\sigma^2 \sum_i (n_i-1) +\sum_i n_i x_i^2 -(\sum_i x_i n_i)^2/\sum_i n_i}{\sum_i n_i (\sum_i n_i-1)}\right)^{1/2}.
\label{eq9}
\end{equation}

For large $\sigma$, this is equivalent to Eq.~\ref{eq6}, i.e.,
the result is dominated by the internal dispersion of each data set.
Choosing a reasonable value of $\sigma$ is thus critical.
We adopt the value of $\sigma$ yielding the smallest realistic samples,
$\min(n_i)=2$. This gives the largest, conservative estimates of the errors.
\end{appendix}
\begin{appendix}
\section{Spectra}
\label{sec:spectra}
  \begin{figure*}
   \includegraphics[width=15.5cm]{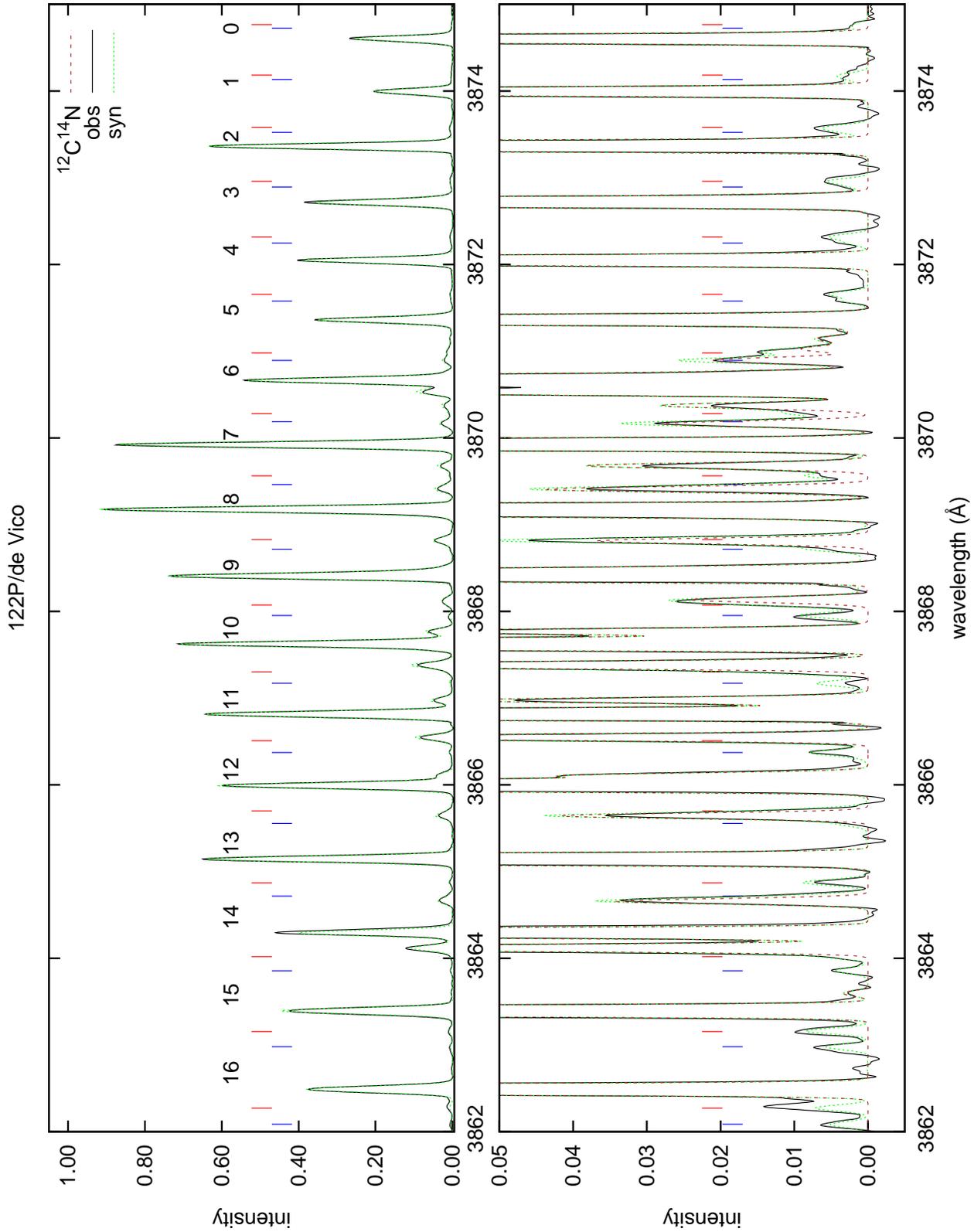}
      \caption{Observed (2DCoud{\'e}) and synthetic (dotted) spectra of comet 122P/de Vico.
      In this and the following graphs, the upper (red) ticks indicate the position 
        of the major R lines of {$^{13}$C$^{14}$N}{}\, the lower (blue) ticks indicate the position
        of the major R lines of {$^{12}$C$^{15}$N}{}. The corresponding quantum numbers  are indicated 
       in the upper panel midway between the strong {$^{12}$C$^{14}$N}{}\ lines and the faint
       isotopic lines. { The intensity scale is in relative units.} 
             }
         \label{plotdeV}
  \end{figure*}
  \clearpage
  \begin{figure*}
   \includegraphics[width=15.5cm]{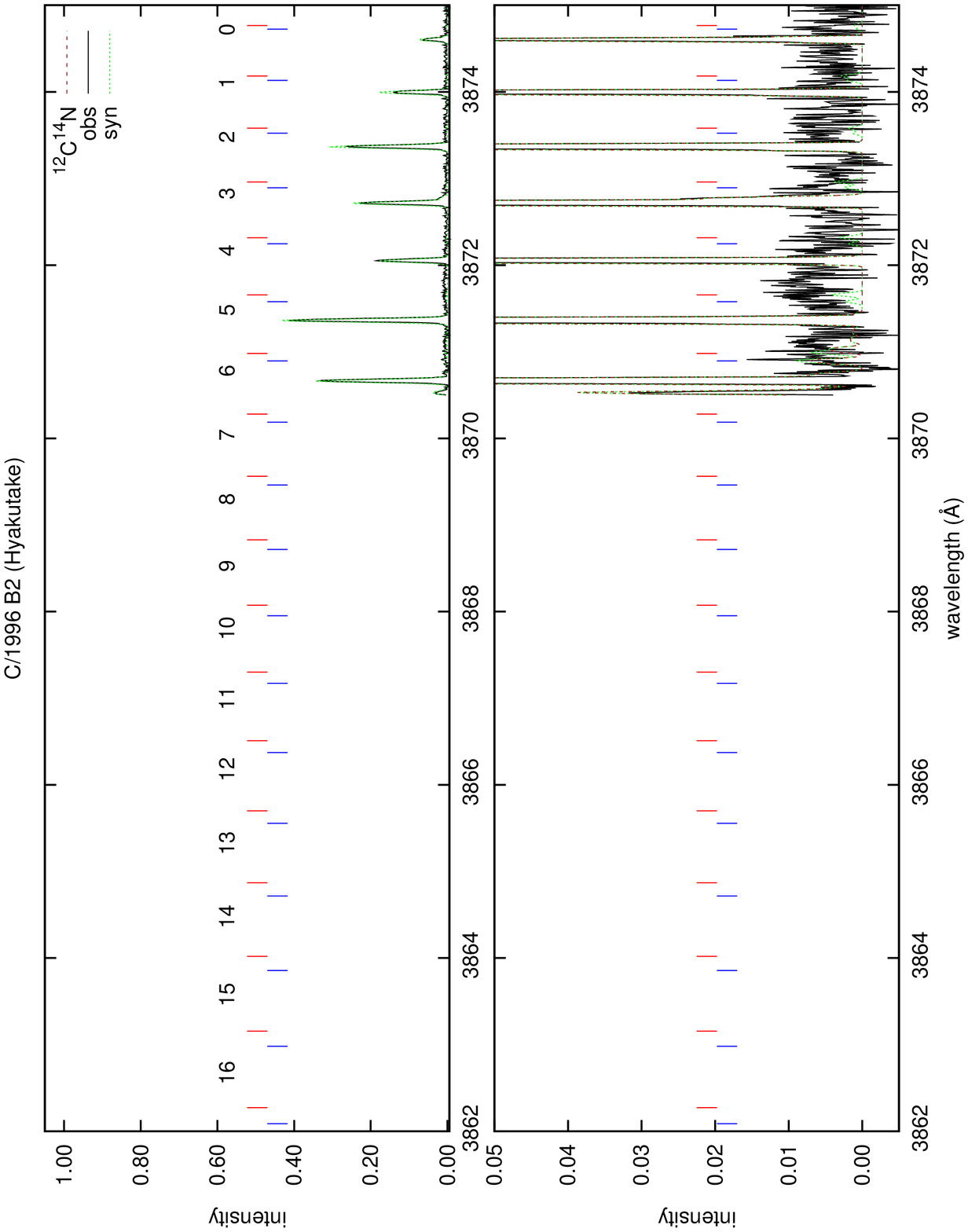}
      \caption{Observed (2DCoud{\'e}) and synthetic (dotted) spectra of comet C/1996 B2 (Hyakutake)
             }
         \label{plothyakhi}
  \end{figure*}
  \clearpage
  \begin{figure*}
   \includegraphics[width=15.5cm]{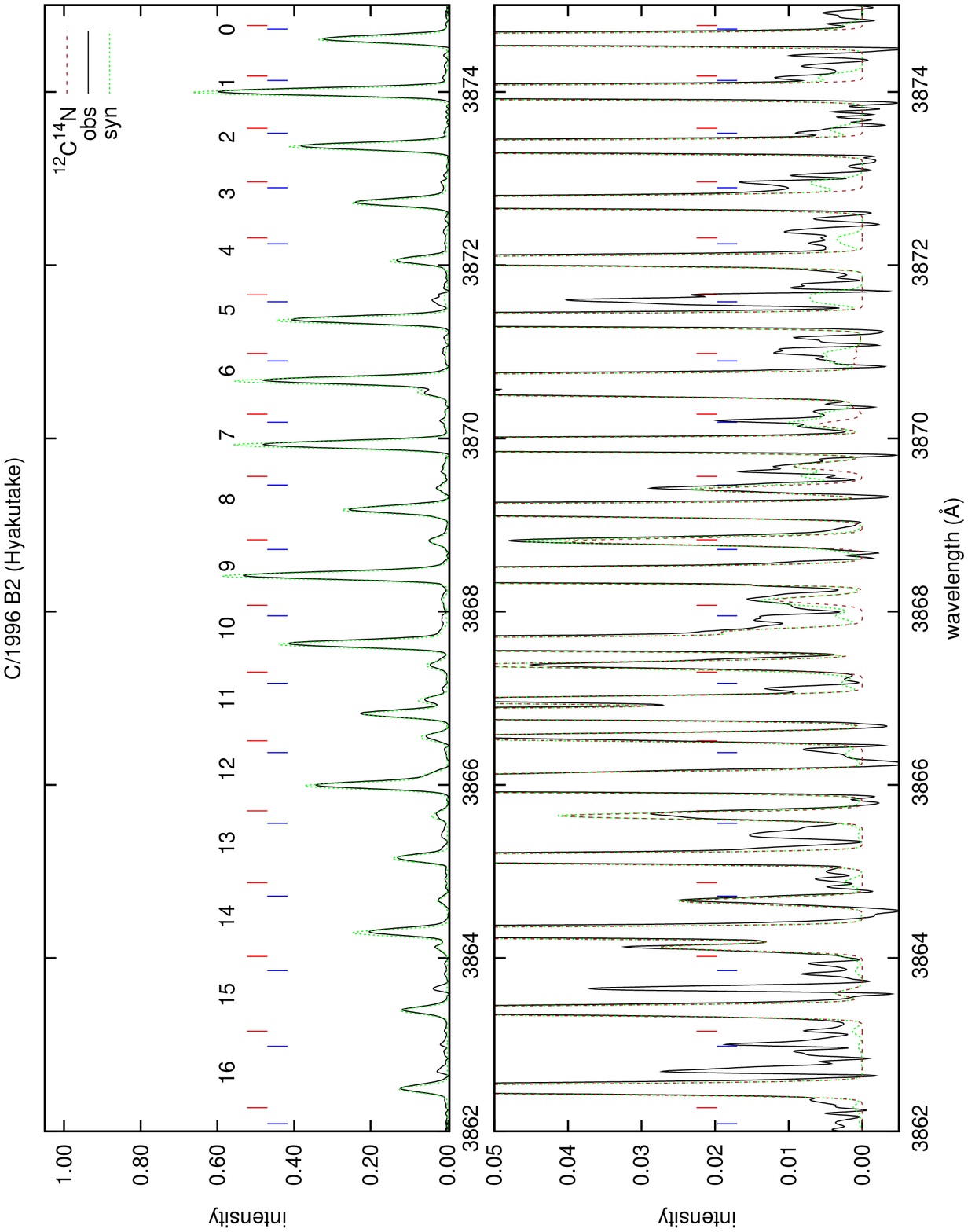}
      \caption{Observed (2DCoud{\'e}) and synthetic (dotted) spectra of comet C/1996 B2 (Hyakutake)
             }
         \label{plothyaklo}
  \end{figure*}
  \clearpage
  \begin{figure*}
   \includegraphics[width=15.5cm]{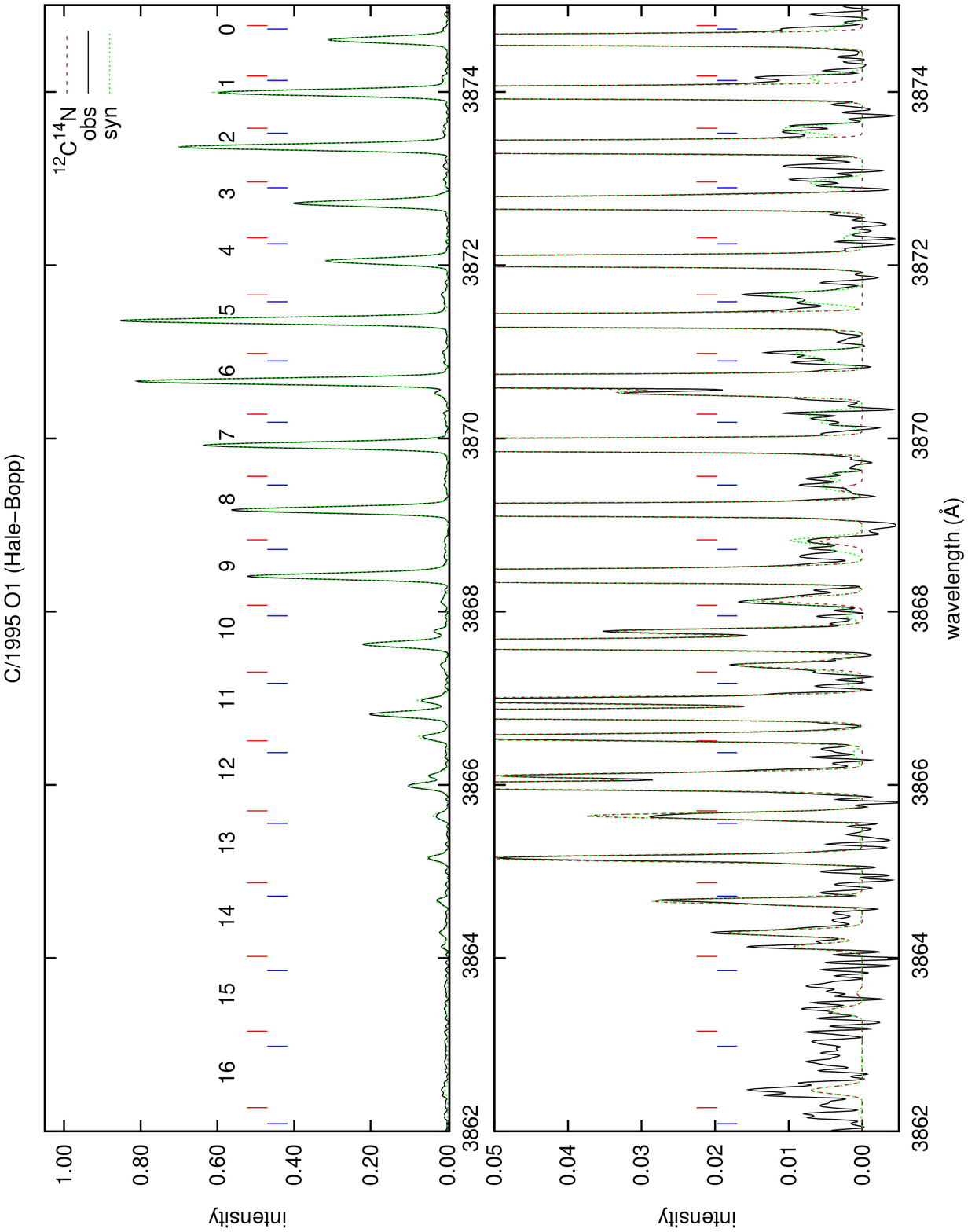}
      \caption{Observed (2DCoud{\'e}) and synthetic (dotted) spectra of comet C/1995 O1 (Hale-Bopp)
             }
         \label{plothb}
  \end{figure*}
  \clearpage
  \begin{figure*}
   \includegraphics[width=15.5cm]{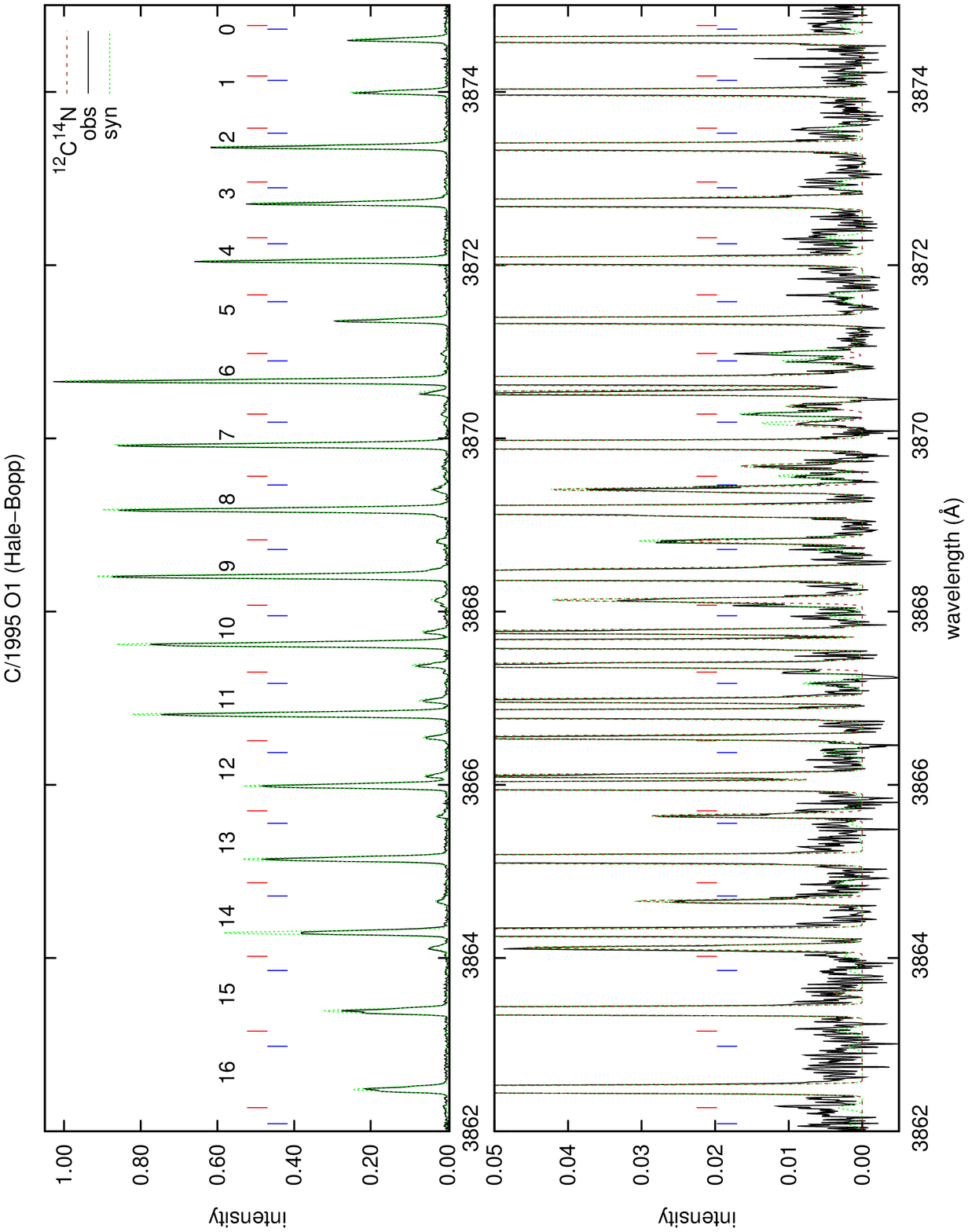}
      \caption{Observed (2DCoud{\'e}) and synthetic (dotted) spectra of comet C/1995 O1 (Hale-Bopp)
             }
         \label{plothbhi}
  \end{figure*}
  \clearpage
  \begin{figure*}
   \includegraphics[width=15.5cm]{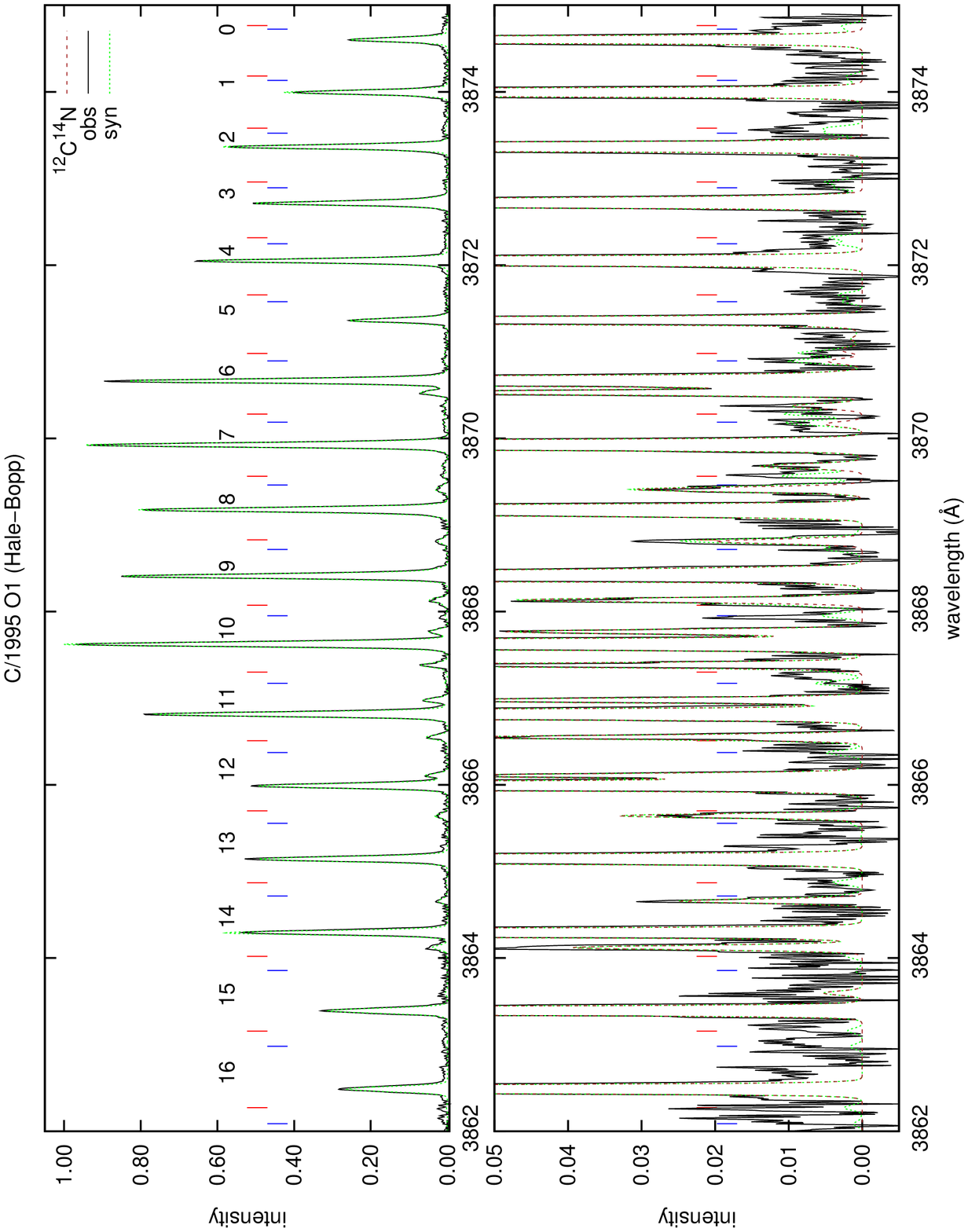}
      \caption{Observed (SOFIN) and synthetic (dotted) spectra of comet C/1995 O1 (Hale-Bopp)
             }
         \label{plothbnot}
  \end{figure*}
  \clearpage
  \begin{figure*}
   \includegraphics[width=15.5cm]{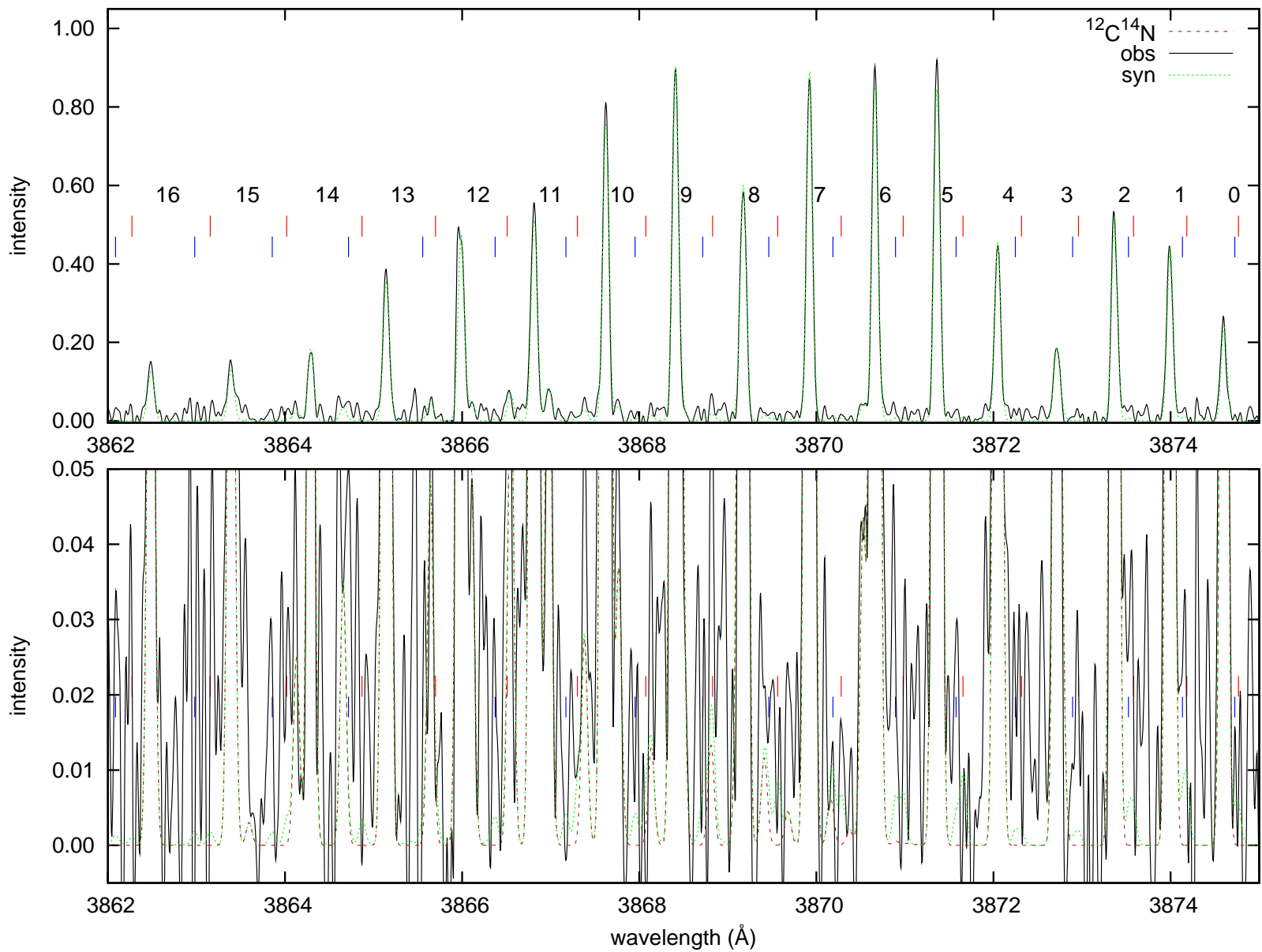}
      \caption{Observed (2DCoud{\'e}) and synthetic (dotted) spectra of comet 55P/Tempel-Tuttle  
             }
         \label{plottt}
  \end{figure*}
  \clearpage
  \begin{figure*}
   \includegraphics[width=15.5cm]{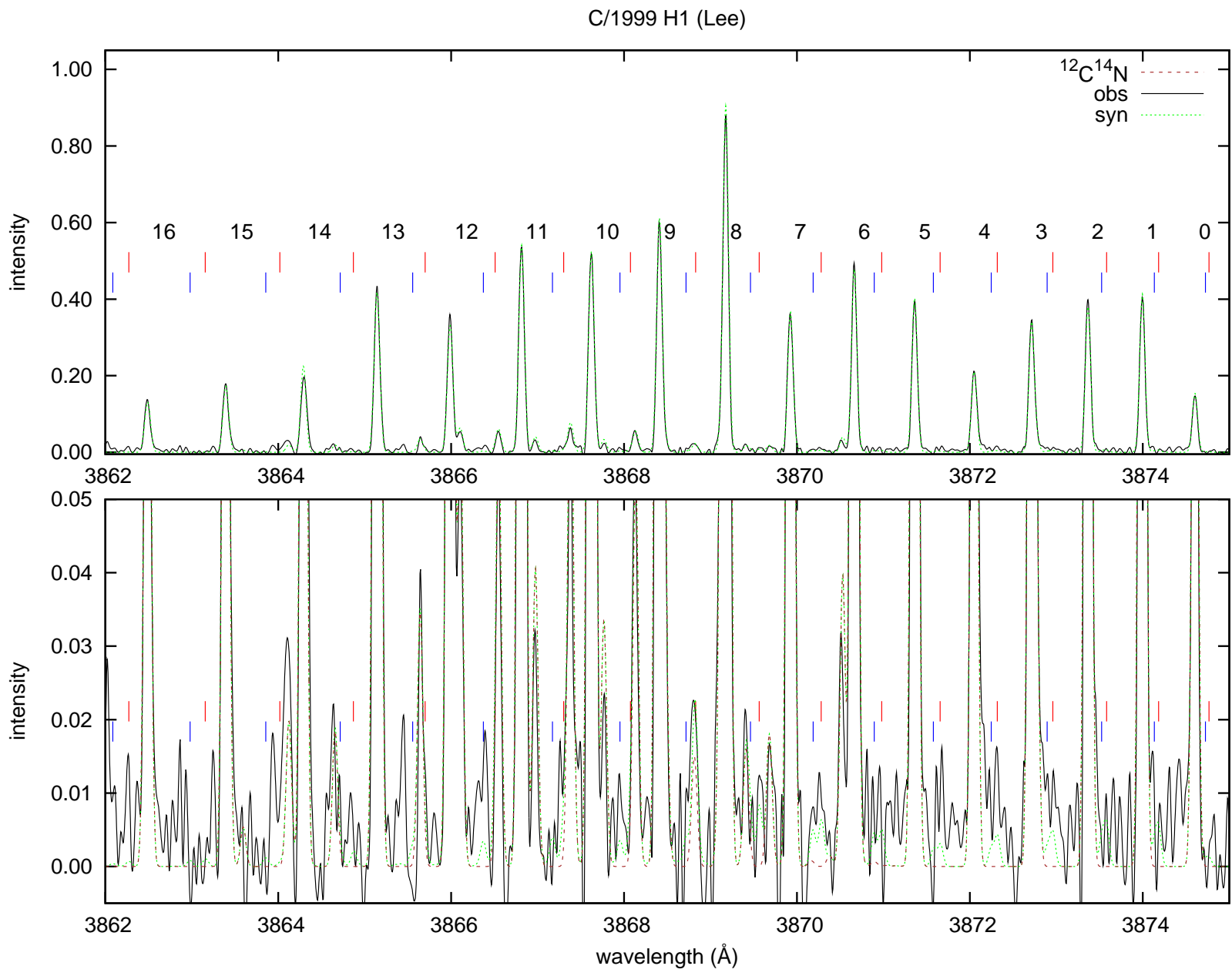}
      \caption{Observed (2DCoud{\'e}) and synthetic (dotted) spectra of comet C/1999 H1 (Lee)
             }
         \label{plotlee}
  \end{figure*}
  \clearpage
  \begin{figure*}
   \includegraphics[width=15.5cm]{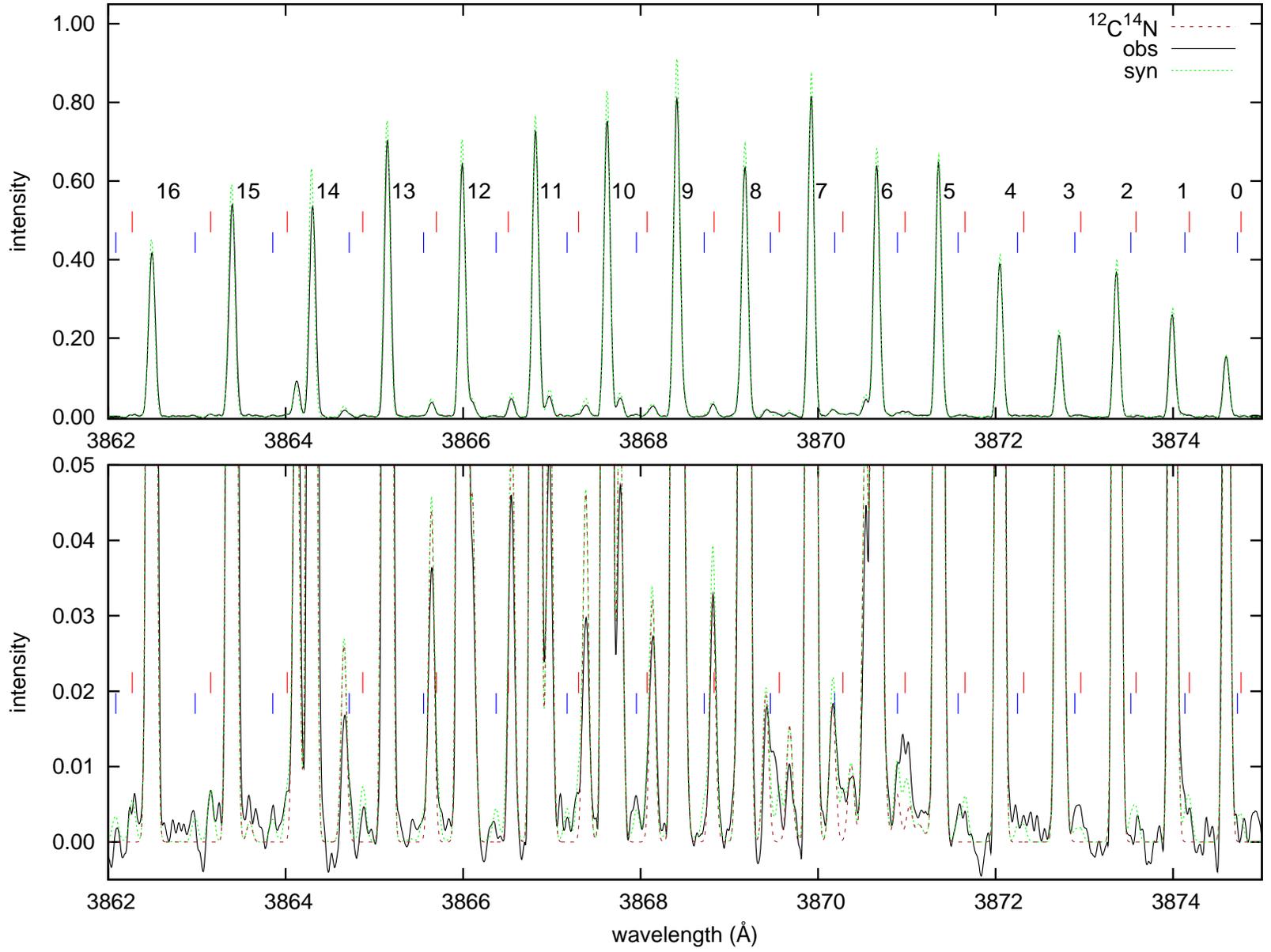}
      \caption{Observed (2DCoud{\'e}) and synthetic (dotted) spectra of comet C/1999 S4 (LINEAR)
             }
         \label{plotlins4}
  \end{figure*}
  \clearpage
  \begin{figure*}
   \includegraphics[width=15.5cm]{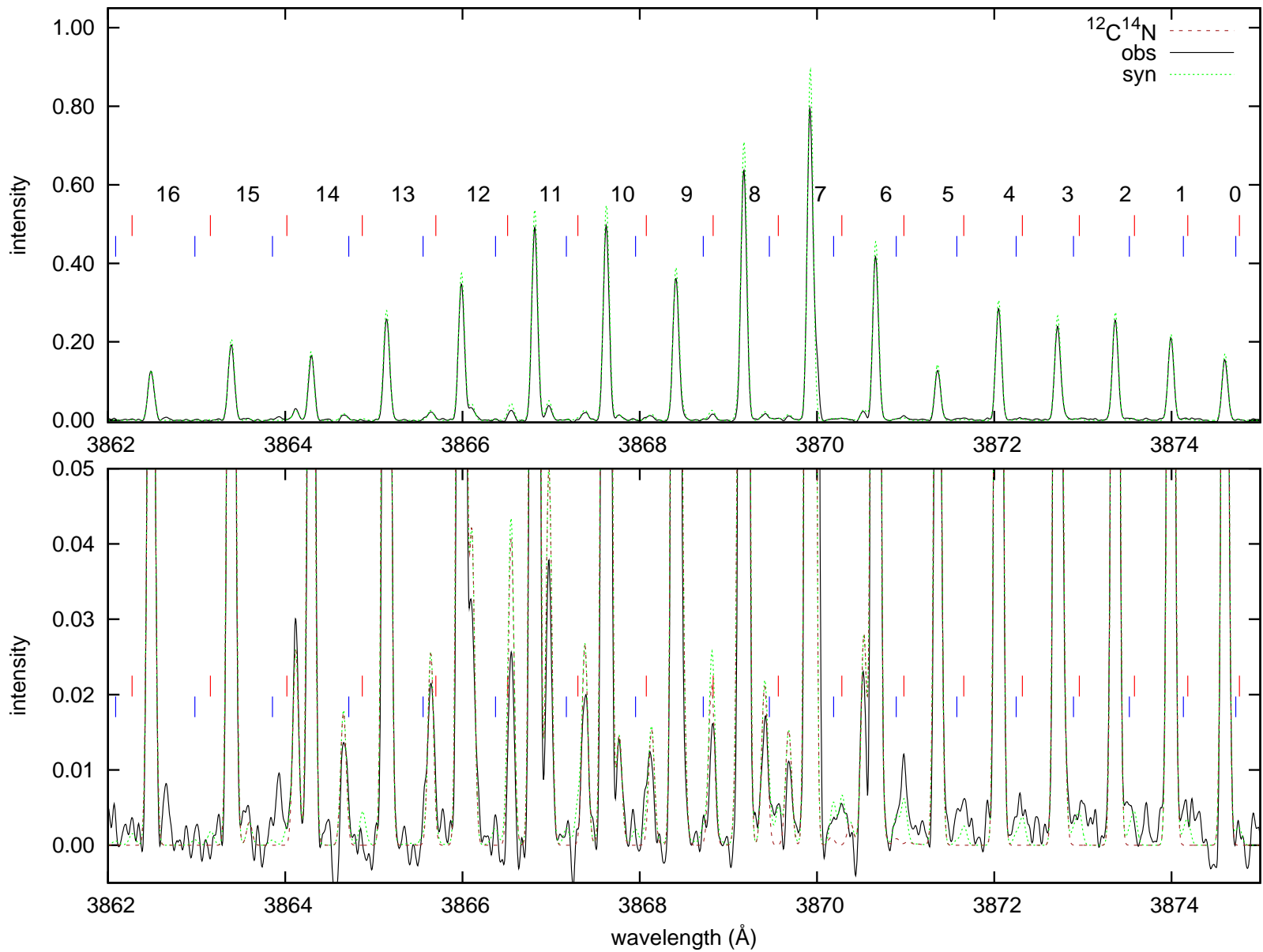}
      \caption{Observed (2DCoud{\'e}) and synthetic (dotted) spectra of comet C/1999 T1 (McNaught-Hartley)
             }
         \label{plott1}
  \end{figure*}
  \clearpage
  \begin{figure*}
   \includegraphics[width=15.5cm]{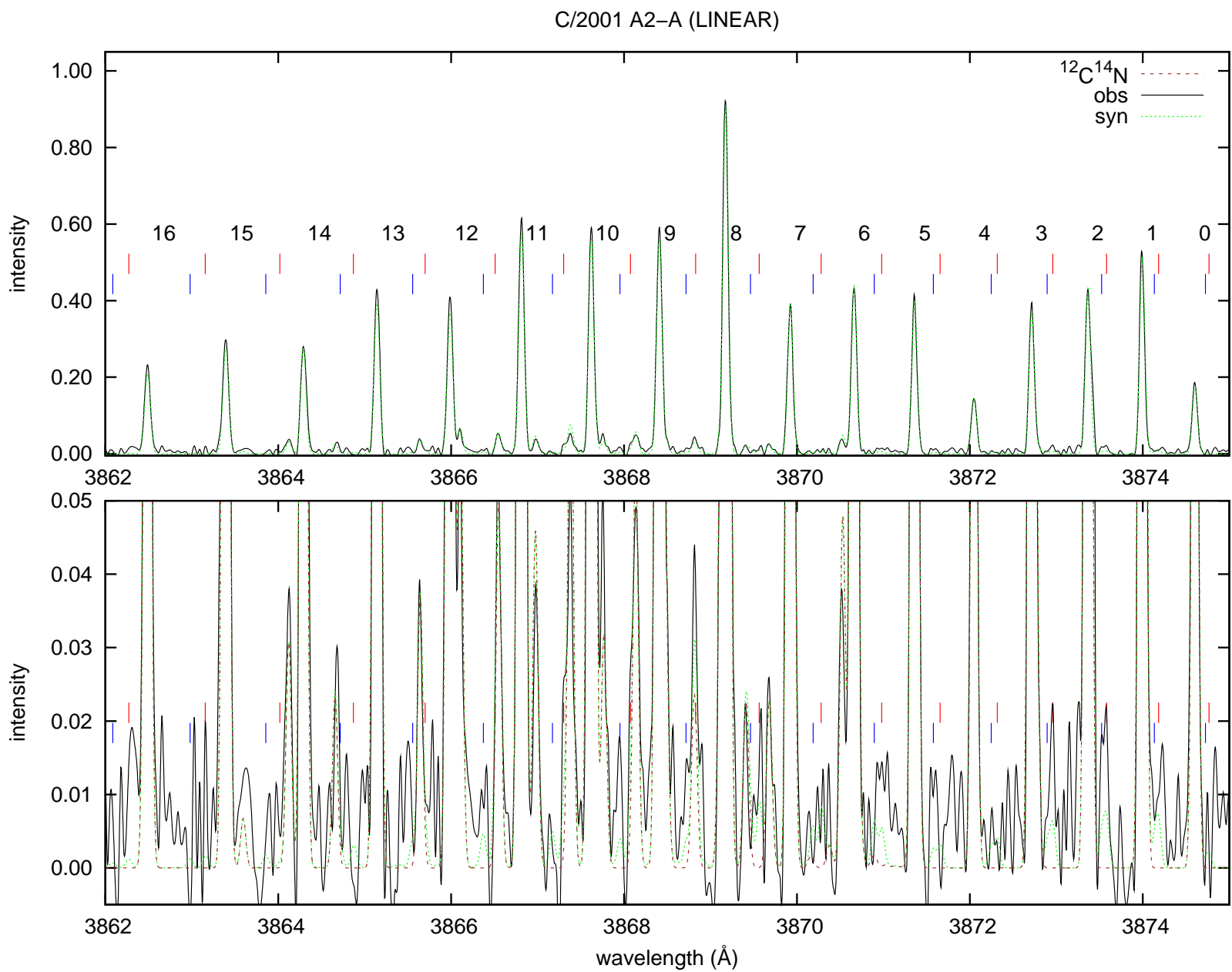}
      \caption{Observed (2DCoud{\'e}) and synthetic (dotted) spectra of comet C/2001 A2-A (LINEAR)
             }
         \label{plota2}
  \end{figure*}
  \clearpage
  \begin{figure*}
   \includegraphics[width=15.5cm]{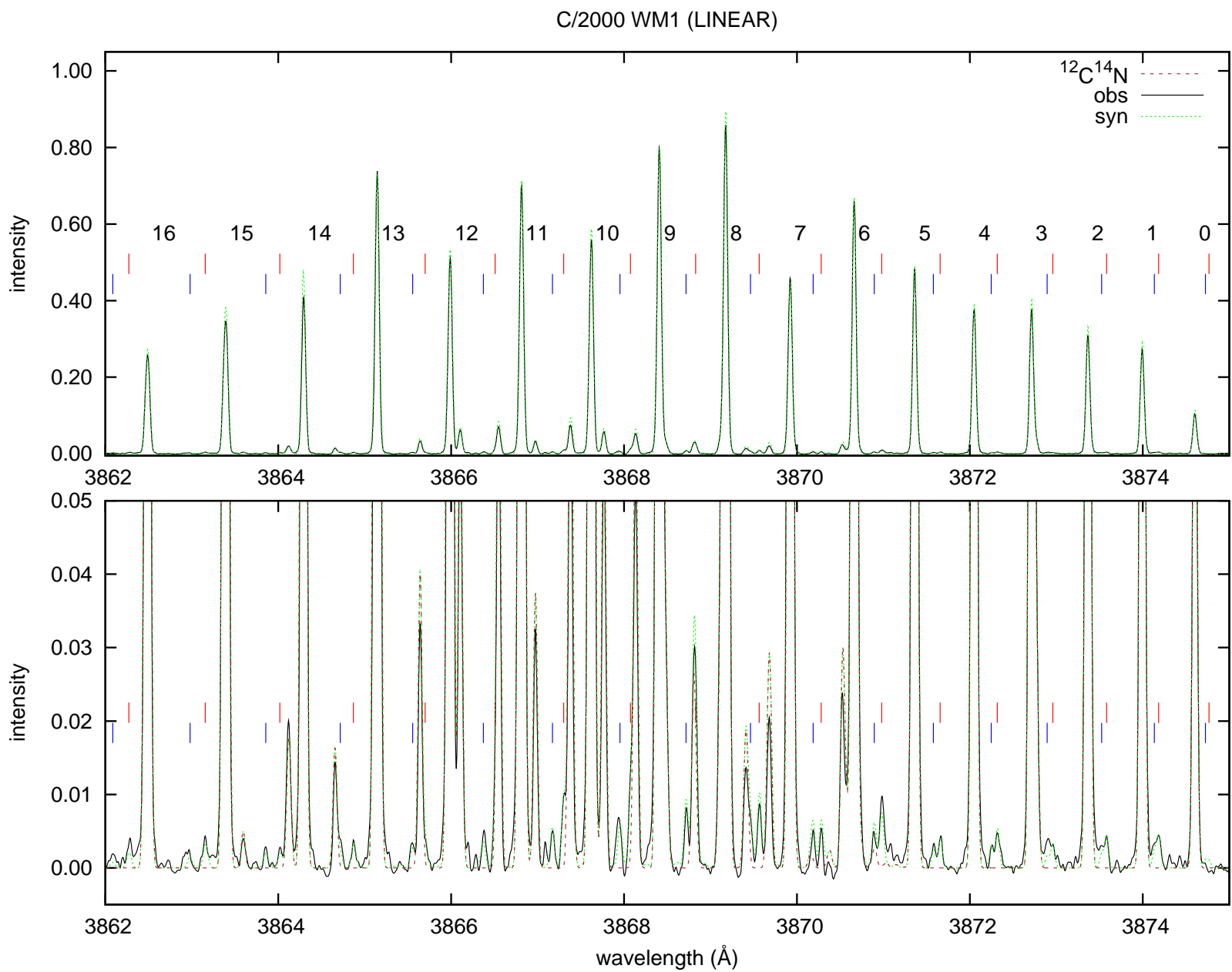}
      \caption{Observed (UVES) and synthetic (dotted) spectra of comet C/2000 WM1 (LINEAR)
             }
         \label{plotwm1}
  \end{figure*}
  \clearpage
  \begin{figure*}
   \includegraphics[width=15.5cm]{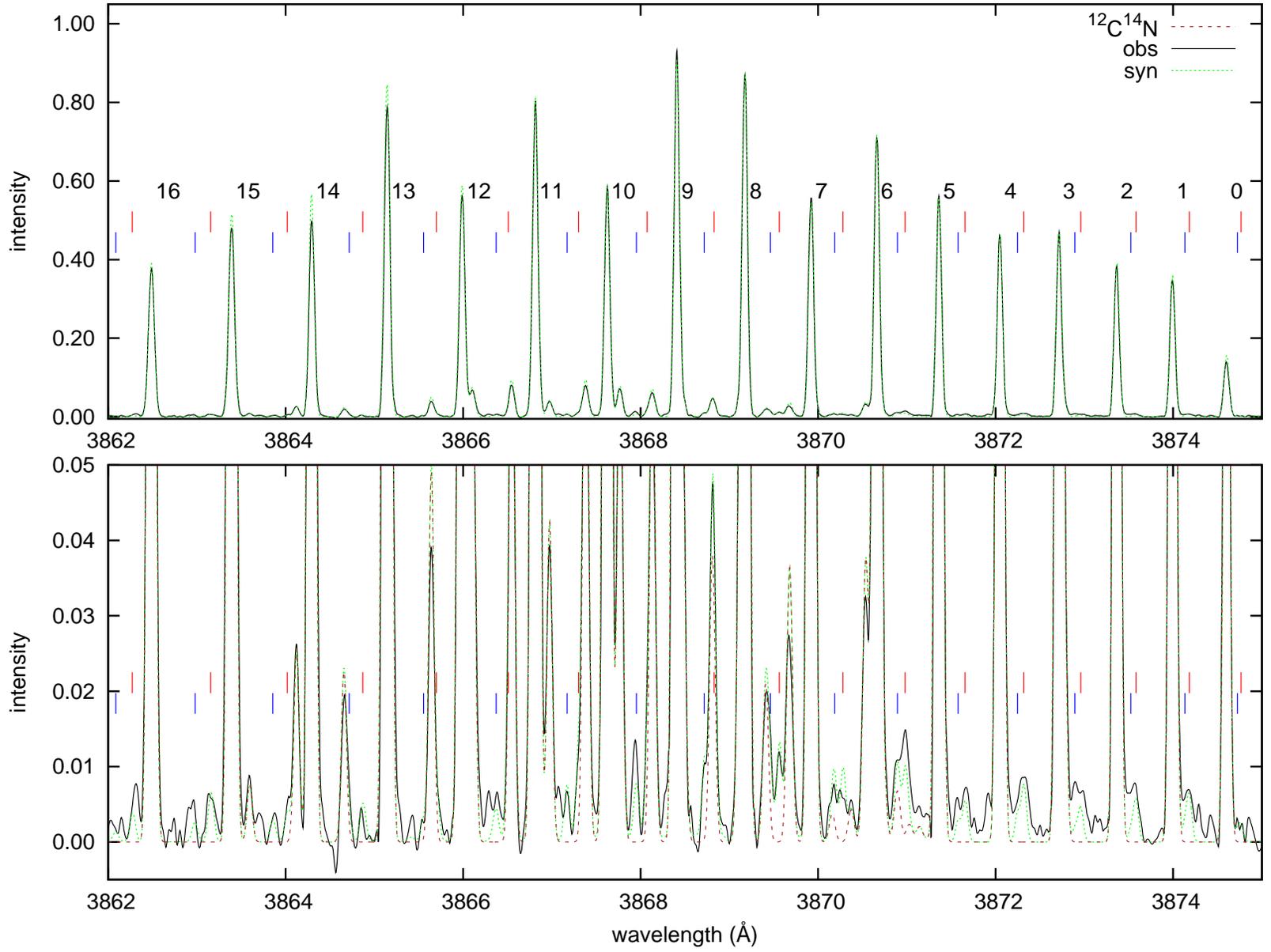}
      \caption{Observed (2DCoud{\'e}) and synthetic (dotted) spectra of comet 153P/Ikeya-Zhang  
             }
         \label{plotiz}
  \end{figure*}
  \clearpage
  \begin{figure*}
   \includegraphics[width=15.5cm]{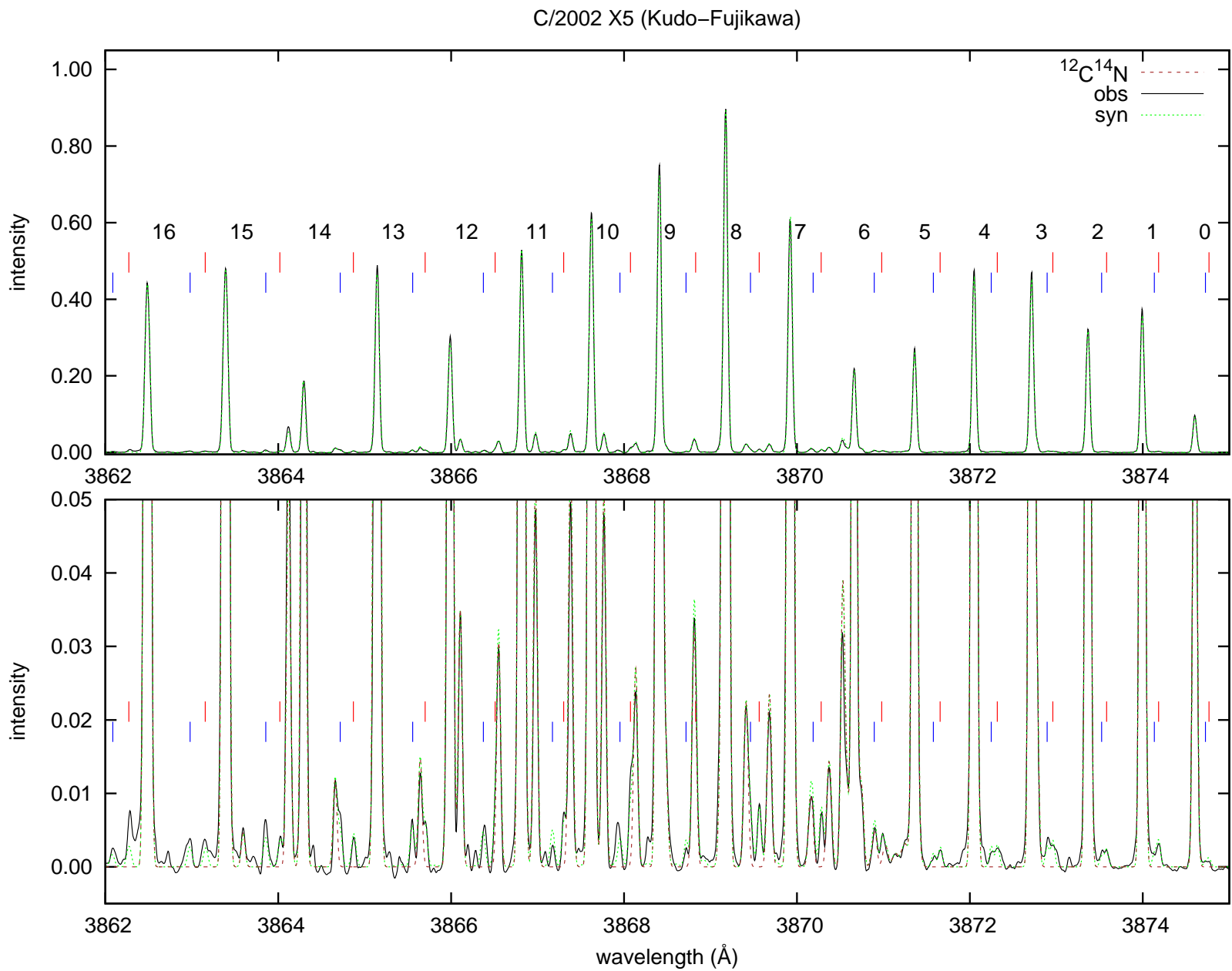}
      \caption{Observed (UVES) and synthetic (dotted) spectra of comet C/2002 X5 (Kudo-Fujikawa)
             }
         \label{plotx5}
  \end{figure*}
  \clearpage
  \begin{figure*}
   \includegraphics[width=15.5cm]{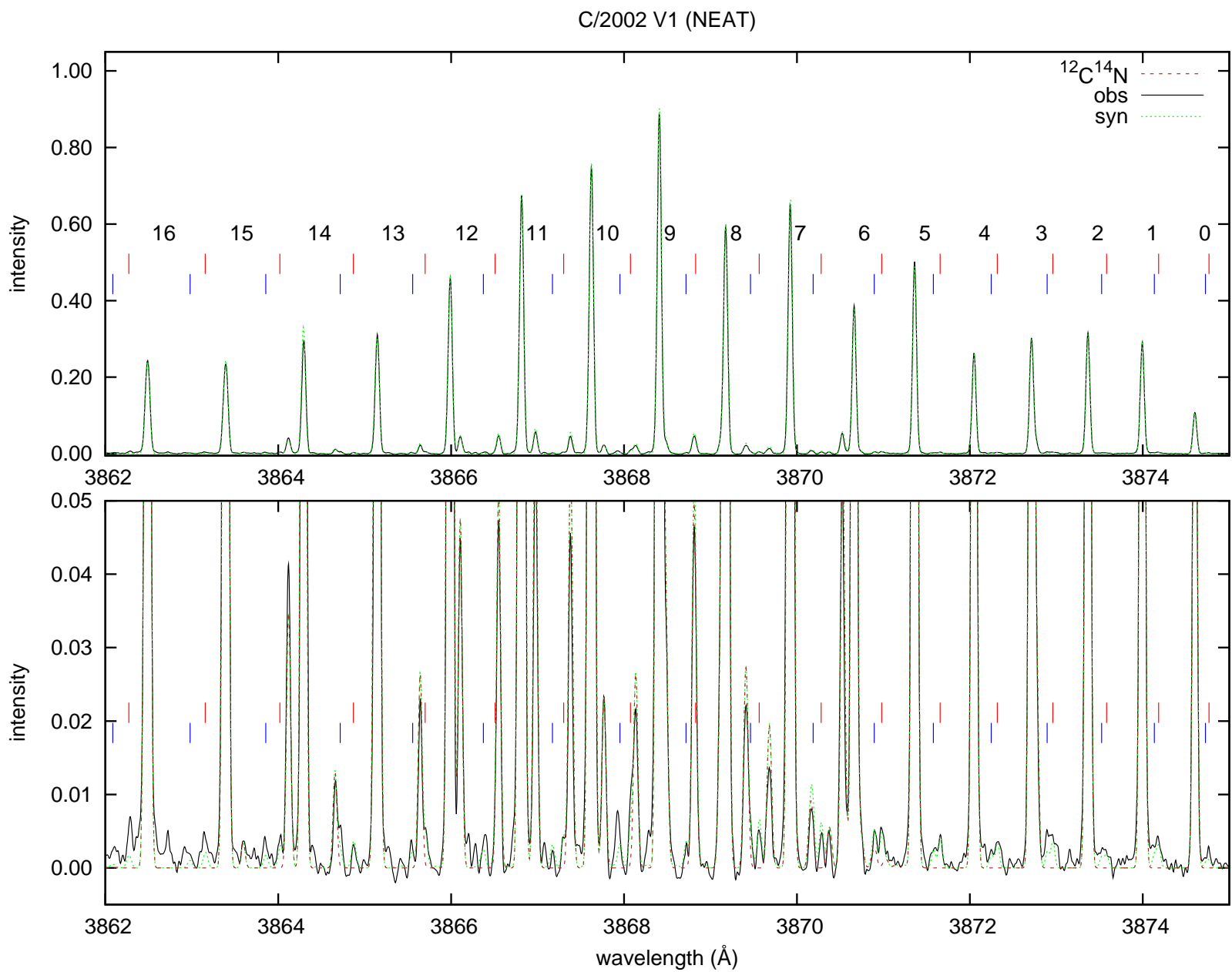}
      \caption{Observed (UVES) and synthetic (dotted) spectra of comet C/2002 V1 (NEAT)
             }
         \label{plotv1}
  \end{figure*}
  \clearpage
  \begin{figure*}
   \includegraphics[width=15.5cm]{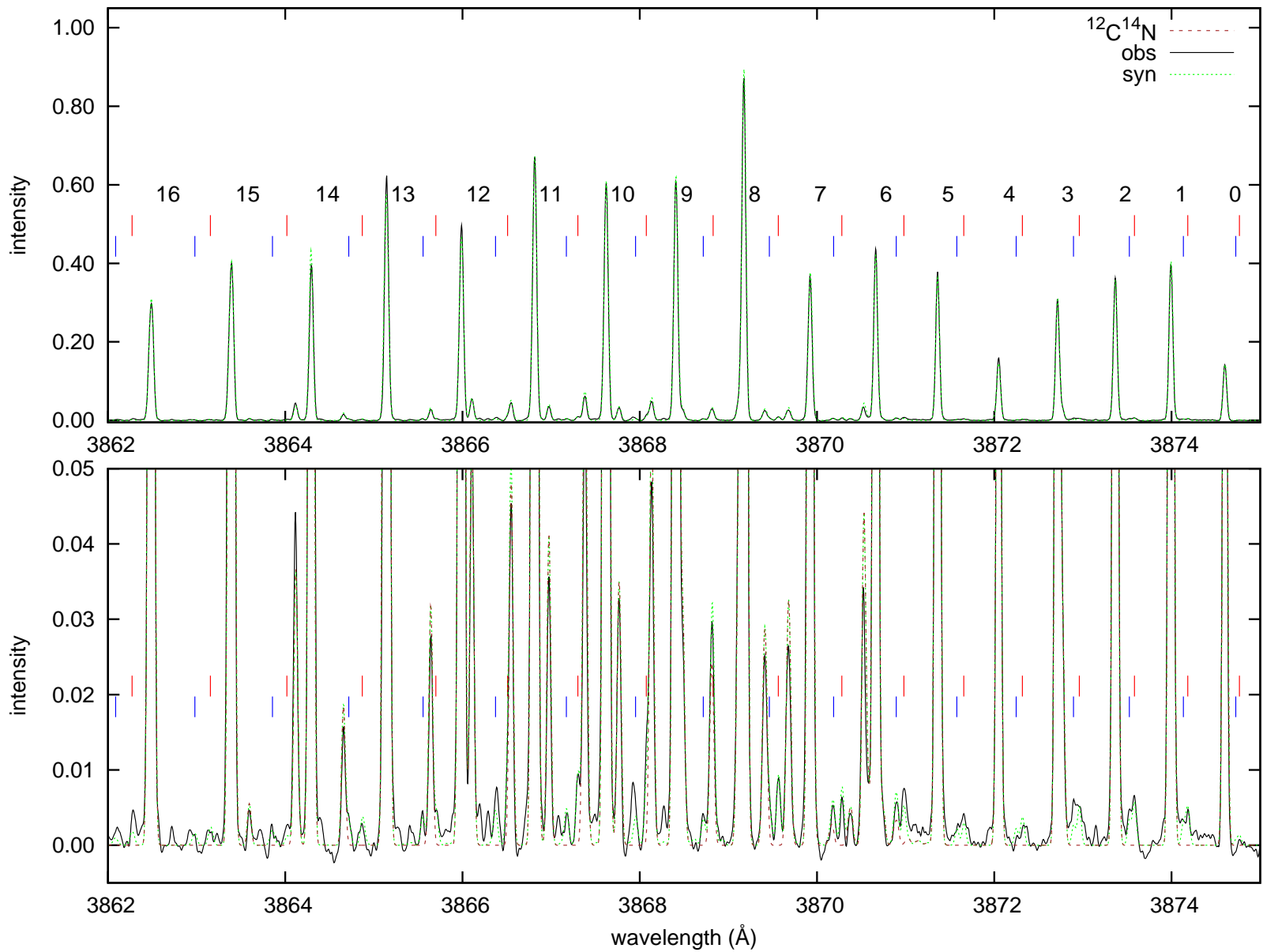}
      \caption{Observed (UVES) and synthetic (dotted) spectra of comet C/2002 Y1 (Juels-Holvorcem)
             }
         \label{ploty1}
  \end{figure*}
  \clearpage
  \begin{figure*}
   \includegraphics[width=15.5cm]{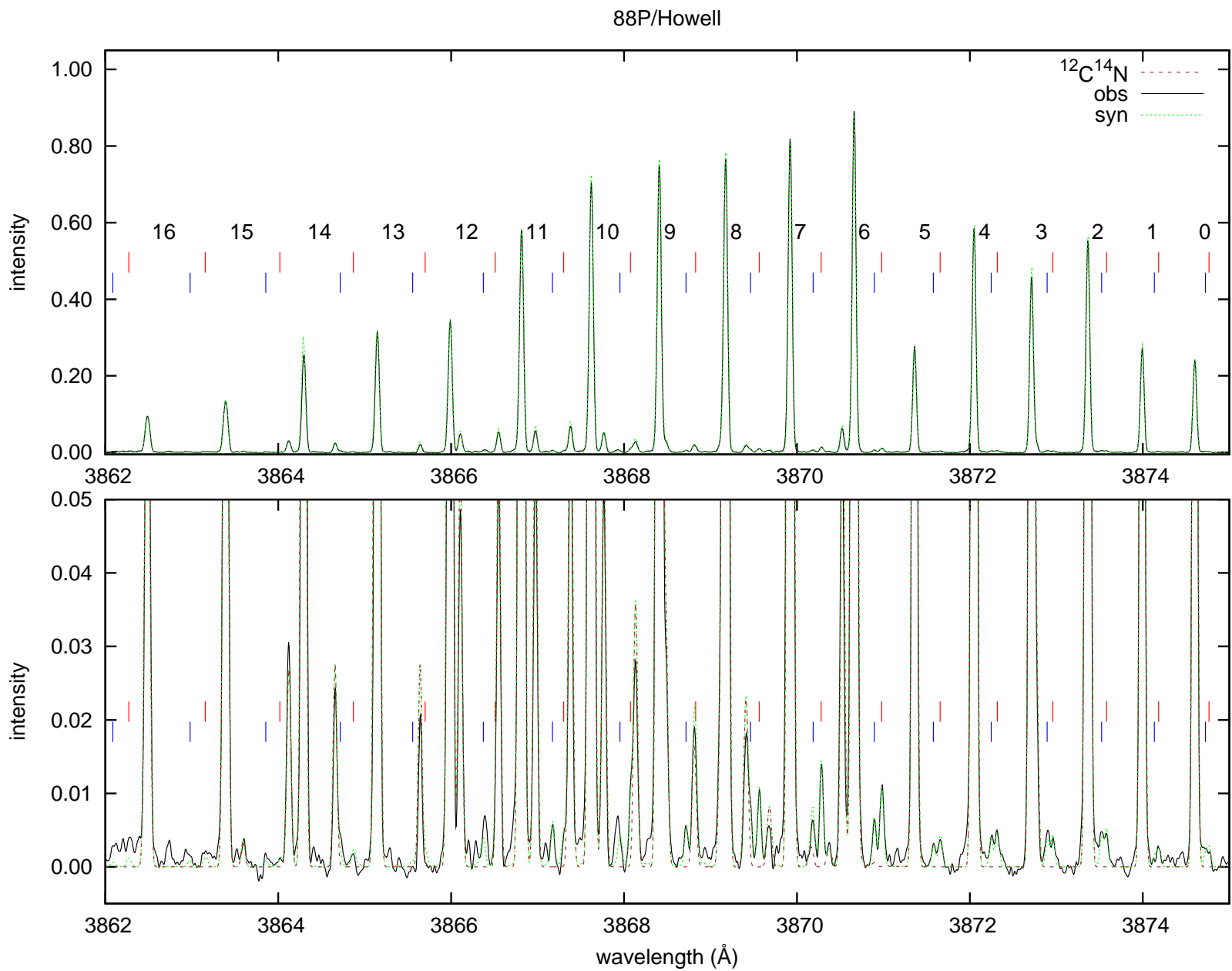}
      \caption{Observed (UVES) and synthetic (dotted) spectra of comet 88P/Howell  
             }
         \label{plothowell}
  \end{figure*}
  \clearpage
  \begin{figure*}
   \includegraphics[width=15.5cm]{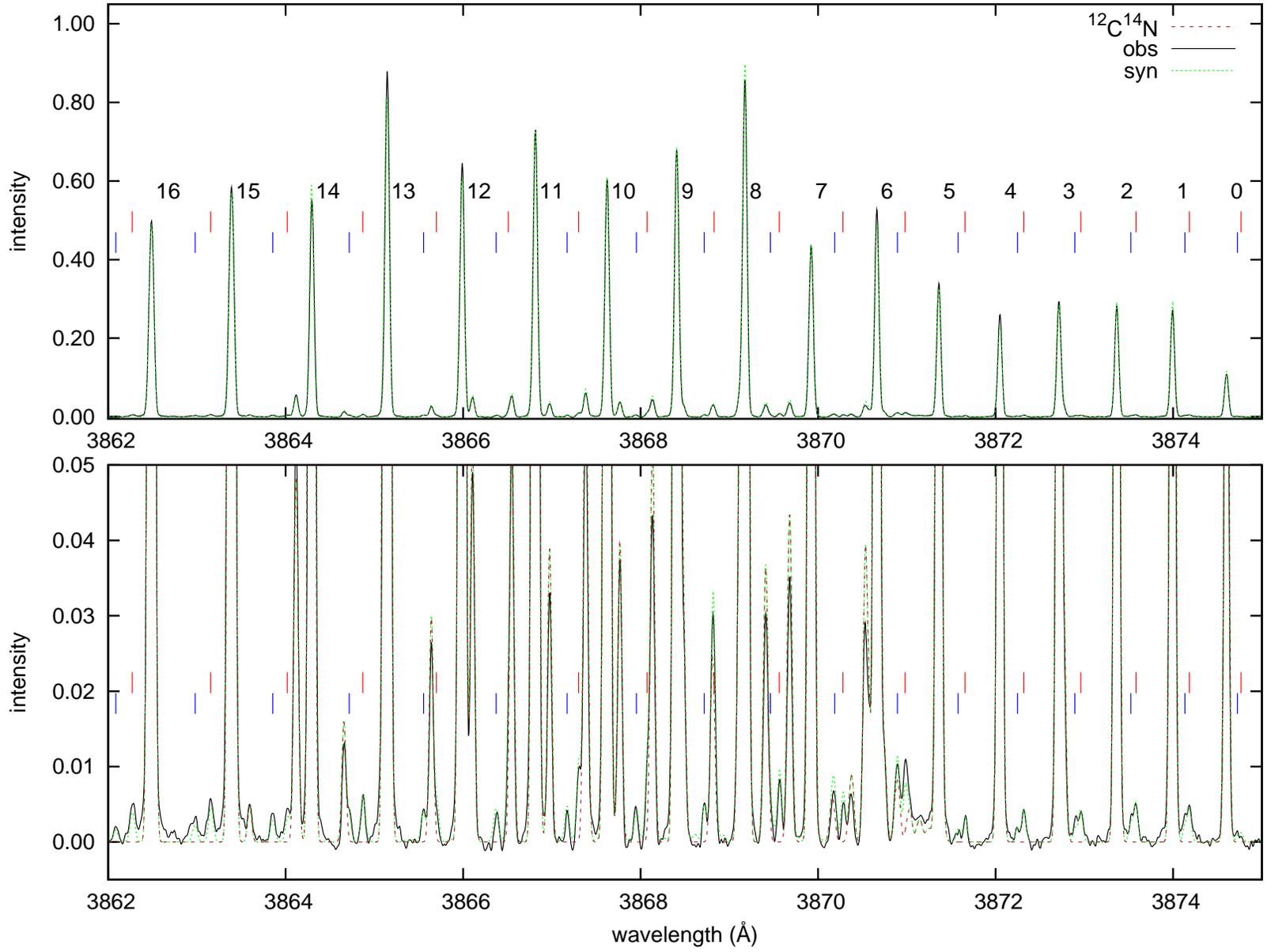}
      \caption{Observed (UVES) and synthetic (dotted) spectra of comet C/2002 T7 (LINEAR)
             }
         \label{plott7}
  \end{figure*}
  \clearpage
  \begin{figure*}
   \includegraphics[width=15.5cm]{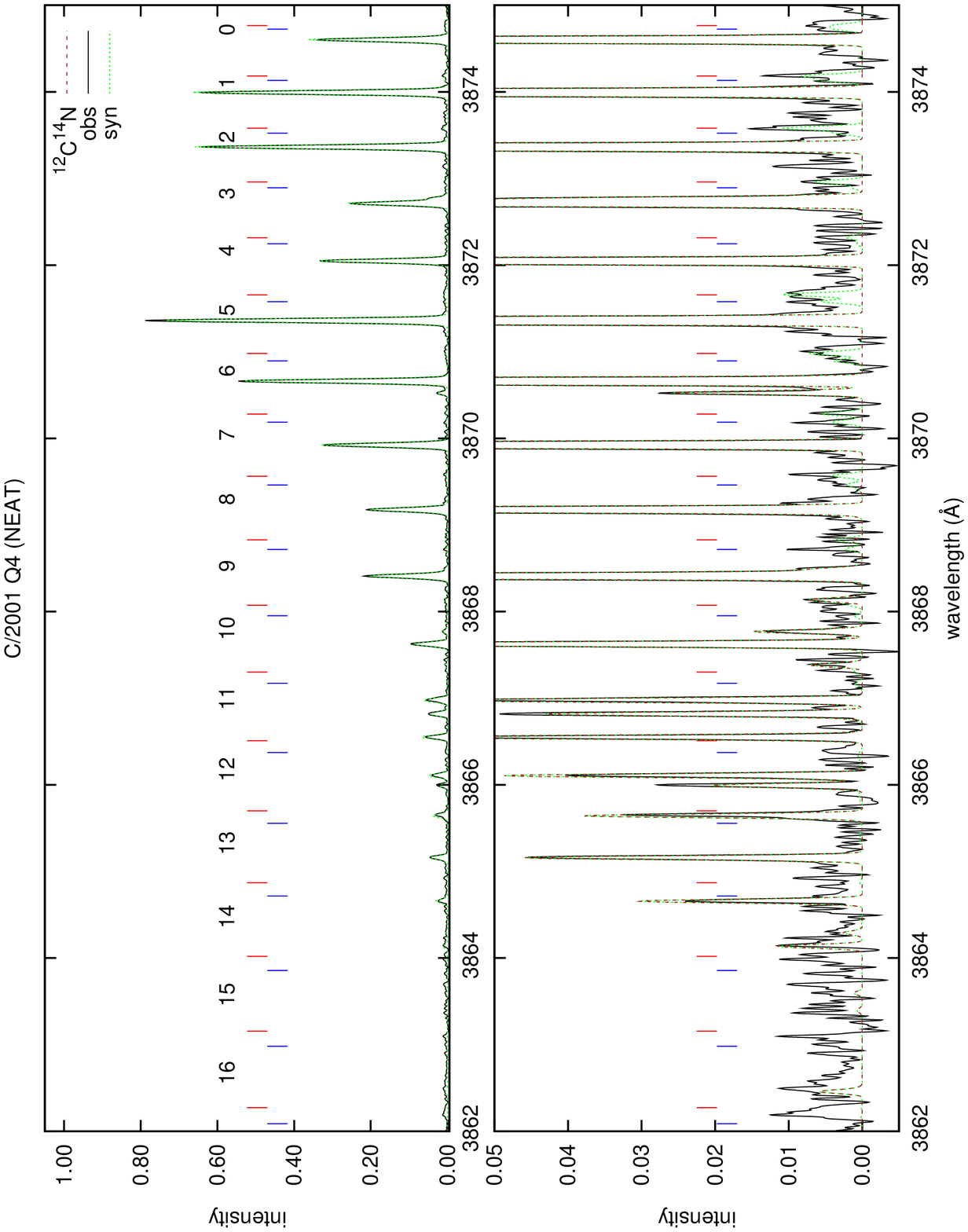}
      \caption{Observed (UVES) and synthetic (dotted) spectra of comet C/2001 Q4 (NEAT)
             }
         \label{plotq4}
  \end{figure*}
  \clearpage
  \begin{figure*}
   \includegraphics[width=15.5cm]{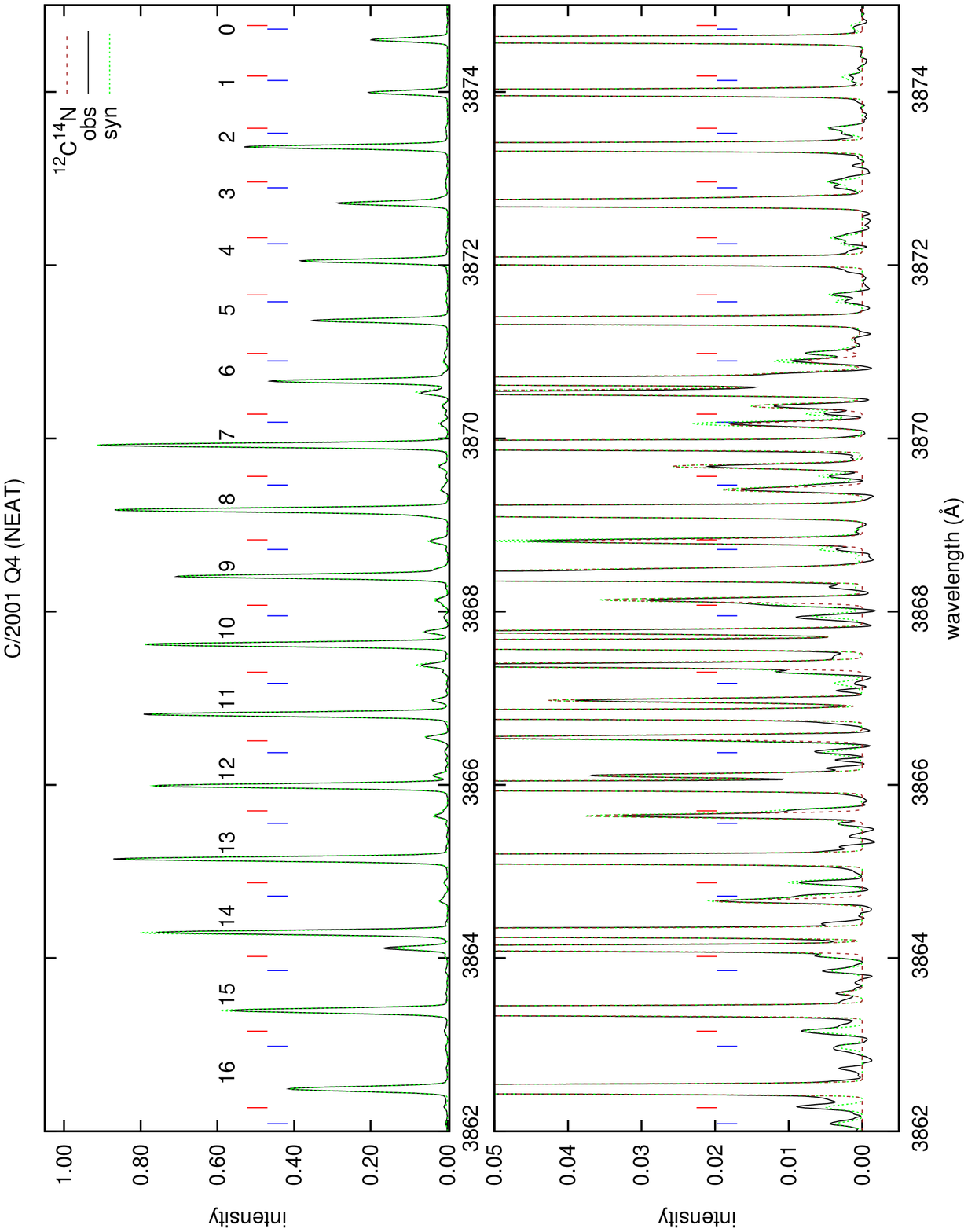}
      \caption{Observed (UVES) and synthetic (dotted) spectra of comet C/2001 Q4 (NEAT)
             }
         \label{plotq4p}
  \end{figure*}
  \clearpage
  \begin{figure*}
   \includegraphics[width=15.5cm]{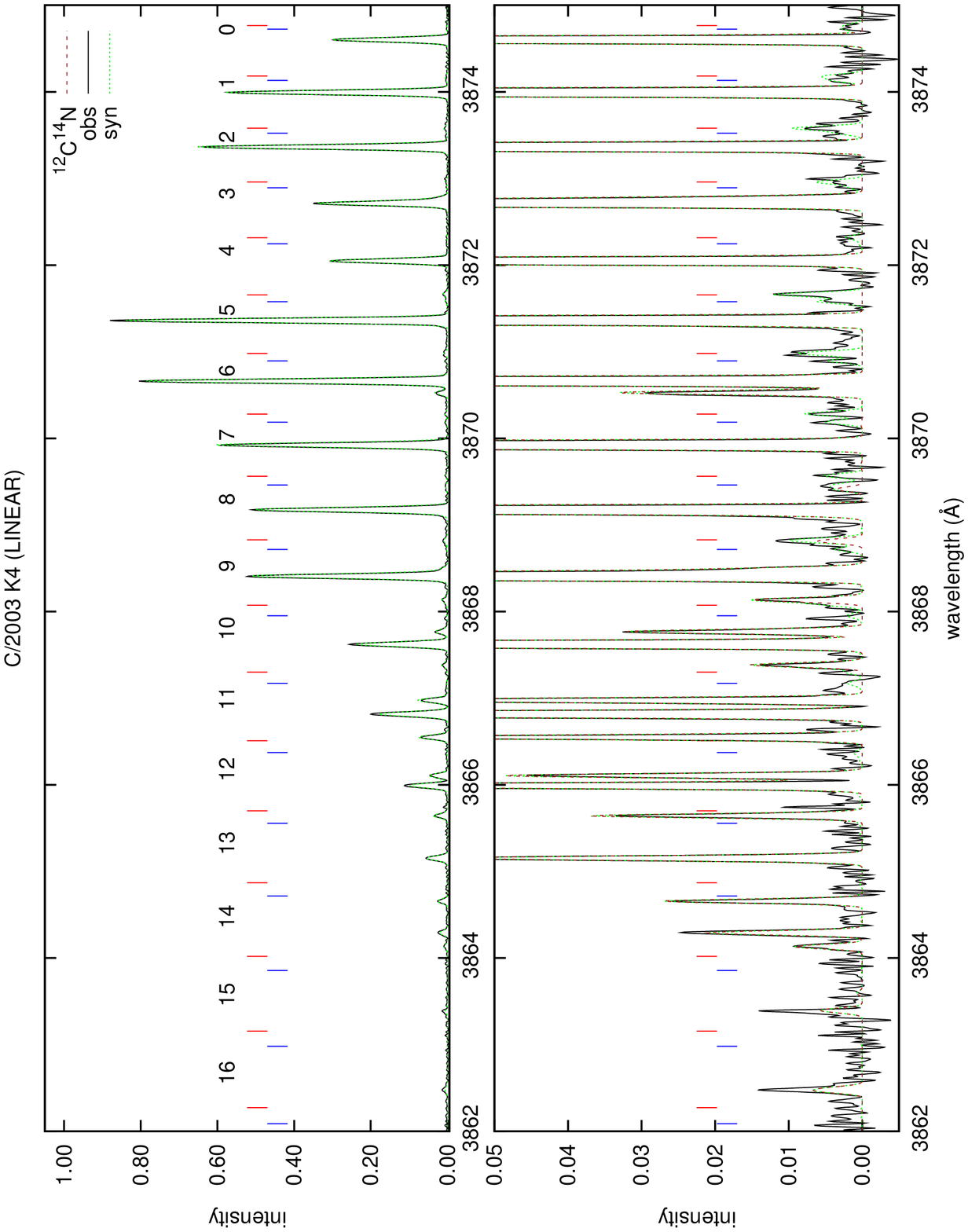}
      \caption{Observed (UVES) and synthetic (dotted) spectra of comet C/2003 K4 (LINEAR)
             }
         \label{plotk4}
  \end{figure*}
  \clearpage
  \begin{figure*}
   \includegraphics[width=15.5cm]{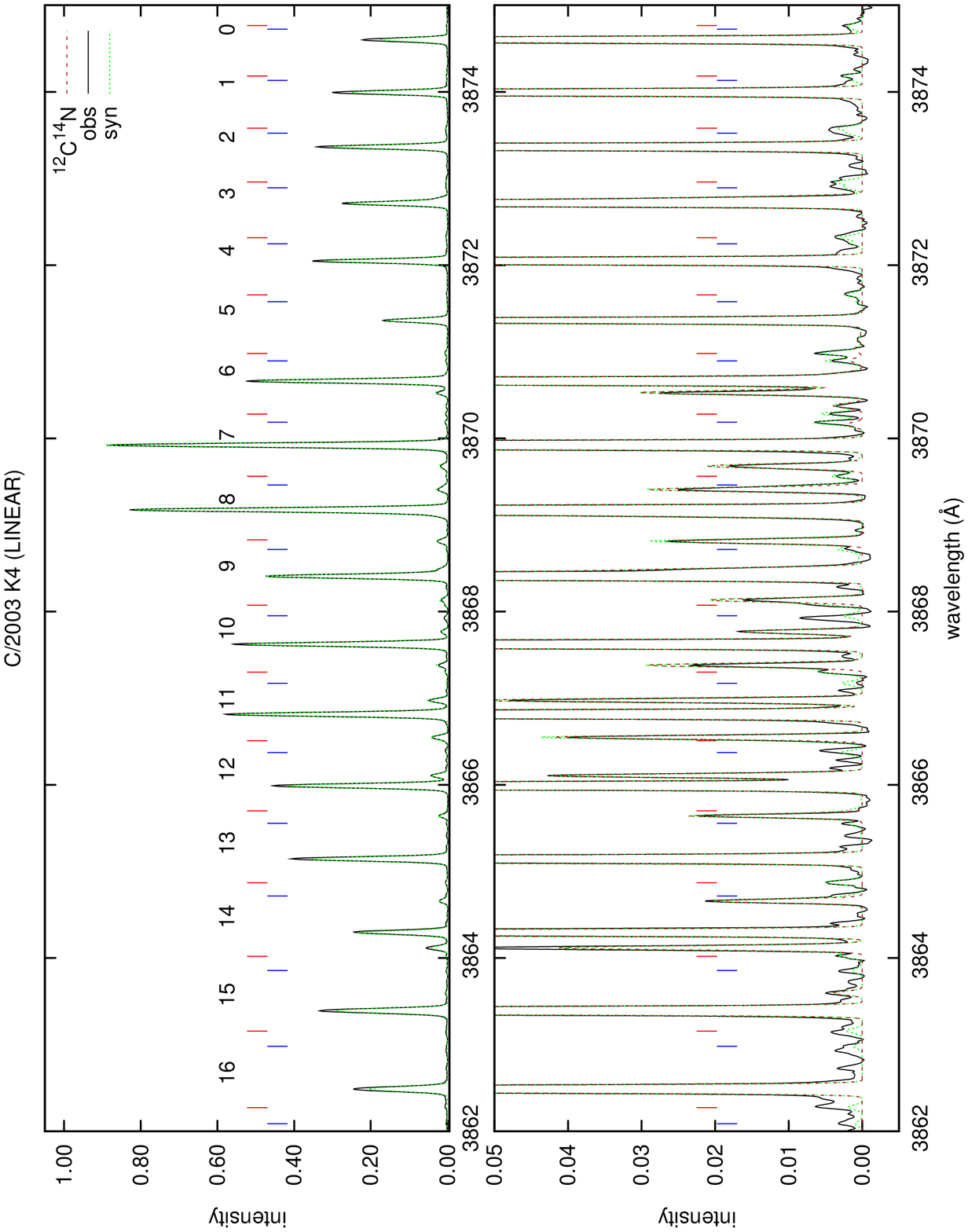}
      \caption{Observed (UVES) and synthetic (dotted) spectra of comet C/2003 K4 (LINEAR)
             }
         \label{plotk4p}
  \end{figure*}
  \clearpage
  \begin{figure*}
   \includegraphics[width=15.5cm]{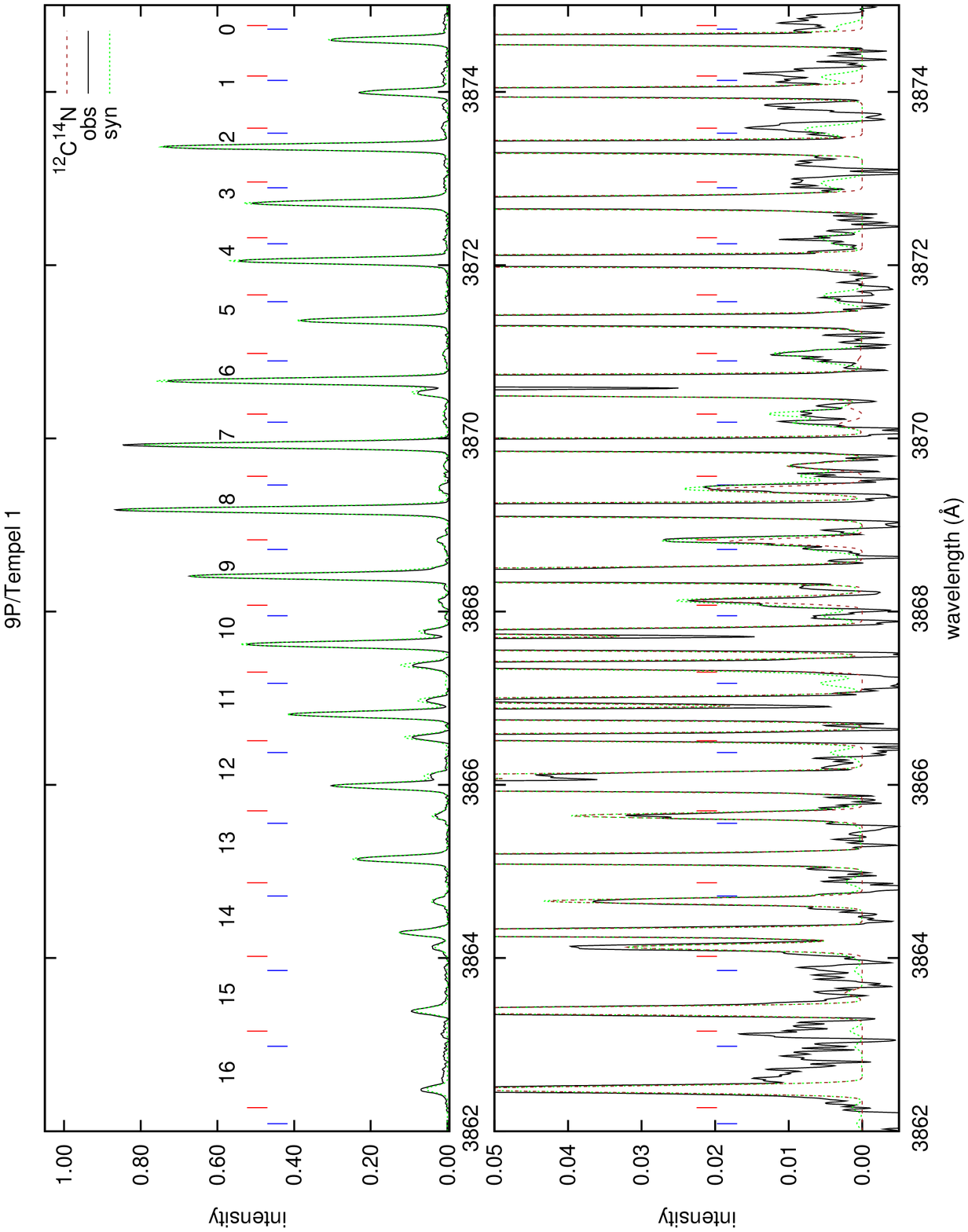}
      \caption{Observed (HIRES) and synthetic (dotted) spectra of comet 9P/Tempel 1 
             }
         \label{plotKeck}
  \end{figure*}
  \clearpage
  \begin{figure*}
   \includegraphics[width=15.5cm]{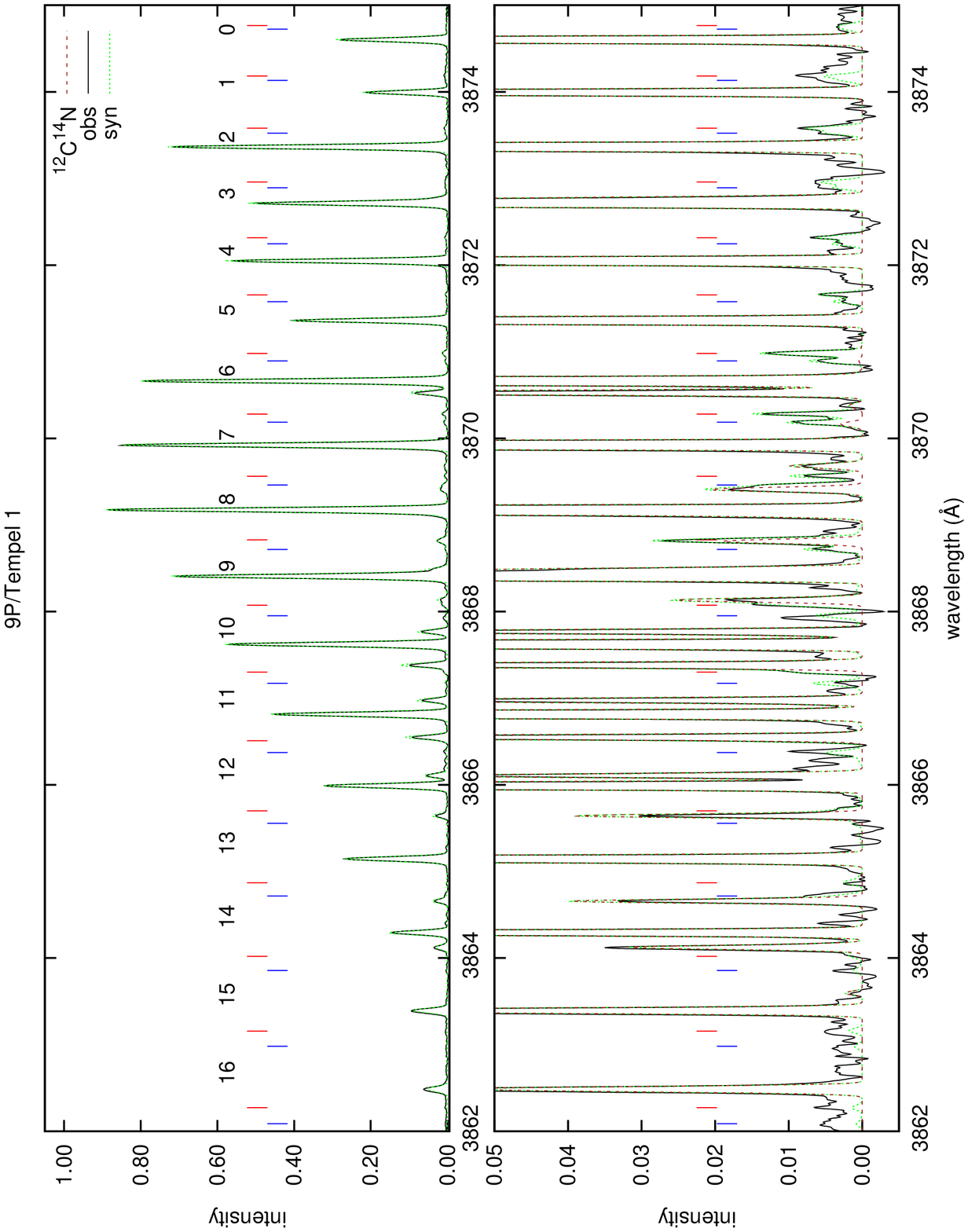}
      \caption{Observed (UVES) and synthetic (dotted) spectra of comet 9P/Tempel 1 
             }
         \label{plottempel1}
  \end{figure*}
  \clearpage
  \begin{figure*}
   \includegraphics[width=15.5cm]{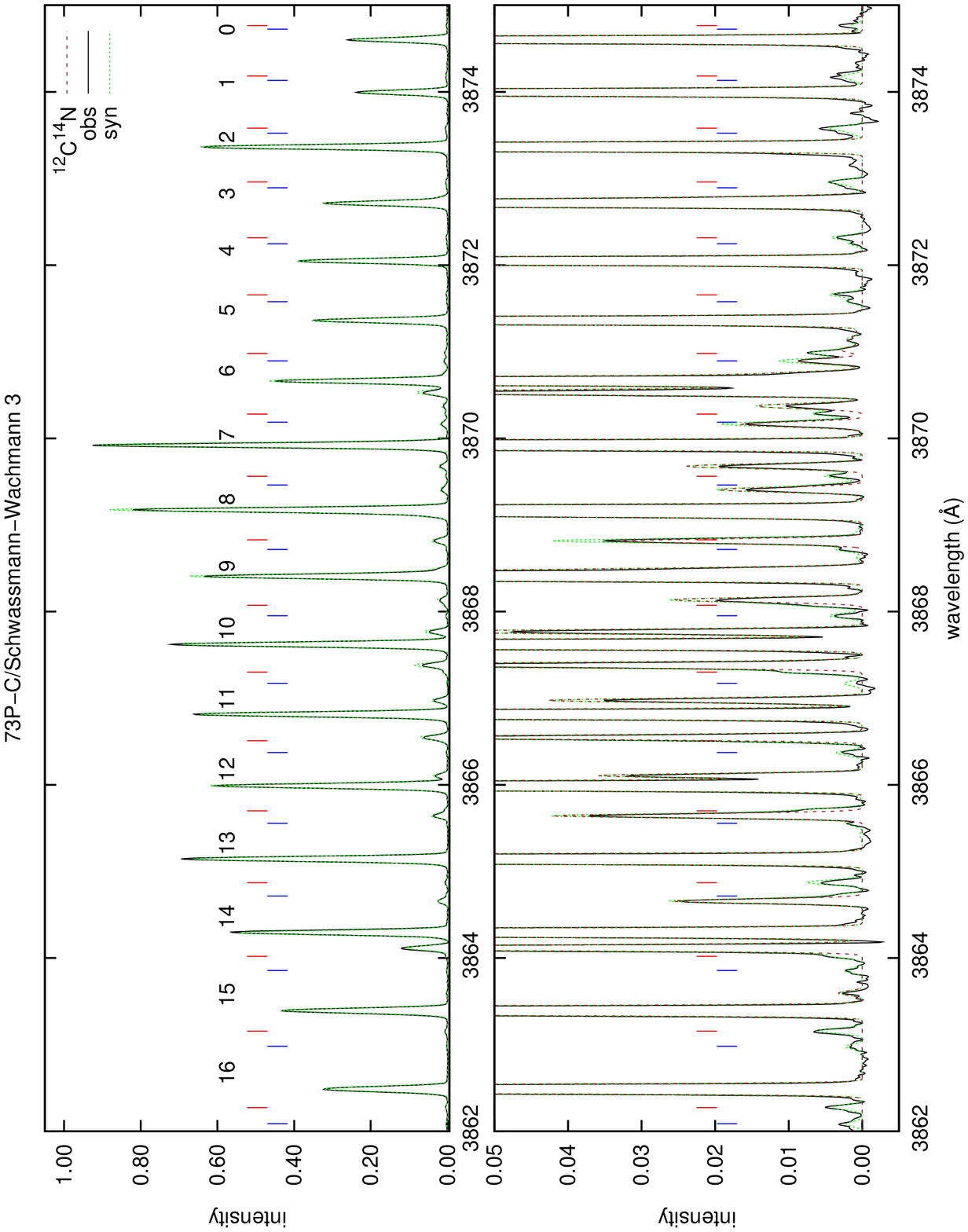}
      \caption{Observed (UVES) and synthetic (dotted) spectra of comet 73P-B/Schwassmann-Wachmann 3 
             }
         \label{plotsw3C}
  \end{figure*}
  \clearpage
  \begin{figure*}
   \includegraphics[width=15.5cm]{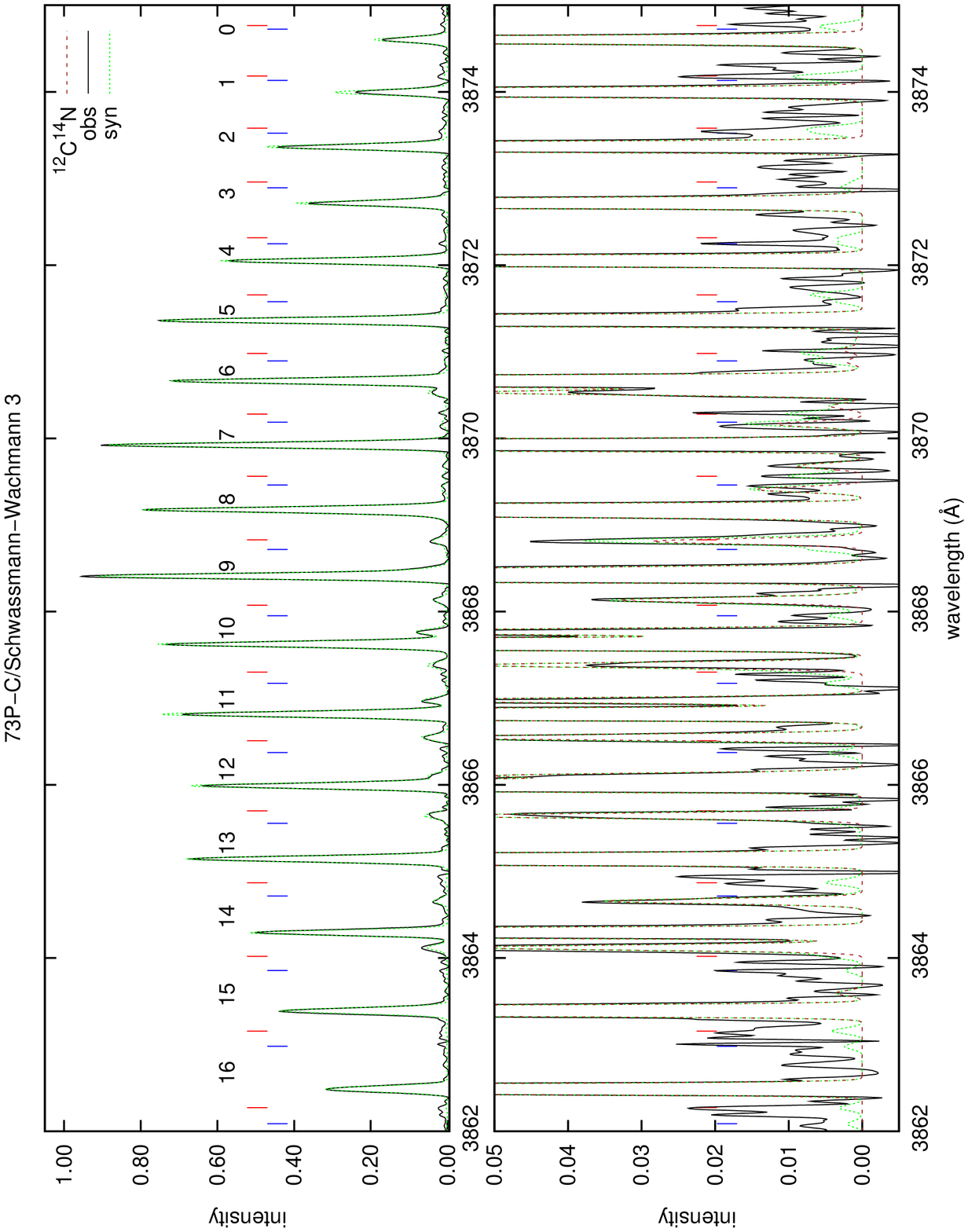}
      \caption{Observed (2DCoud{\'e}) and synthetic (dotted) spectra of comet 73P-B/Schwassmann-Wachmann 3 
             }
         \label{plotsw3MDC}
  \end{figure*}
  \clearpage
  \begin{figure*}
   \includegraphics[width=15.5cm]{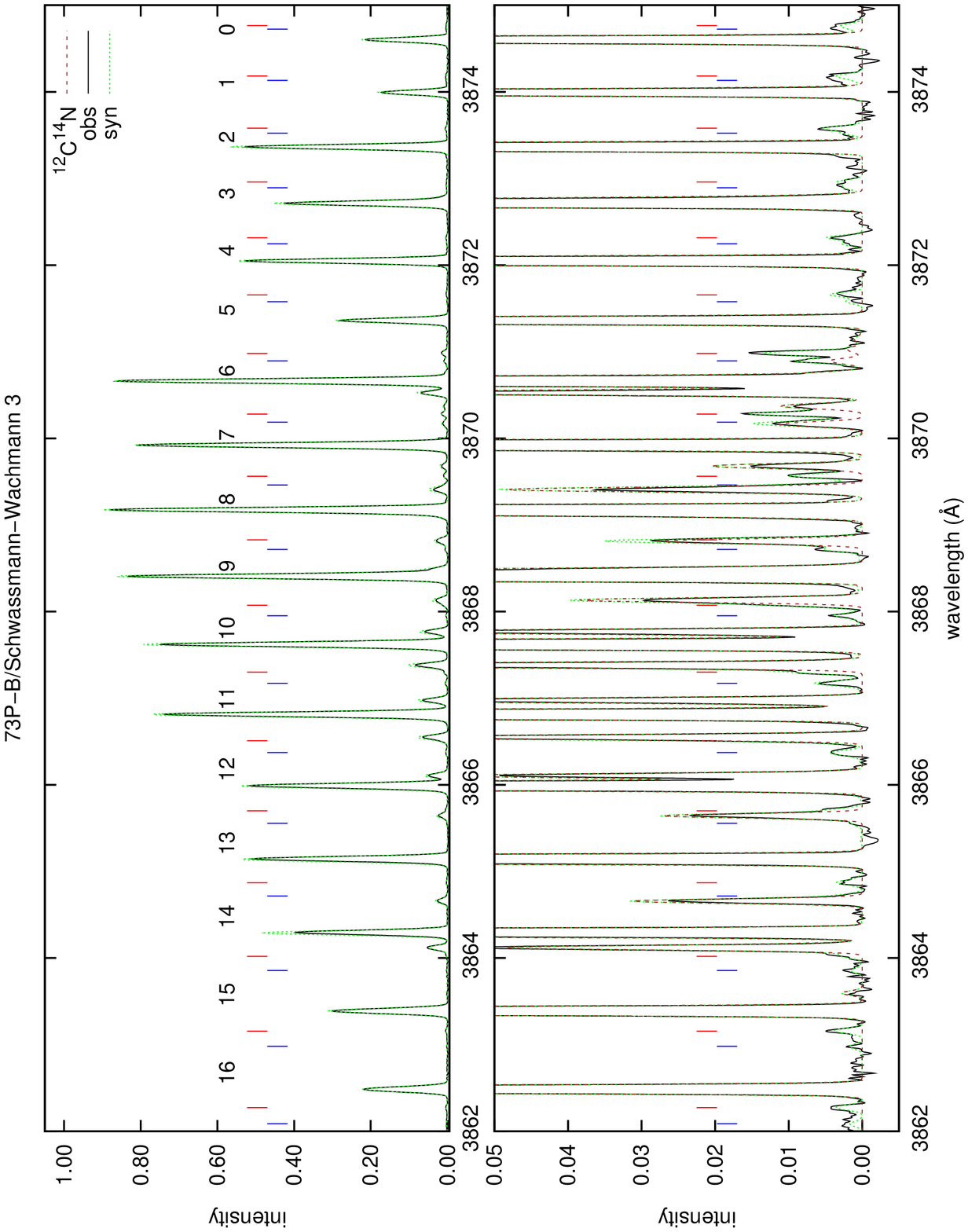}
      \caption{Observed (UVES) and synthetic (dotted) spectra of comet 73P-C/Schwassmann-Wachmann 3 
             }
         \label{plotsw3B}
  \end{figure*}
  \clearpage
  \begin{figure*}
   \includegraphics[width=15.5cm]{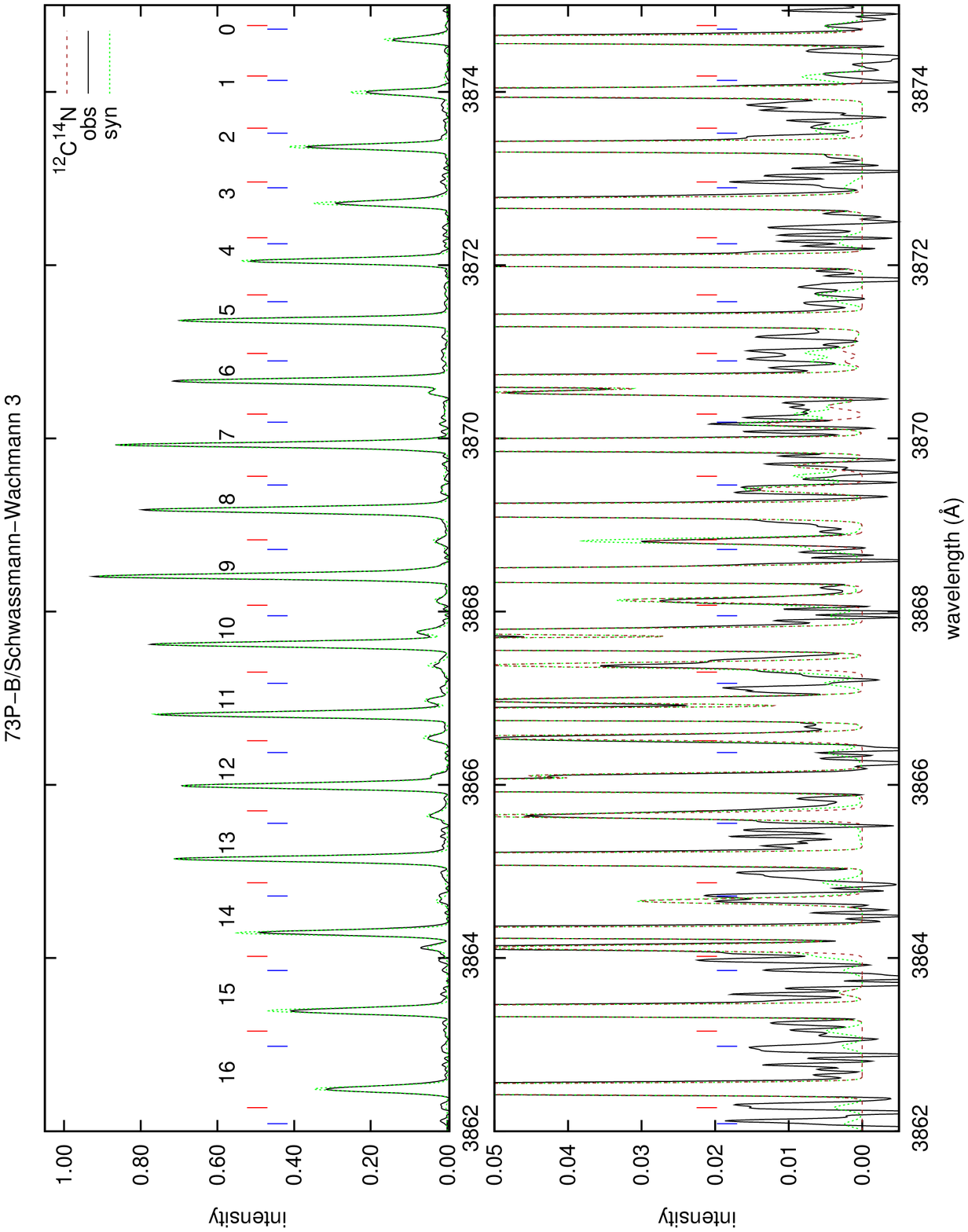}
      \caption{Observed (2DCoud{\'e}) and synthetic (dotted) spectra of comet 73P-C/Schwassmann-Wachmann 3 
             }
         \label{plotsw3MDB}
  \end{figure*}
  \clearpage
  \begin{figure*}
   \includegraphics[width=15.5cm]{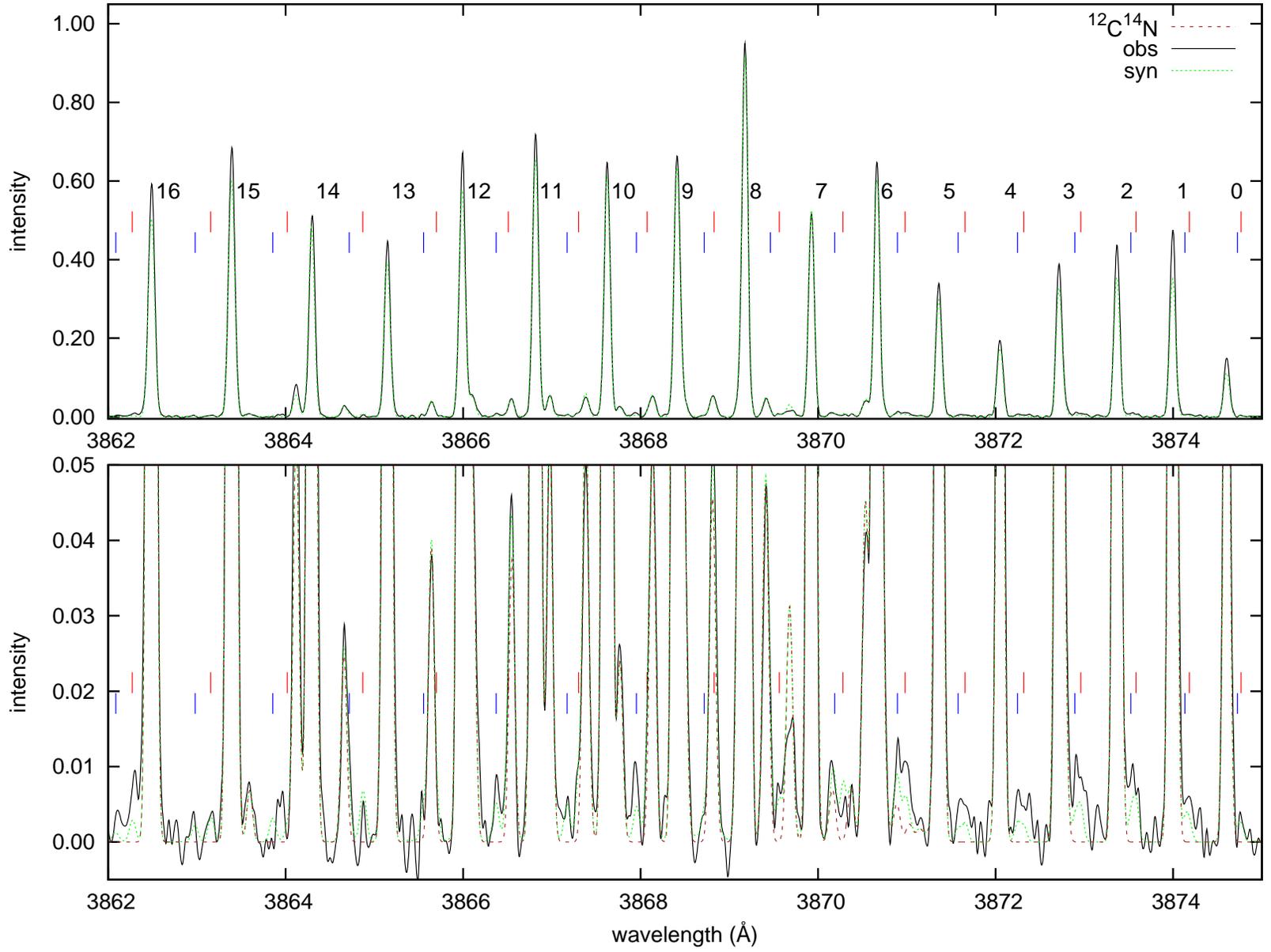}
      \caption{Observed (2DCoud{\'e}) and synthetic (dotted) spectra of comet C/2006 M4 (SWAN)
             }
         \label{plotswan}
  \end{figure*}
  \clearpage
  \begin{figure*}
   \includegraphics[width=15.5cm]{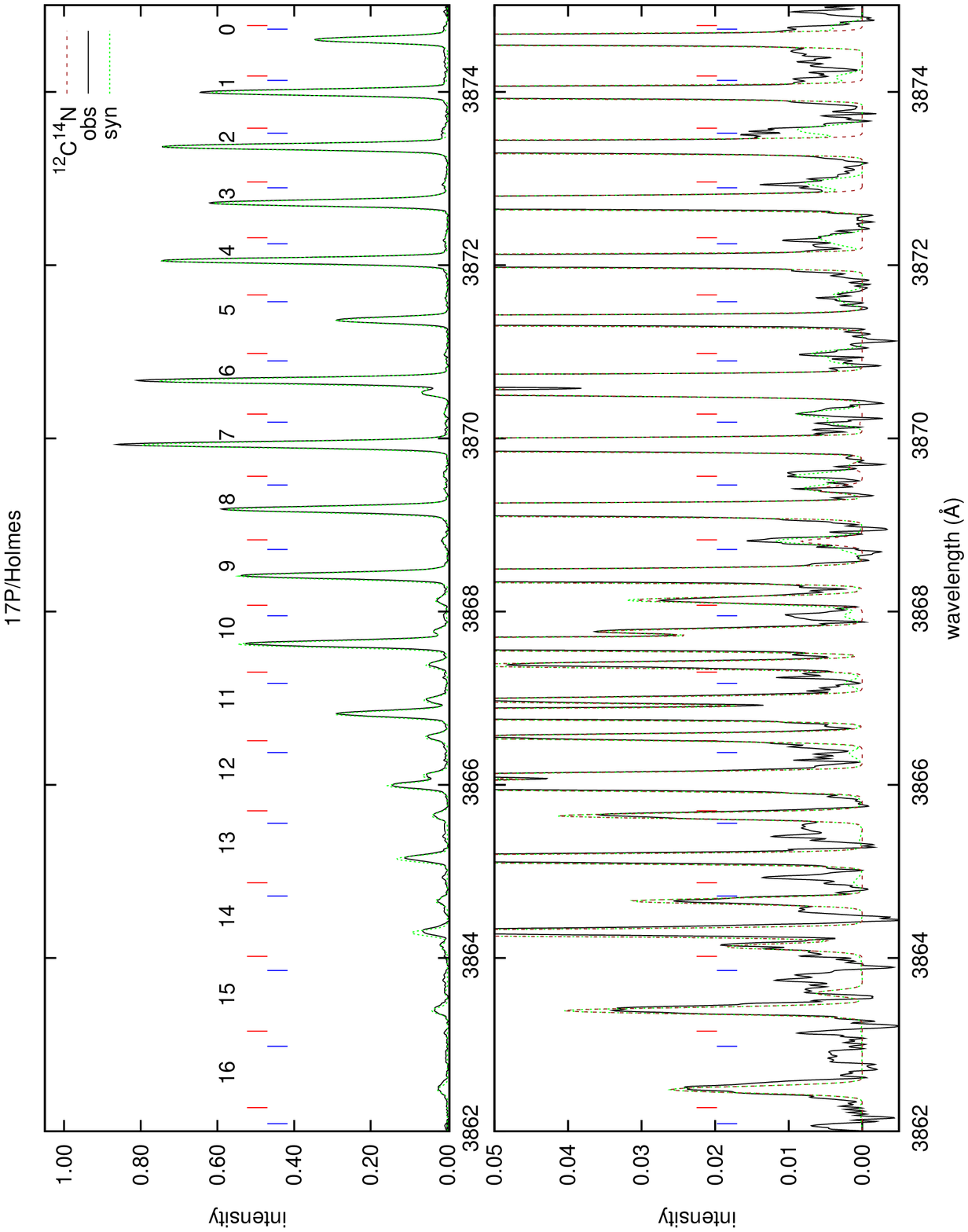}
      \caption{Observed (HIRES) and synthetic (dotted) spectra of comet 17P/Holmes  
             }
         \label{plotholmesKeck}
  \end{figure*}
  \clearpage
  \begin{figure*}
   \includegraphics[width=15.5cm]{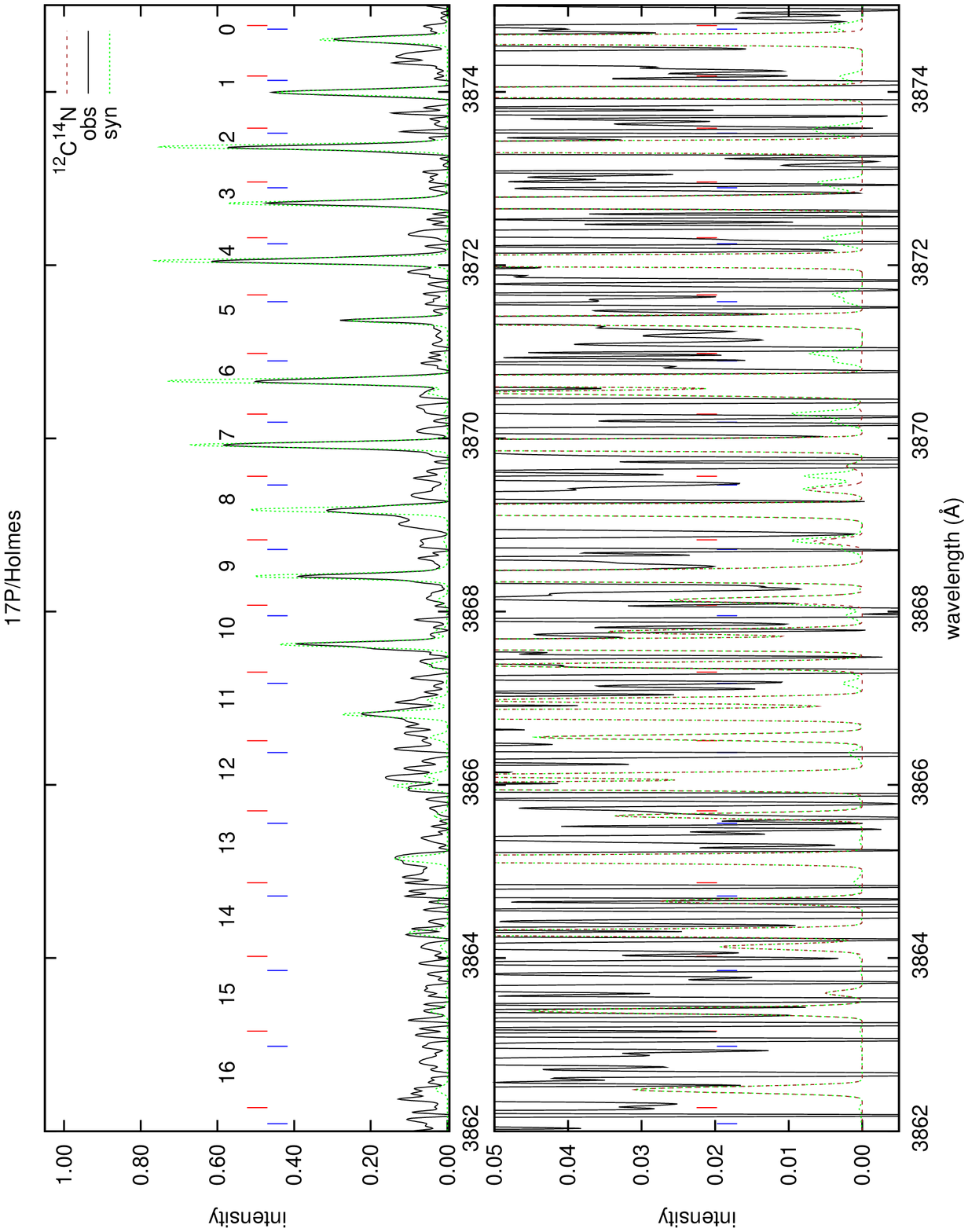}
      \caption{Observed (2DCoud{\'e}) and synthetic (dotted) spectra of comet 17P/Holmes  
             }
         \label{plotholmesMD}
  \end{figure*}
  \clearpage
  \begin{figure*}
   \includegraphics[width=15.5cm]{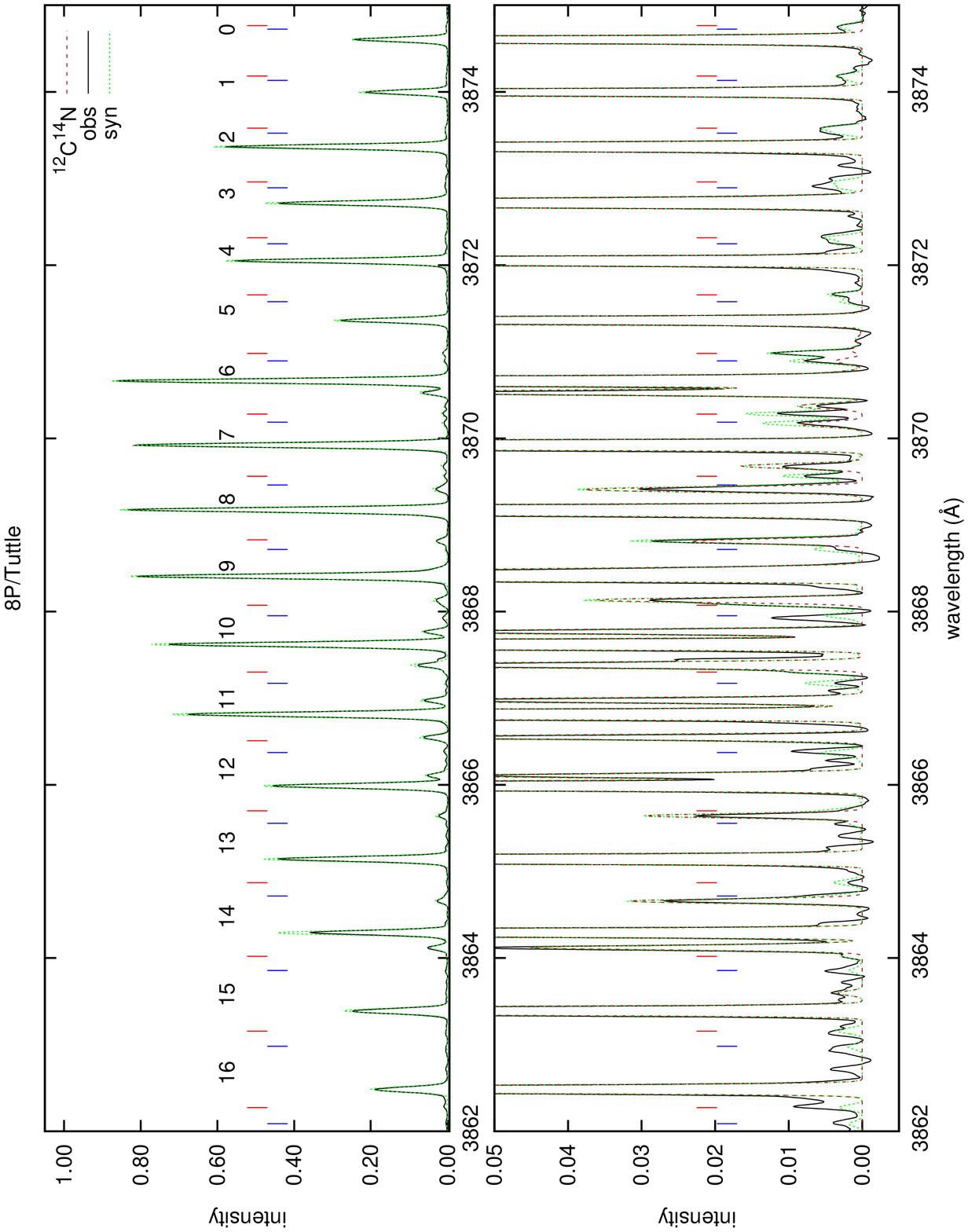}
      \caption{Observed (2DCoud{\'e}) and synthetic (dotted) spectra of comet 8P/Tuttle  
             }
         \label{plottuttle}
  \end{figure*}
  \clearpage
  \begin{figure*}
   \includegraphics[width=15.5cm]{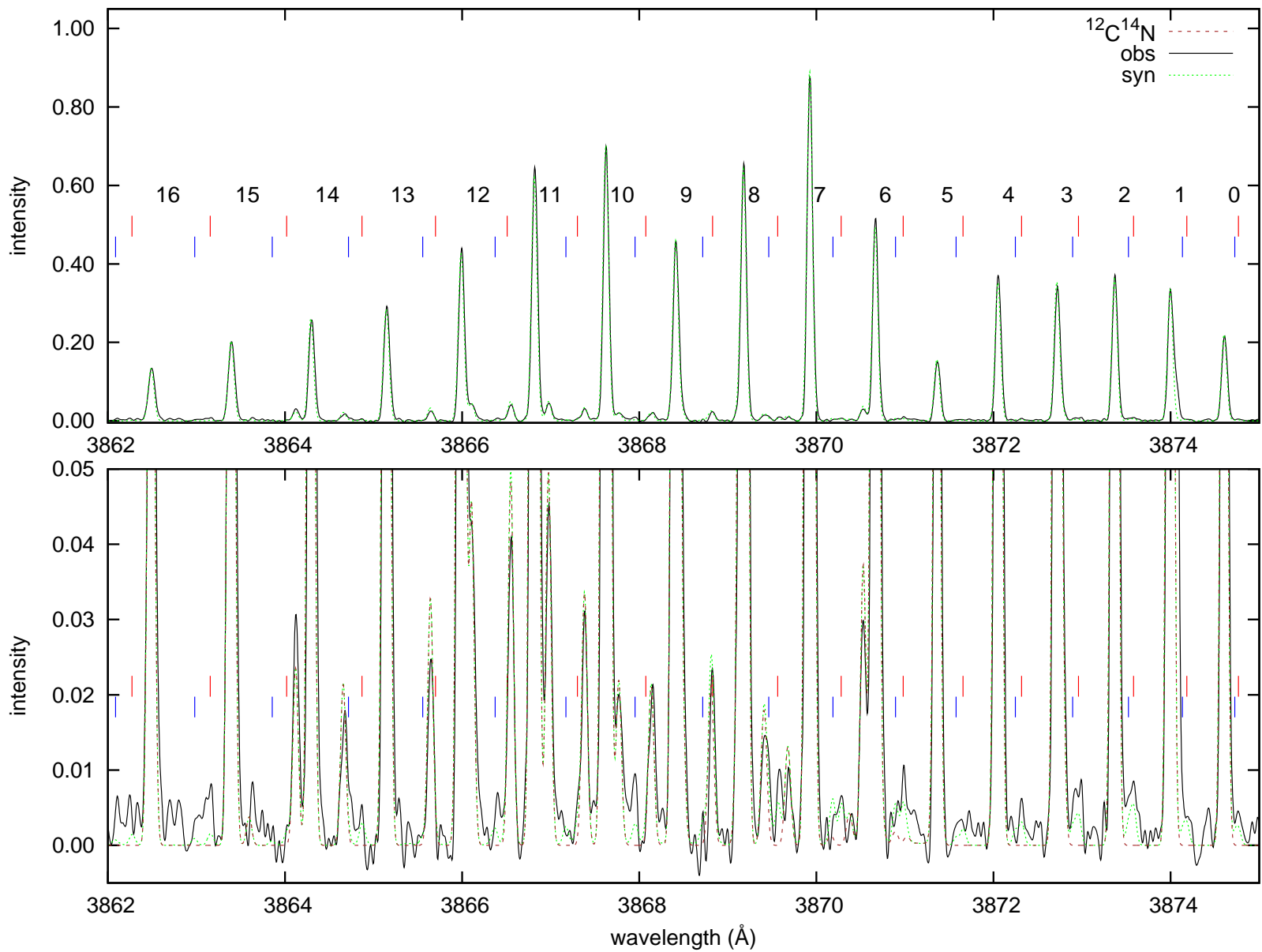}
      \caption{Observed (2DCoud{\'e}) and synthetic (dotted) spectra of comet C/2007 N3 (Lulin)
             }
         \label{plotlulin}
  \end{figure*}
\end{appendix}
\clearpage
\begin{appendix}
\section{Synthetic spectra}
\label{sec:synth}
\begin{figure}
   \resizebox{\hsize}{!}{\includegraphics{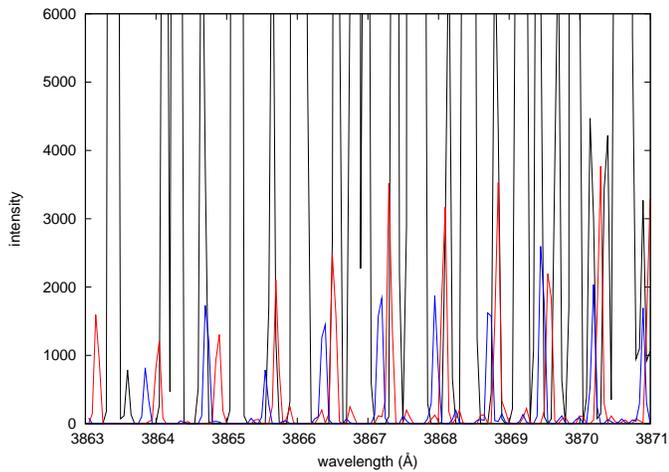}}
\caption{Small region of the B-X 0-0 band of the three CN isotopologues. 
The synthetic spectra were computed using isotopic abundances of 1, 1/89 and 1/145 for {$^{12}$C$^{14}$N}{}\ (black), 
{$^{13}$C$^{14}$N}{}\ (red) and {$^{12}$C$^{15}$N}{}\ (blue), 
respectively, in the absence of collisional effects,
and with $r=1$~AU, $\dot{r}=0$. The Gaussian profile has FWHM$=0.05$~\AA.
The intensity scale is arbitrary{, but coherent with the three following plots.}}
\label{plotCN3863_3871}
\end{figure}
\begin{figure} 
   \resizebox{\hsize}{!}{\includegraphics{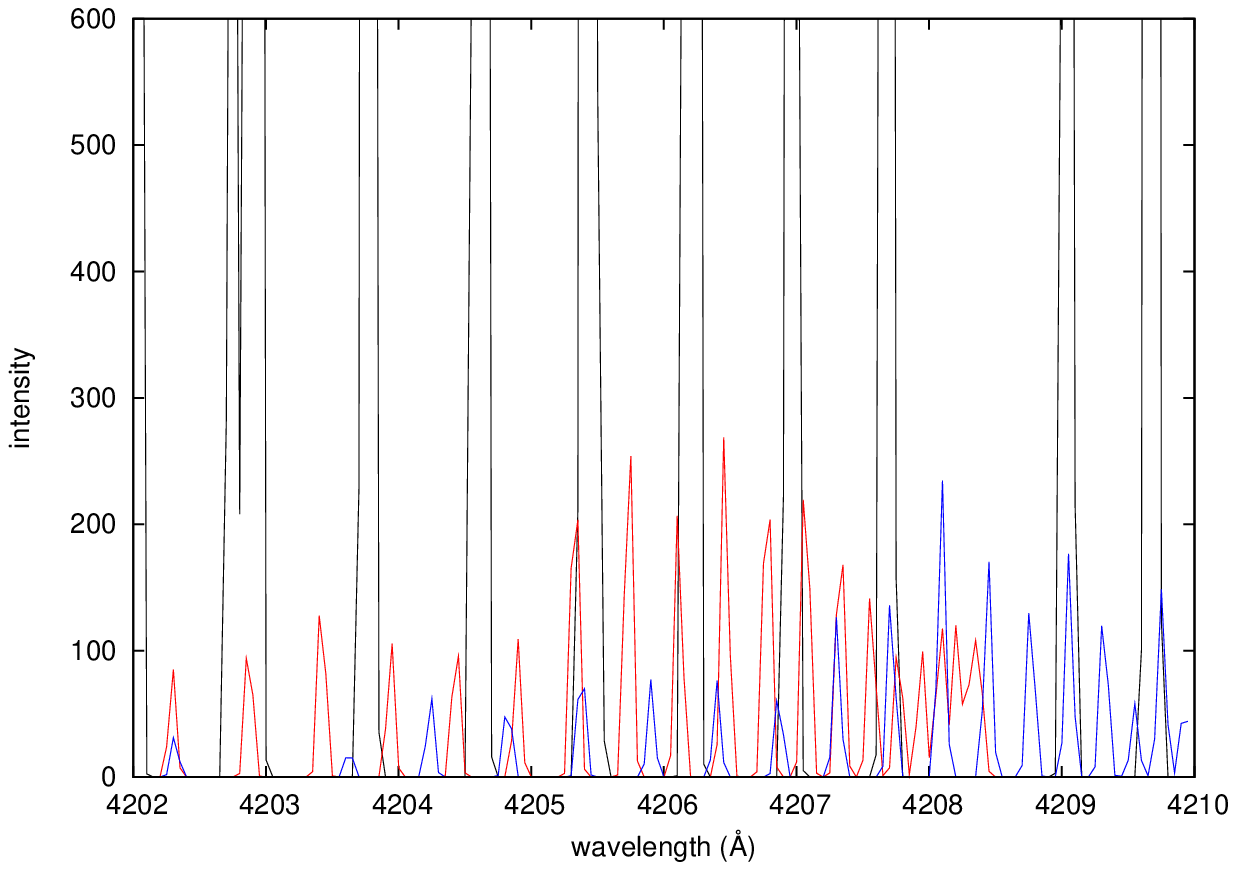}}
\caption{Same as Fig.~\ref{plotCN3863_3871} for a region of the B-X 0-1 band.}
\label{plotCN4202_4210}
\end{figure}
\begin{figure} 
   \resizebox{\hsize}{!}{\includegraphics{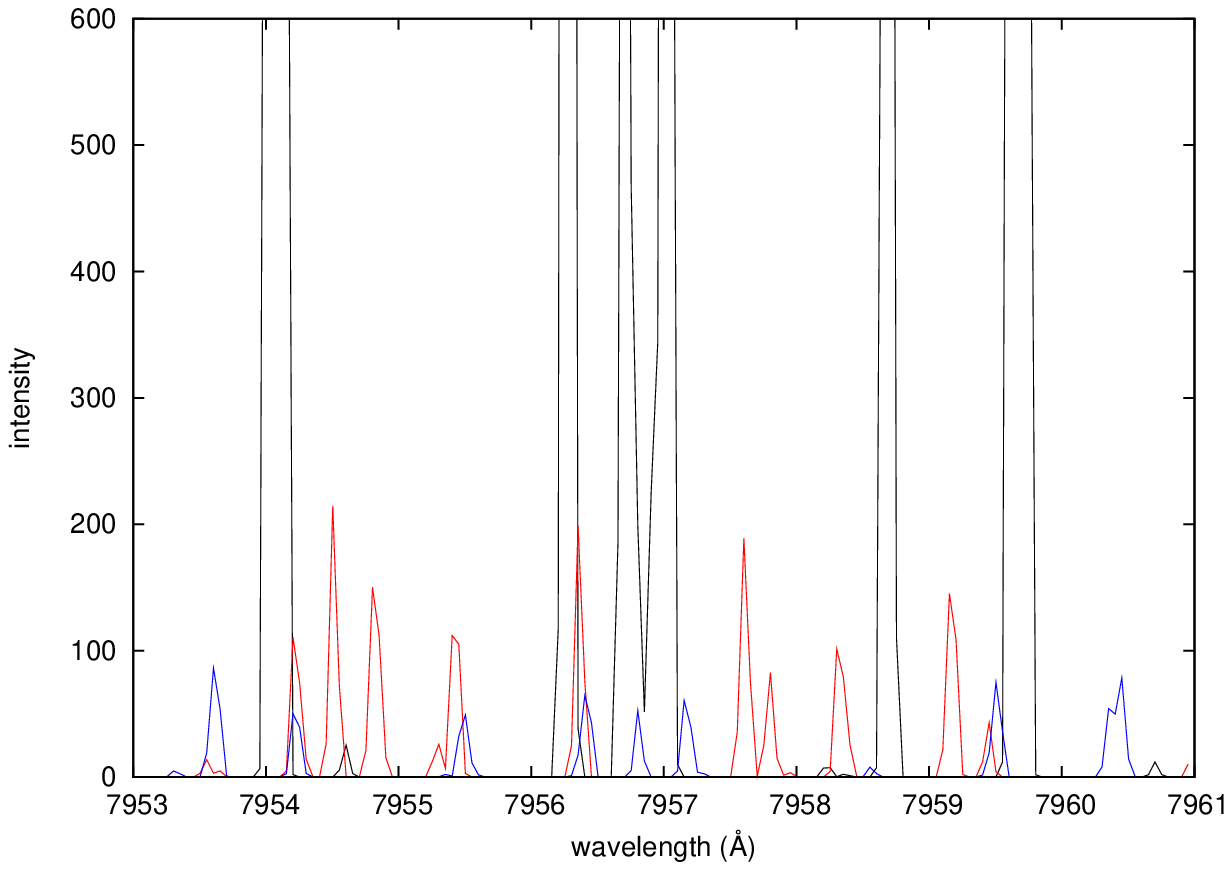}}
\caption{Same as Fig.~\ref{plotCN3863_3871} for a region of the A-X 2-0 band.}
\label{plotCN7953_7961}
\end{figure}
\begin{figure} 
   \resizebox{\hsize}{!}{\includegraphics{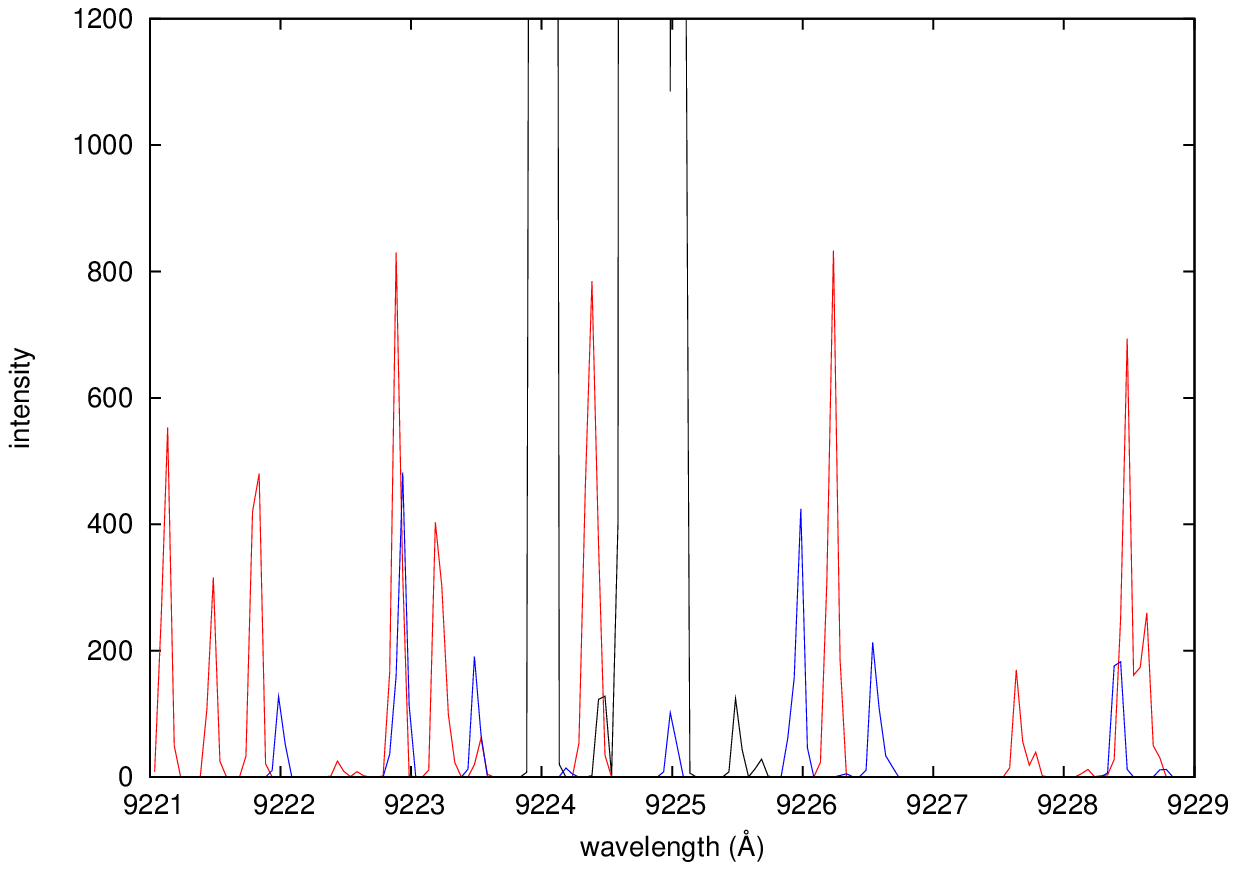}}
\caption{Same as Fig.~\ref{plotCN3863_3871} for a region of the A-X 1-0 band.}
\label{plotCN9221_9229}
\end{figure}
\end{appendix}
\end{document}